\documentclass[aps,prd,preprint,superscriptaddress,tightenlines,nofootinbib,floatfix]{revtex4}
\usepackage{amssymb}
\usepackage{bm}
\usepackage{graphicx}
\usepackage{amsfonts}
\usepackage[mathscr]{eucal}
\usepackage{latexsym}
\usepackage{epsfig}
\DeclareGraphicsExtensions{eps,ps}

\newcommand{\postscript}[2]{\setlength{\epsfxsize}{#2\hsize}
   \centerline{\epsfbox{#1}}}
\newcommand{\beq}{\begin{equation}}
\newcommand{\eeq}{\end{equation}}
\newcommand{\eps}{\epsilon}
\newcommand{\En}{E_n}
\newcommand{\Fn}{F_n}
\newcommand{\Enu}{E_{\overline{\nu}}}

\newcommand{\Enubar}{\epsilon_{\overline{\nu}}}
\newcommand{\thnu}{\overline{\theta}_{\overline{\nu}}}
\newcommand{\cth}{\cos\overline{\theta}_{\overline{\nu}}}
\newcommand{\Fnu}{F_{\overline{\nu}}}
\newcommand{\tbar}{\overline{\tau}_n}

\newcommand{\nuebar}{\overline{\nu}_e}
\newcommand{\nutau}{\nu_\tau}
\newcommand{\nue}{\nu_e}
\newcommand{\numu}{\nu_\mu}
\newcommand{\Enue}{E_{\nu_e}}
\newcommand{\Enumu}{E_{\nu_\mu}}

\newcommand{\Enuj}{E_{\nu_j}}

\newcommand{\Fnue}{F_{\nu_e}}
\newcommand{\Fnumu}{F_{\nu_\mu}}

\newcommand{\Fnuj}{F_{\nu_j}}
\newcommand{\Epho}{E_\gamma}
\newcommand{\Epi}{E_\pi}
\newcommand{\Fpi}{F_\pi}
\newcommand{\Fpho}{F_\gamma}

\newcommand{\gev}{\text{GeV}}

\newcommand{\etal}{{\em et al.}}

\newcommand{\eqref}[1]{Eq.~(\ref{#1})}

\def\emin{E_{\rm min}}
\def\emax{E_{\rm max}}
\def\deg{^{\circ}}

\def\deg{\ifmmode{^{\circ}}\else ${^{\circ}}$\fi}
\def\bi{\begin{itemize}}
\def\ei{\end{itemize}}
\def\ed{\end{document}}

\def\pri{^{\, \prime}}

\def\cf#1{\ifmmode{\cal #1}\else${\cal #1}$\fi}

\def\be{\begin{equation}}
\def\ee{\end{equation}}
\def\beas{\begin{eqnarray*}}
\def\eea{\end{eqnarray}}
\def\bea{\begin{eqnarray}}
\def\eeas{\end{eqnarray*}}

\def\gev{\ifmmode{\mbox{GeV}}\else GeV\fi}
\def\es{{\rm erg\ s}^{-1}}

\def\lf{L_{41}}
\def\ton{T_{\rm on}}

\begin{document}
\baselineskip=15.5pt
\pagestyle{plain}
\setcounter{page}{1}

\vskip -1.3cm

\rightline{\small{\tt NUB-3244/Th-04}}
\rightline{\tt astro-ph/0402371}

\vspace{.5cm}

\begin{center}

{\large  {\bf Astrophysical Origins of Ultrahigh Energy Cosmic
Rays}}

\vskip 1.cm
{Diego F. Torres$^a$ and Luis A. Anchordoqui$^b$}

\medskip
\medskip
{$^a${\it Lawrence Livermore National Laboratory, 7000 East Ave., L-413,\\
Livermore, CA 94550, USA}}

\vspace{.25cm}

{$^b${\it Department of Physics, Northeastern University\\
Boston, MA 02115, USA}}
\vspace{.25cm}

{\tt dtorres@igpp.ucllnl.org}\\
{\tt l.anchordoqui@neu.edu}
\vspace{.5cm}

\end{center}

\setcounter{footnote}{0}

\begin{center}
{\large {\bf Abstract}}
\end{center}

In the first part of this review we discuss the basic
observational features at the end of the cosmic ray energy
spectrum. We also present there the main characteristics of each
of the experiments involved in the detection of these particles.
We then briefly discuss the status of the chemical composition and
the distribution of arrival directions of cosmic rays. After that,
we examine the energy losses during propagation, introducing  the
Greisen-Zaptsepin-Kuzmin (GZK) cutoff, and discuss the level of
confidence with which each  experiment have detected particles
beyond the GZK energy limit. In the second part of the review, we
discuss astrophysical environments able to accelerate particles up
to such high energies, including active galactic nuclei, large
scale galactic wind termination shocks, relativistic jets and
hot-spots of Fanaroff-Riley radiogalaxies, pulsars, magnetars,
quasar remnants, starbursts, colliding galaxies, and gamma ray
burst fireballs. In the third part of the review we provide a
brief summary of scenarios which try to explain the super-GZK
events with the help of new physics beyond the standard model. In
the last section, we give an overview on neutrino telescopes and
existing  limits on the energy spectrum and discuss some of the
prospects for a new (multi-particle) astronomy. Finally, we
outline how extraterrestrial neutrino fluxes can be used to probe
new physics beyond the electroweak scale.

\begin{center}
{\small{\it  Solicited Review Article Prepared for Reports on
Progress in Physics}}
\end{center}

\noindent

\newpage

\begin{center}
{\large {\bf PREFACE}}
\end{center}

\vspace{2cm}

Reviewing cosmic ray physics is a risky business. Theoretical models
continue to appear at an amazing rate, both in the astrophysical
and more exotic domains. We warn the reader: we do not (nor we
could) intend to make an homogeneous coverage of all the ideas of
our fellow colleagues. Reading differently focused reviews on the issue
(quoted here along the way) is, in our view, the best approach to such a wide
topic of research.

\newpage

\tableofcontents

\newpage

\section{There's Something About Cosmic Ray Observations}

\subsection{Experiments and future projects}

The cosmic ray (CR) spectrum spans over roughly 11 decades of energy.
Continuously
running monitoring using sophisticated equipment on high altitude balloons
and ingenious installations on the Earth's surface
encompass a plummeting flux that goes down from $10^4$ m$^{-2}$ s$^{-1}$
at $\sim 10^9$~eV to $10^{-2}$ km$^{-2}$ yr$^{-1}$ at $\sim 10^{20}$~eV.
Its shape is remarkably featureless, with little deviation from a constant
power law across this large energy range. The small change in slope, from
$\propto E^{-2.7}$ to $\propto E^{-3.0}$, near $10^{15.5}$~eV is known as
the ``knee''. The spectrum steepens further to $E^{-3.3}$ above the ``dip''
($\approx 10^{17.7}$~eV), and then flattens to $E^{-2.7}$ at the ``ankle''
($\approx 10^{19}$~eV). Within statistical uncertainty of current
observations, which is large above $10^{20}$~eV, the upper end of the
spectrum is consistent with a simple extrapolation at that slope to the
highest energies, possibly with a slight accumulation around $10^{19.5}$~eV
(For  recent surveys of experimental data the reader is referred
to~\cite{Anchordoqui:2002hs,Nagano:ve,Watson:kk,Bertou:2000ip,Yoshida:1998jr,Sokolsky:rz}).

It is a lucky coincidence that at the energy ($\sim 10^{14}$~eV) where direct
measurement of CRs becomes limited by detector area and exposure
time, the resulting air showers that such particles produce when they strike
the upper atmosphere become big enough to be detectable at
ground level. There are several techniques that can be employed in
the process of detection:

{\it (i)} Direct detection of shower particles is the most
commonly used method, and involves constructing an array of sensors spread
over a large area to sample particle densities as the shower arrives at the
Earth's surface. The pioneering development of the air shower techniques 
(and the first use of
plastic scintillation detectors for the dual use of measuring arrival
directions and particle densities)  was started at the Agassiz Station of the
Harvard College Observatory, a work carried out between 1954 and 
1957~\cite{aga1,aga2,aga3}. The existence of primary particles with energies
greater than $10^{18}$~eV was established by the observation of one shower
with more than $10^9$ particles. Soon afterwards, this technique flourished
with measurements of ultra high energy cosmic rays (UHECRs) with the 
Volcano Ranch
experiment in the 60 s~\cite{Linsley:prl,Linsley:pr,Linsley}, 
as well as with several other arrays, such
as Haverah Park in England~\cite{Lawrence:cc}, 
Yakutsk in Russia~\cite{Khristiansen,Ivanov:2003iv},
the Sydney University Giant Airshower Recorder in
Australia (SUGAR)~\cite{Winn:un}, and the
Akeno Giant Air Shower Array (AGASA) in Japan~\cite{Chiba:1991nf,Ohoka:ww}.

{\it (ii)} Another well-established method
of detection involves measurement of the longitudinal
development (number of particles versus atmospheric depth)
of the extensive air shower (EAS) by sensing the fluorescence
light produced via interactions of the charged particles in the atmosphere.
The emitted light is typically in the 300 - 400~nm ultraviolet
range to which the atmosphere is quite transparent. Under favorable
atmospheric conditions, EASs can be detected at distances as large as
20~km,
about 2 attenuation lengths in a standard desert atmosphere at ground
level. However, observations can only be done on clear Moon{\it less} nights,
resulting in an average 10\% duty cycle.
The fluorescence technique has so far been implemented only in the Dugway
desert (Utah). Following a
successful trial at Volcano Ranch~\cite{Bergeson:nw} the
group from the University of Utah built
a device containing
two separated Fly's Eyes~\cite{Baltrusaitis:mx,Baltrusaitis:ce}. The two-eye
configuration monitored the sky from 1986 until 1993. As an up-scaled
version of Fly's Eye, the High Resolution (HiRes) Fly's Eye
detector begun operations in May 1997~\cite{Corbato:fq,Abu-Zayyad:uu}.
In monocular mode, the effective
acceptance of this instrument is $\sim 350 (1000)$~km$^2$~sr at
$10^{19}\, (10^{20})$~eV, on average about 6 times the Fly's Eye acceptance,
and the threshold energy is $10^{17}$~eV. This takes into account a  10\%
duty cycle.

{\it (iii)} A more recently
proposed technique uses radar echos from the column of ionized air
produced by the shower. This idea
suggested already in 1940~\cite{Blackett}, has been recently
re-explored~\cite{Gorham:2000da,Gorham:2000fp} as either an
independent method to study air showers, or as
a complement to existing fluorescence and surface detectors.
A proposal has recently been put forth
to evaluate the method using the Jicamarca radar system near Lima,
Peru~\cite{Vinogradova:2000fr}.

In order to increase the statistics at the high end of the spectrum
significantly, two projects are now under preparation:

{\it (i)} The Pierre Auger Observatory (PAO), currently under construction in
Argentina,
is the first experiment designed to work in a hybrid mode incorporating both
a ground-based array of 1600 particle detectors spread over 3000~km$^2$ with
fluorescence telescopes placed on the boundaries of the surface
array~\cite{Abraham}. A second array will be set up in the Northern
hemisphere to cover the whole sky. Such a full-sky coverage is very important
to allow sensitive anisotropy analysis. The overall aperture (2 sites) for
CRs with primary zenith angle $< 60^\circ$ and primary energy
$> 10^{19}$~eV is $\approx 1.4 \times 10^4$ km$^2$ sr.

{\it (ii)} The mission ``Extreme Universe Space Observatory'' (EUSO) will
observe the fluorescence signal of CRs,
with energy $> 4 \times 10^{19}$~eV, looking downward from the International
Space Station to the dark side of the Earth
atmosphere~\cite{Catalano:mm,Scarsi:fy}. The characteristic
wide angle optics of the instrument (with opening field of view
$\pm 30^\circ$ at an average orbit altitude
of $\approx 400$~km) yields a geometric aperture of
$\approx 5 \times 10^5$~km$^2$ sr, taking into account a 10\% duty cycle. The
monocular stand-alone
configuration of the telescope  will serve as a
pathfinder mission to develop the required technology to observe the
fluorescent trails of EASs from space.

The experimental input for $\gamma$-ray physics will be further enriched by dedicated \v{C}erenkov detectors like HESS~\cite{Hinton:2004eu}, MAGIC~\cite{Baixeras:hd}, CANGAROO~\cite{Mori:hv}, and 
VERITAS~\cite{Weekes:2001pd}, as well as satellites like GLAST~\cite{Gehrels:2000bv,Gehrels:ri} and 
AGILE~\cite{Longo:db}.

\subsection{Primary species}

When a CR enters the Earth atmosphere it collides with a
nucleus of an air atom, producing a roughly conical cascade of billions
of elementary particles which reaches the ground in the form of a giant
``saucer'' traveling at nearly the speed of light.

Unfortunately, because of the highly indirect method of
measurement, extracting precise information from EASs has
proved to be exceedingly difficult. The most fundamental problem  is that the
first generations of particles in the cascade are subject to
large inherent fluctuations and consequently this limits the event-by-event
energy resolution of the experiments. In addition, the center-of-mass energy
of the first few cascade steps is well beyond any reached in collider
experiments. Therefore, one needs to rely on hadronic interaction models that
attempt to extrapolate, using different mixtures of theory and phenomenology,
our understanding of particle physics. At present,
the different approaches used to
model the underlying physics of $p\bar{p}$ collisions show clear
differences in multiplicity predictions which increase with rising
energy~\cite{Anchordoqui:1998nq,Ranft:2000mz,Alvarez-Muniz:2002ne}.
Therefore, distinguishing
between a proton and a nucleus shower is extremely difficult at the
highest energies~\cite{Knapp:2002vs,Anchordoqui:2004xb}.

Fortunately, photon and hadron primaries can be distinguished by comparing
 the rate of vertical to inclined showers, a technique which
 exploits the attenuation of the electromagnetic shower
 component for large slant depths. Comparing the predicted
rate to the rate observed by Haverah Park for showers in the range
$60^\circ< \theta < 80^\circ$, Ave et al.~\cite{Ave:2000nd}
conclude that above $10^{19}$ eV, less
than 48\% of the primary CRs can be photons  and
above $4 \times 10^{19}$ eV less than 50\%  can be
photons. Both of these statements are made at the 95\% CL.

The longitudinal
development has a well defined maximum, usually referred to as $X_{\rm max}$,
which increases with primary energy as more cascade generations are required
to degrade the secondary particle energies. Evaluating $X_{\rm max}$ is a
fundamental part of many of the composition studies done by detecting air
showers. For showers of a given total energy, heavier nuclei have smaller
$X_{\rm max}$ because the shower is already subdivided into $A$ nucleons
when it enters the atmosphere. Specifically, the way the average depth of
maximum $\langle X_{\rm max} \rangle$ changes with
energy depends on the primary composition and particle interactions
according to
\begin{equation}
\langle X_{\rm max} \rangle = D_e \ln \left( \frac{E}{E_0} \right),
\end{equation}
where $D_e$ is the so-called ``elongation rate'' and $E_0$ is a
characteristic energy that depends on the primary
composition~\cite{Linsley:gh}. Therefore,
since
$\langle X_{\rm max} \rangle$ and $D_e$ can be determined directly from the
longitudinal shower profiles measured with a fluorescence
detector,
$E_0$ and thus the composition, can be extracted after estimating $E$ from
the total fluorescence yield. Indeed, the parameter often
measured is $D_{10}$, the rate of change of $\langle X_{\rm max} \rangle$ per
{\it decade} of energy.

Another important observable which can be related to primary energy and
chemical composition is the total number of muons $N_\mu$ reaching ground
level. For vertical proton showers, numerical simulations~\cite{Hillas}
indicate that the muon production is related to the energy of the primary
via~\cite{Winn:un}
\begin{equation}
E = 1.64 \times 10^{18} \,\,\left(\frac{N^p_\mu}{10^7} \right)^{1.073}
\,\,{\rm eV}.
\label{Hillas}
\end{equation}
Thus, modeling a shower produced by a nucleus with energy $E_A$ as the
collection of $A$ proton showers, each with energy $A^{-1}$ of the nucleus
energy, leads -- using Eq.~(\ref{Hillas}) -- to
$N^A_\mu \propto A (E_A/A)^{0.93}$~\cite{Anchordoqui:2003gm}.
Consequently, one expects a CR nucleus to produce about $A^{0.07}$
more muons than a proton. This implies that an iron nucleus
produces a shower with around 30\% more muons than a proton shower of the
same energy.

\begin{figure} [t]
\postscript{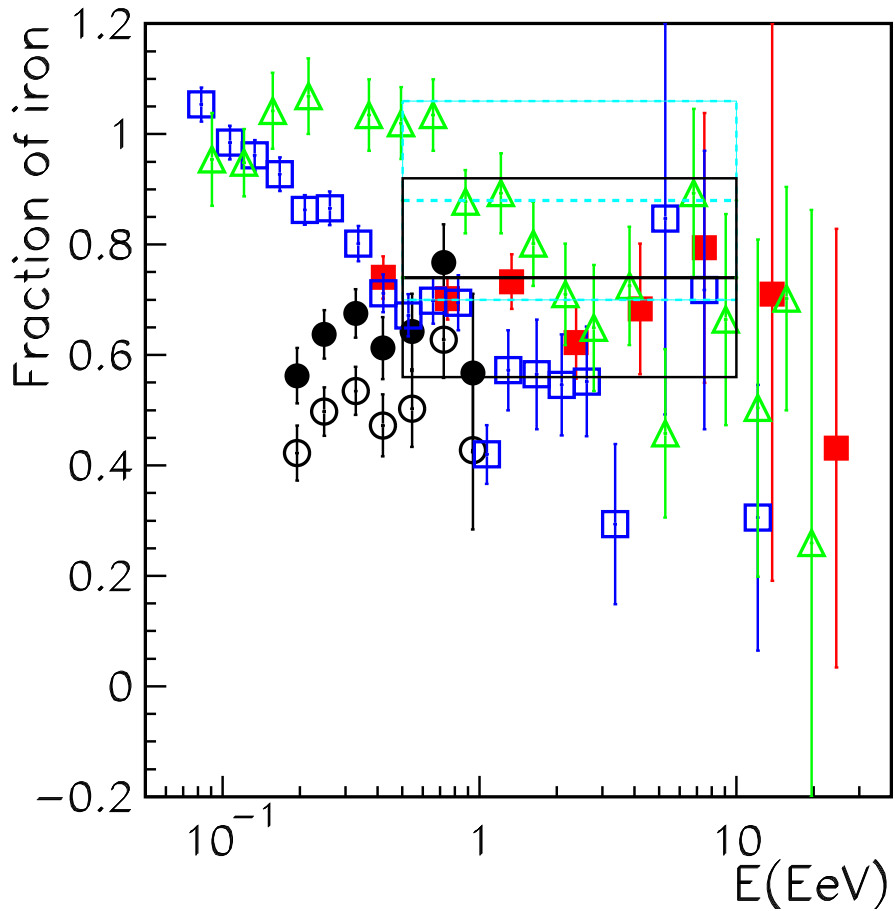}{0.80}
\caption{Predicted fraction of iron nuclei in the CR beam at the
top of the atmosphere from various experiments: Fly's Eye ($\triangle$),
AGASA A100 ({\tiny $\blacksquare$}), AGASA A1 ({\tiny $\square$})
using {\sc sibyll
1.5} as the hadronic interaction event generator~\cite{Dawson:1998kk}
and Haverah Park~\cite{Ave:2002gc},
using {\sc qgsjet98} ({\large $\bullet$}) and {\sc qgsjet01} ({\large
$\circ$}) to process the hadronic collisions. The solid (dashed) line
rectangle indicates the mean composition
 with the corresponding error estimated using the Volcano Ranch
data and {\sc qgsjet98} ({\sc qgsjet01}); the systematic shift in the
fraction of iron induced by the hadronic event
generator is 14\%~\cite{Dova:2003ng}. } \label{VR}
\end{figure}

The analysis of the elongation rate and the spread
in $X_{\rm max}$ at a given energy reported by the Fly's Eye
Collaboration suggests a change from an iron dominated composition at
$10^{17.5}$~eV to a proton dominated composition near
$10^{19}$~eV~\cite{Bird:yi,Gaisser:ix}. Such behavior of $D_e$ is in
agreement with an earlier analysis from Haverah
Park~\cite{Walker:sa}. However, the variation of the density of muons
with energy reported by the Akeno Collaboration favours a composition that
remains mixed over the $10^{18} - 10^{19}$~eV
decade~\cite{Hayashida:95}. More recently, Wibig and Wolfendale~\cite{Wibig}
reanalyzed the Fly's Eye data considering not only  proton and iron
components (as in~\cite{Bird:yi}) but a larger number of atomic mass
hypotheses. Additionally, they adopted a different hadronic model
that shifts the prediction of $X_{\rm max}$ for primary protons of
$10^{18}$ eV from 730 g cm$^{-2}$~\cite{Bird:yi} to 751 g cm$^{-2}$. The
difference, although apparently small, has a significant effect on the mass
composition inferred from the data. The study indicates that at the highest
energies ($10^{18.5} - 10^{19}$~eV and somewhat above) there is a
significant fraction of primaries with charge greater than unity. This result
is more in accord with the conclusions of the Akeno group than those of
the Fly's Eye group. Very recently, the Volcano Ranch data was re-analyzed
taking into account a bi-modal proton-iron model~\cite{Dova:2003an}. The best
fit gives a mixture with $75 \pm 5 \%$ of iron, with corresponding percentage
of protons.
A summary of the diferent bi-modal analyses is shown in Fig.~\ref{VR}.
Within statistical errors and systematic uncertainties introduced by
hadronic interaction models, the data seem to indicate that iron is
the dominant
component of CRs between $\sim 10^{17}$~eV and $\sim 10^{19}$~eV.
Nonetheless, in view
of the low statistics at the end of the spectrum and the wide variety of
uncertainties in these experiments, one may conservatively say that this
is not a closed issue.

\subsection{Distribution of arrival directions}
\label{DoAD}

The distribution of arrival directions is perhaps the most helpful
observable in yielding clues about the CR origin. On the one hand, if cosmic
rays cluster within
a small angular region (see e.g.~\cite{Hayashida:bc}) or show
directional alignment with powerful compact
objects (see e.g.~\cite{Farrar:1998we}), one might be able to
associate them with isolated sources in the sky.
On the other hand, if the distribution of arrival directions exhibits a
large-scale anisotropy, this could indicate whether or not certain classes of
sources are associated with large-scale structures (such
as the Galactic plane or the Galactic halo).

Cosmic ray air shower detectors which experience stable operation over a
period of a year or more can have a uniform exposure in right ascension,
$\alpha$. A traditional technique to
search for large-scale anisotropies is then to fit the right ascension
distribution of events to a sine
wave with period $2\pi/m$ ($m^{\rm th}$ harmonic) to determine the
components ($x, y$) of the Rayleigh
vector~\cite{Linsley:aniso}
\begin{equation}
x = \frac{2}{N} \sum_{i=1}^{N} \, \cos(m\, \alpha_i) \,, \,\,\,\,\,y =
\frac{2}{N} \sum_{i=1}^{N} \,\,
\, \sin( m\, \alpha_i)\,.
\end{equation}
The $m^{\rm th}$ harmonic amplitude of $N$ measurements of $\alpha_i$ is
given by the Rayleigh vector
length ${\cal R}~=~(x^2~+~y^2)^{1/2}$. The expected length of such a vector
for values randomly
sampled from a uniform phase distribution is ${\cal R}_0~=~2/\sqrt{N}$.
The chance probability
of obtaining an amplitude with length larger than that measured is
$p(\geq~{\cal R})~=~e^{-k_0},$ where $k_0~=~{\cal R}^2/{\cal R}_0^2.$
To give a specific example, a vector of length $k_0~\geq~6.6$ would be
required to
claim an observation whose probability of arising from random fluctuation
was 0.0013 (a ``$3\sigma$'' result)~\cite{Sokolsky:rz}.

AGASA has revealed a correlation of the arrival direction of the cosmic
rays to the Galactic Plane (GP) at the $4\sigma$
level~\cite{Hayashida:1998qb}. The energy bin width which gives the
maximum
$k_0$-value corresponds to the region $10^{17.9}$~eV -- $10^{18.3}$~eV
where $k_0 = 11.1,$
yielding a chance probability of $p(\geq~{\cal R}_{_{E\sim {\rm
EeV}}}^{^{\rm AGASA}}) \approx 1.5 \times 10^{-5}.$ The GP excess, which
is roughly 4\% of
the diffuse flux, is mostly
concentrated in the direction of the Cygnus region, with a second
spot towards the Galactic Center (GC)~\cite{Teshimaicrc}. Evidence at
the 3.2$\sigma$ level for GP
enhancement in a similar energy range has also been reported by the HiRes
Collaboration~\cite{Bird:1998nu}. The existence of a point-like excess in
the
direction of the GC has been confirmed via independent
analysis~\cite{Bellido:2000tr} of data collected with SUGAR.\footnote{Interestingly, sub-TeV $\gamma$-ray 
emission from the direction of the GC has been observed using the CANGAROO-II Imaging Atmospheric \v{C}erenkov Telescope~\cite{Tsuchiya:2004wv}.}  This is a
remarkable level of
agreement among experiments using a variety of techniques.

At lower energies ($\sim$~PeV), the Rayleigh analysis shows no evidence of
anisotropy~\cite{Antoni:2003jm}.
Hence, the excess from the GP is very suggestive of neutrons as candidate
primaries,
because the directional signal requires relatively-stable neutral
primaries, and time-dilated neutrons can reach
the Earth from typical Galactic distances when the neutron energy exceeds
$10^{18}$~eV. Arguably, if the Galactic messengers
are neutrons, then those with energies below $10^{18}$~eV will decay in
flight, providing a flux of cosmic antineutrinos
above 1 TeV that should be observable at  kilometer-scale neutrino
telescopes~\cite{Anchordoqui:2003vc}. A measurement of the
$\bar \nu$-flux will supply a strong confirmation of the GP neutron
hypothesis.

For the ultra high energy~($\agt~10^{19.6}$~eV) regime, all
experiments to date have reported
$k_0 \ll
6.6,$ $\forall m <5$~\cite{Edge:rr,Winn:up,Cassiday:zh,Takeda:1999sg}.\footnote{For the
Fly's Eye data-sample
the Rayleigh vector was computed using weighted showers, because it has
had a nonuniform exposure in sideral time. A shower's weight depends on
the hour of its sideral arrival
time, and the 24 different weights are such that every time bin has the
same weighted number of showers.}
This does not imply an isotropic
distribution, but it merely means that available data are too sparse to
claim a statistically significant
measurement of anisotropy.
In other words, there may exist anisotropies at a level too low to
discern given existing statistics~\cite{Evans:2001rv}.

The right harmonic analyses are completely blind to intensity
variations which depend only on declination, $\delta$.  Combining
anisotropy searches in $\alpha$ over a range of declinations could
dilute the results, since significant but out of phase Rayleigh
vectors from different declination bands can cancel each other
out.  Moreover, the analysis methods that consider distributions
in one celestial coordinate, while integrating away the second,
have proved to be potentially misleading~\cite{Wdowczyk:rb}. An
unambiguous interpretation of anisotropy data requires two
ingredients: {\it exposure to the full celestial sphere and
analysis in terms of both celestial coordinates.} In this
direction, a recent study~\cite{Anchordoqui:2003bx} of the angular
power spectrum of the distribution of arrival directions of CRs
with energy $> 10^{19.6}$~eV, as seen by the AGASA and SUGAR
experiments, shows no departures from either homogeneity or
isotropy on an angular scale greater than $10^\circ$. Finally, the
recently analyzed HiRes data is also statistically consistent with
an isotropic distribution~\cite{Abbasi:2003tk}.

All in all, the simplest interpretation of the existing data is that,
beyond the ankle,
a new population of extragalactic CRs emerges to dominate the more steeply
falling Galactic population. Moreover, there are two extreme explanations
for
the near observed isotropy beyond $10^{19.6}$~eV: one is to argue a
cosmological
origin for these events, and the other is that we have nearby sources
(say, within the Local Supercluster) with a tangled magnetic field in the
Galaxy, and beyond, which bends the particle orbits, camouflaging the exact
location of the sources.

Although there seems to be a remarkable agreement among experiment on
predictions about isotropy on large scale
structure, this is certainly not the case when
considering the two-point correlation function on a small angular
scale.  The analyses carried out by 
AGASA Collaboration seem to indicate
that the pairing of events on the celestial sky could be occurring at
higher than chance coincidence~\cite{Hayashida:bc,Hayashida:2000zr}.
Specifically, when showers with separation angle less
than the angular resolution $\theta_{\rm min}= 2.5^\circ$ are
paired up, AGASA finds five doublets and one triplet
among the 58 events reported with mean energy above $10^{19.6}$~eV.
The probability of observing these clusters by chance coincidence under an
isotropic distribution was quoted as smaller than 1\%. A third independent
analysis~\cite{Anchordoqui:2001qk}, using the
Goldberg--Weiler formalism~\cite{Goldberg:2000zq}, confirmed the result
reported by AGASA Collaboration
and further showed that the chance probability is extremely sensitive to
the angular binning. The ``world'' data set has also been
studied~\cite{Uchihori:1999gu}.
Six doublets and two triplets out of 92 events with energies
$ >  10^{19.6}$~eV were found, with the chance
probability being less than 1\% in the restricted region within
$\pm 10^\circ$ of the super-Galactic plane. The angular
two-point correlation function of a
combined data sample of AGASA ($E > 4.8 \times 10^{19}$~eV) and Yakutsk
($E> 2.4 \times 10^{19}$~eV) was analyzed~\cite{Tinyakov:2001ic}. For a
uniform distribution of sources, the probability of chance clustering is
reported to be as small as $4 \times 10^{-6}$. Far from
confirming
what seemed a fascinating discovery, the recent analysis reported
by the HiRes Collaboration showed that the data is consistent with no
small-scale anisotropy among the highest energy 
events~\cite{Abbasi:2004ib,Abbasi:2004dx}.

The discovery of such clusters would be a tremendous breakthrough for the
field, but the case for them is not yet proven.  To calculate a meaningful
statistical significance in such an analysis, it is important to define
the
search procedure {\it a priori} in order to ensure it is not inadvertently
devised especially to suit the particular data set after having
studied it. In the analyses carried out by
AGASA Collaboration~\cite{Hayashida:bc,Hayashida:2000zr},
for instance, the angular bin size was not defined
ahead of time.
Very recently, with the aim to avoid accidental bias on the
number of trials performed in selecting the angular bin, the original
claim of AGASA Collaboration~\cite{Hayashida:bc} was re-examined
considering
only the events observed after the claim~\cite{Finley:2003ur}. This study
showed that the
evidence for clustering in the AGASA data set is weaker than was
previously
claimed, and consistent with the null hypothesis of isotropically
distributed
arrival directions.

Summing up, the clustering on small angular scale at the upper end of the
spectrum remains an open question, and the increase in statistics and improved
resolution attainable with PAO is awaited to solve the issue.

\subsection{Propagation of UHECRs}

In this section we briefly summarize the relevant interactions that
CRs suffer on their trip to Earth. For a more detailed discussion the reader
is refer to~\cite{Anchordoqui:2002hs,Bhattacharjee:1998qc,Sigl:2000vf,Stecker:2003wm,Anchordoqui:gap}.

\subsubsection{The GZK-cutoff}

Ever since the discovery of the cosmic microwave background (CMB)
standard physics implies there would be a cuttoff in the
observed CR-spectrum. In the mid-60's
Greisen, Zatsepin, and Kuzmin (GZK)~\cite{Greisen:1966jv,Zatsepin:1966jv}
pointed out that this photonic molasses
makes the universe opaque to protons of sufficiently high
energy, i.e., protons with energies beyond the photopion
production threshold,
\begin{equation}
E_{p\gamma_{\rm CMB}}^{\rm th} = \frac{m_\pi \,
(m_p + m_\pi/2)}{{\cal E}_{\rm CMB}}
\approx 6.8 \times 10^{19}\,
\left(\frac{{\cal E}_{\rm CMB}}{10^{-3}~{\rm eV}}\right)^{-1} \,\,{\rm eV}\,,
\label{pgammat}
\end{equation}
where $m_p$ ($m_\pi$) denotes the proton (pion) mass and ${\cal E}_{\rm CMB}
\sim 10^{-3}$~eV is a typical CMB photon energy. After pion production, the
proton (or perhaps, instead, a neutron) emerges with at least
50\% of the incoming energy. This implies that the nucleon energy changes by
an $e$-folding after a
propagation distance~$\lesssim (\sigma_{p\gamma}\,n_\gamma\,y)^{-1}
\sim 15$~Mpc. Here,
$n_\gamma \approx 410$~cm$^{-3}$
is the number density of the CMB photons, $\sigma_{p \gamma} > 0.1$~mb is the
photopion production cross section, and $y$ is the average energy fraction
(in the laboratory system) lost by a nucleon per interaction. Energy losses
due to pair production become relevant below $\sim 10^{19}$~eV.
For heavy nuclei, the giant dipole resonance can be
excited at similar total energies and hence, for example, iron nuclei do not
survive fragmentation over comparable distances. Additionally, the survival
probability for extremely high energy ($\approx 10^{20}$~eV)
$\gamma$-rays (propagating on magnetic fields~$\gg 10^{-11}$~G) to
a distance $d$, \mbox{$p(>d) \approx \exp[-d/6.6~{\rm Mpc}]$}, becomes less
than $10^{-4}$ after traversing a distance of 50~Mpc.\footnote{It should be 
stressed that if the extragalactic magnetic field is $< 10^{-12}$~G, photons with energy $\gg 10^{21}$~eV can reach us without significant energy loss from distant (redshift $z \agt 0.03$) sources~\cite{Kalashev:2001qp}.}

\begin{figure} [t]
\postscript{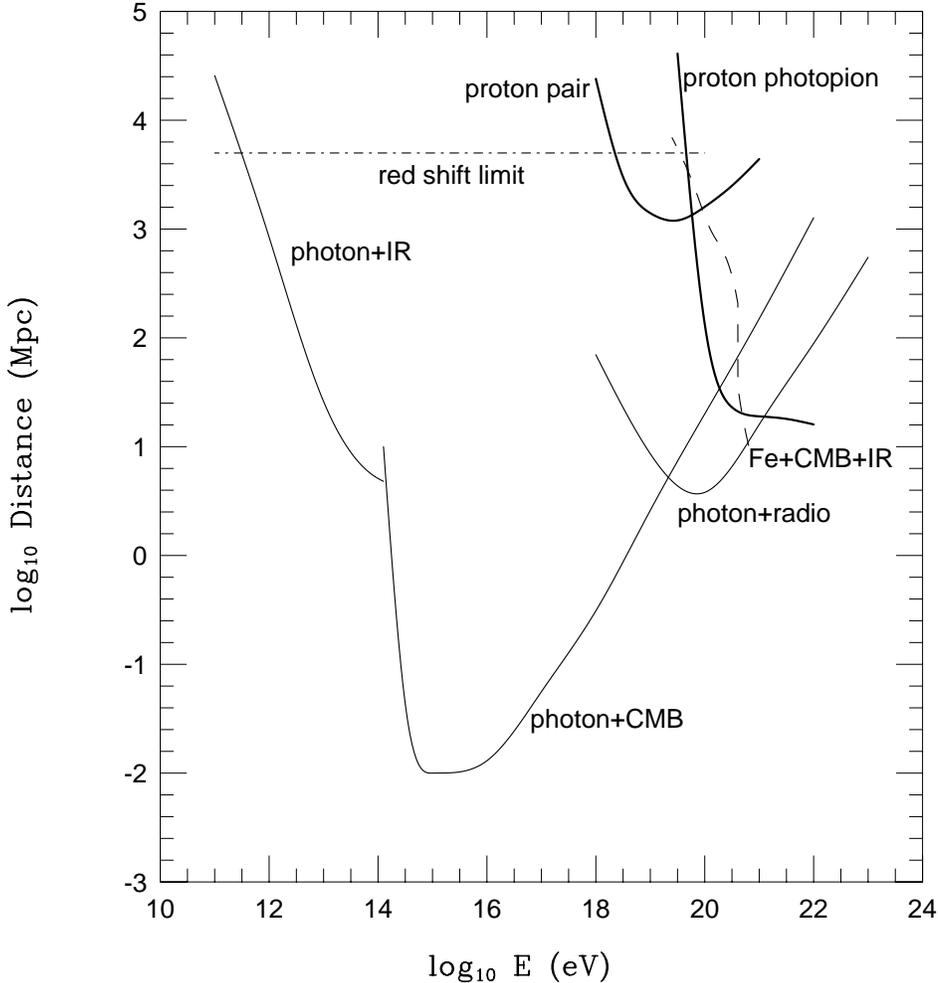}{0.80}
\caption{Attenuation length of $\gamma$'s, $p$'s, and $^{56}$Fe's in various
background radiations as a function of energy. The 3 lowest and left-most
thin solid curves refer to $\gamma$-rays, showing the attenuation by
infra-red, microwave, and radio backgrounds. The upper, right-most thick
solid curves refer to propagation of protons in the CMB, showing separately
the effect of pair production and photopion production.  The dashed-dotted
line indicates the adiabatic fractional
energy loss at the present cosmological epoch (see e.g. Appendix B of
Ref.~\cite{Anchordoqui:2002hs}). The dashed curve illustrates the attenuation
of iron nuclei.} \label{gzk}
\end{figure}

In recent years, several studies on the propagation of CRs (including both
analytical analyses and numerical simulations) have been carried
out~\cite{Stecker:68,Hill:1983mk,Berezinsky:wi,Stecker:1989ti,Aharonian:90,Yoshida:pt,Aharonian:nn,Protheroe:1995ft,Anchordoqui:gs,Anchordoqui:1996ru,Lee:1996fp,Achterberg:1999vr,Stanev:2000fb,Fodor:2000yi,Anchordoqui:2000ad,Blumenthal:nn,Stecker:fw,Puget:nz,Anchordoqui:1997rn,Anchordoqui:1998ig,Epele:1998ia,Stecker:1998ib}. A summary of the UHECR attenuation lengths for the above mentioned
processes (as derived in these analyses) is given in Fig.~\ref{gzk}. It is
easily seen that our horizon shrinks dramatically for energies
$\gtrsim 10^{20}$~eV. Therefore, if
UHECRs originate at cosmological distances, the net effect
of their interactions would yield a pile-up of particles around
$4 - 5 \times  10^{19}$~eV with the spectrum droping sharply thereafter.
As one can infer from Fig.~\ref{gzk}, the subtleties of the spectral
shape
depend on the nature of the primary species, yielding some ambiguity in the
precise definition of the ``GZK cutoff''. In what follows we
consider an event to supersed the cutoff if the lower energy limit at
the 95\%~CL
exceeds $7 \times 10^{19}$~eV. This conforms closely to the strong criteria
outlined in Ref.~\cite{Farrar:1999fw}.

\subsubsection{Propagation of CRs in a magnetized neighborhood
of the Galaxy}

In addition to the interactions with the radiation fields permeating the
universe, CRs suffer deflections on extragalactic and
Galactic magnetic fields.

Over the last few years, it has become evident
that the observed
near-isotropy of arrival directions can be easily explained if our Local
Supercluster contains a large scale magnetic field  which provides
sufficient bending to the CR trajectories~\cite{Lemoine:1999ys,Farrar:1999bw}.
Intergalactic field strengths and coherence 
lengths are not well established, but it is plausible to assume that fields
have coherent directions on scales $\ell \approx 0.5 - 1$~Mpc. The Larmor
radius of a CR of charge $Ze$ propagating in a
magnetic field $B_{\rm nG} \equiv B/10^{-9}$~G is given by
\begin{equation}
r_{\rm L} \approx \frac{100 \,\,E_{20}}{Z\,\,B_{\rm nG}}\,\,{\rm Mpc}\,\,,
\label{Larmor}
\end{equation}
where $E_{20}$ is the particle's energy in units of $10^{20}$~eV.
For $r_{\rm L} \gg \ell$ the motion is not very different from a quasilinear
trajectory,
with small deflections away from the straight line path given by
\begin{equation}
\theta (E) \approx 0.3^\circ \,\, \frac{L_{\rm Mpc}\,Z\,B_{ {\rm nG}}}{E_{20}}\,,
\label{t1}
\end{equation}
where $L_{\rm Mpc}$ is the propagation distance in units of Mpc.
As the Larmor
radius starts
approaching $\ell$ the particles begin to diffuse.

Diffusion has two distinctive regimes. Particles that are trapped
inside magnetic subdomains (of size $\ell_{\rm Mpc} \equiv
\ell/{\rm Mpc}$) follow Kolmogorov diffusion. In such a case, the
functional dependence of energy of the difussion coefficient  is
found to be~\cite{Blasi:1998xp}
\begin{equation}
D(E) \approx 0.0048\,
\left(\frac{E_{20} \, \ell^2_{\rm Mpc}}{Z\, B_{{\rm nG}}}\right)^{1/3}\,\,
{\rm Mpc}^2/{\rm Myr}\,.
\label{D(E)}
\end{equation}
With rising energy, $r_{\rm L} \rightarrow \ell,$ and there is a transition to
Bohm diffusion. The diffusion coefficient in this regime is of
order the Larmor radius times velocity ($\sim c$)~\cite{Sigl:1998dd}.
In this case the accumulated deflection angle from the direction of the
source, can be estimated assuming that the particles make a random
walk in the magnetic field~\cite{Waxman:1996zn}
\begin{equation}
\theta (E) \approx 0.54^\circ\, \,
\frac{\ell_{\rm Mpc}^{1/2}\,L_{\rm Mpc}^{1/2}\,Z \, B_{\rm nG}}{E_{20}}\,.
\label{t2}
\end{equation}

Surprisingly
little is actually known about the extragalactic magnetic field strength.
There are some measurements of diffuse radio
emission from the bridge area between Coma and Abell superclusters
that under assumptions of equipartition allows an estimate of
$0.2 - 0.6~\mu$G for the magnetic field in this
region~\cite{Kim}.\footnote{Fields of ${\cal O}(\mu{\rm G})$ are also
indicated in
a more extensive study of 16 low redshift clusters~\cite{Clarke}.} Such a
strong magnetic field (which is
compatible  with existing upper limits on Faraday rotation
measurements~\cite{Kronberg:1993vk}) could be possibly understood
if the bridge region lies along a filament or sheet of large scale
structures~\cite{Ryu}.  Faraday rotation
measurements~\cite{Vallee:360,Kronberg:1993vk} have thus far served to set
upper bounds of ${\cal O}(10^{-9}-10^{-8})$ G on extragalactic magnetic
fields on various scales~\cite{Kronberg:1993vk,Blasi:1999hu}, as have
the limits on distortion
of the CMB \cite{Barrow:1997mj,Jedamzik:1999bm}. The Faraday rotation
measurements sample extragalactic field strengths of any origin out to
quasar distances, while the CMB analyses set limits on primordial magnetic
fields. Finally, there are some hints suggesting that the extragalactic
field strength can be increased in the neighborhood of the Milky
Way, $B_{{\rm nG}} > 10$~\cite{Anchordoqui:2001bs}. Now,
using Eq.~(\ref{Larmor}), one can easily see that because of
the large uncertainty on the magnetic field strength, ${\cal O}({\rm nG}) -
 {\cal O}(\mu{\rm G}),$ all 3 different regimes discussed above are
likely to describe UHECR propagation.

If CRs propagate diffusively, the radius of the sphere for potential proton
sources becomes significantly reduced. This is because one expects negligible
contribution to the flux from times prior to the arrival time of the
diffusion front, and so
the average time delay in the low energy region,
$
\tau_{\rm delay} \approx d^2/[4 D(E)],
$
must be smaller than the age of the source, or else the age of the
universe (if no source within the GZK radius is active today, but
such sources have been active in the past). Note that the diffuse
propagation of UHE protons requires  magnetic fields $\sim 1\mu$G.
Therefore, for typical coherence lengths of extragalactic magnetic
fields the time delay of CRs with $E \approx
10^{18.7}$~eV cannot exceed $\tau_{\rm delay} \alt 14$~Gyr,
yielding a radius of $d \sim 30$~Mpc. In the case CR sources
are active today, the radius for potential sources is even
smaller $d \sim 5$~Mpc.

On the other hand,  the sphere of potential nucleus-emitting-sources is
severely
constrained by the GZK cutoff: straightforward calculation, using the
attenuation length given in Fig.~\ref{gzk}, shows that less than
1\% of iron nuclei (or any surviving fragment of their spallations)
can survive more
than $3 \times 10^{14}$~s with an energy $\agt 10^{20.5}$~eV.
Therefore, the assumption that UHECRs are heavy nuclei implies
ordered extragalactic magnetic fields $B_{\rm nG} \alt 15 - 20$,
or else nuclei would be trapped inside magnetic subdomains
suffering catastrophic spallations.

The large scale structure of the Galactic magnetic field
carries
substantial uncertainties as well, because the position of the solar
system does not allow global measurements. The average field
strength can be directly determined from pulsar observations of
the rotation and dispersion measures average along
the line of sight to the pulsar with a weigh proportional to the
local free electron density,
$\langle B_{||} \rangle \approx 2
\mu$G~\cite{Manchester,Thomson,Han,Indrani}.(We use the
standard, though ambiguous notation, in which $B$ refers to either the
Galactic or
extragalactic magnetic field, depending on the context.)
Measurements of polarized
synchrotron radiation as well as Faraday rotation of the radiation
emitted from pulsars and extragalactic radio sources revealed that
the global structure of the magnetic field in the disk of our
Galaxy could be well described by spiral fields with $2 \pi$
(axisymmetric, ASS) or $\pi$ (bisymmetric, BSS)
symmetry~\cite{Beck}. In the direction perpendicular to the
Galactic plane the fields are either symmetric (S) or antisymmetric (A).
Discrimination between these models is
complicated. Field reversals are certainly observed (in the
Crux-Scutum arm at 5.5 kpc from the Galactic center, the
Carina-Sagittarius arm at 6.5 kpc, the Perseus arm at 10 kpc, and
possibly another beyond~\cite{Han:1999vi}).
However, as discussed by Vall\'ee~\cite{Vallee}, turbulent dynamo
theory can explain field reversals at distances up to $\sim$~15~kpc
within the ASS configuration.

More accurately,
the field strength in the Galactic plane ($z=0$) for the ASS
model is generally described by~\cite{Stanev:1996qj,Harari:1999it}
\begin{equation}
B (\rho, \theta) = B_0(\rho) \, \cos^2 [\,\theta - \beta \ln (\rho/\xi_0)]\,,
\end{equation}
and for the BSS
\begin{equation}
B (\rho, \theta) = B_0(\rho) \, \cos [\,\theta - \beta \ln (\rho/\xi_0)]\,,
\end{equation}
where $\theta$ is the azimuthal coordinate around the Galactic center
(clockwise as seen from the north Galactic pole), $\rho$ is the galactocentric
radial cylindrical coordinate, and
\begin{equation}
B_0(\rho) = \frac{3 r_0}{\rho} \, {\rm tanh}^3 (\rho/\rho_1)\,\,\,\mu{\rm G}\,.
\end{equation}
Here, $\xi_0 = 10.55$~kpc stands for the galactocentric distance
of the maximum of the field in our spiral arm, $\beta = 1/\tan p$
(with the pitch angle, $p =-10^\circ$), $r_0 = 8.5$~kpc is the
Sun's distance to the Galactic center, and $\rho_1 = 2$~kpc. The
$\theta$ and $\rho$ coordinates of the field are correspondingly,
\begin{equation}
B_\theta = B(\rho, \, \theta) \,\cos p\,\,, \hspace{1cm} B_\rho = B(\rho, \theta) \, \sin p\,.
\end{equation}
The field strength above and below the Galactic plane (i.e., the dependence on $z$)  has a
contribution coming from the disk and another from the halo: {(\it i)}
for A models
\begin{equation}
B_{\rm A} (\rho, \theta, z)  =  B(\rho, \theta) \,\,\, {\rm tanh} (z/z_3)\,
\left( \frac{1}{2 \cosh (z/z_1)} +
\frac{1}{2 \cosh (z/z_2)}\right) \,\,,
\label{oko}
\end{equation}
{\it (ii)} for S models,
$B_{\rm S} = B_{\rm A} (\rho, \theta, z)/ {\rm tanh} (z/z_3)$;
where $z_1 = 0.3$~kpc, $z_2 = 4$~kpc and $z_3 = 20$~pc.
With this in mind, the Galactic magnetic
field  produce significant bending to the CR orbits  if
$E_{20}/Z = 0.03$~\cite{Harari:1999it}.

\subsection{GZK-end of the cosmic ray spectrum?}

A first hint of a puzzle surfaced in the  highest energy Fly's Eye
event~ \cite{Bird:1994uy} which has no apparent progenitor
within the Local Supercluster~\cite{Elbert:1994zv}. Subsequent observations
with the AGASA experiment~\cite{Takeda:1998ps}
carried strong indication that
the cutoff was somehow circumvented in the absence of plausible nearby
sources.

\begin{figure} [t]
\postscript{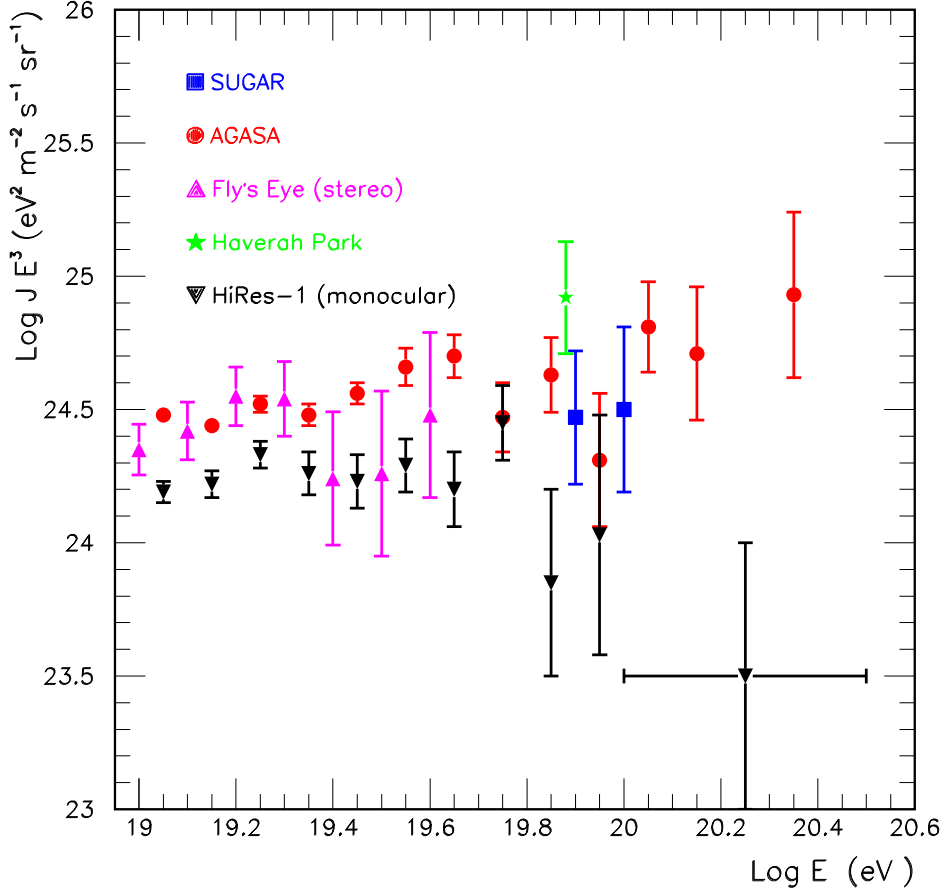}{0.80}
\caption{Upper end of the cosmic ray energy spectrum as observed by
 AGASA~\cite{Takeda:2002at}, Fly's Eye\cite{Bird:wp}, Haverah Park~\cite{Ave:2001hq}, HiRes~\cite{Abu-Zayyad:2002sf}, and
SUGAR~\cite{Anchordoqui:2003gm}.} \label{azukar}
\end{figure}

The big disappointment of 2002 was the CR-flux reported by the HiRes
Collaboration~\cite{Abu-Zayyad:2002ta,Abu-Zayyad:2002sf}, which is in sharp
disagreement with AGASA data~\cite{Takeda:2002at}. The discrepancy
between the two estimated fluxes is shown 
in Fig.~\ref{azukar}. One can argue correctly
that the statistical significance of the discrepancy is small, although such
an assesment requires a conspiracy between the two groups to bend their maximal
systematic errors in opposite directions. Moreover,
an analysis~\cite{Bahcall:2002wi} of the combined data reported by the HiRes,
the Fly's Eye, and the Yakutsk collaborations is supportive of the existence
of the GZK cutoff at the  $>5\sigma$  ($>3.7\sigma$) level. The deviation
from GZK depends on the set of data used as a basis for power law
extrapolation from lower energies. An additional input for this analysis
was the recent claim~\cite{Watson} that there may be technical problems
with the Yakutsk data collection. More recently, fingerprints of super-GZK CRs
have been found~\cite{Anchordoqui:2003gm} by reanalyzing the SUGAR
data~\cite{Winn:un}. However, as one can see in Fig.~\ref{azukar}, the number
of events is not enough to weight in on one side or the other with respect to
the GZK question.

\section{Vanilla Sky: UHECR Generation within the Standard Lore}

\subsection{A brief low energy perspective}
\label{vanilla1}

Supernova remnants (SNRs) are thought to be the main source of
both CR ions and electrons with energies below the knee. The particle
acceleration mechanism
in individual SNRs is usually assumed to be diffusive shock
acceleration, which naturally leads to a power-law population of
relativistic particles. In the standard version of this mechanism
(e.g. \cite{Bell}), particles are scattered by magnetohydrodynamic
waves repeatedly through the shock front. If they encounter an
enhancement of molecular density, the pion channel can lead to
observable amounts of $\gamma$-rays (see Ref. \cite{Torres:2002af}
for a review, and references therein for details). Electrons
suffer synchrotron losses, producing the non-thermal emission from
radio to X-rays usually seen in shell-type SNRs. The maximum
energy achieved depends on the shock speed and age as well as on
any competing loss processes. In young SNRs, electrons can easily
reach energies in excess of 1 TeV, where they produce X-rays by
synchrotron mechanism (see, for example, \cite{R96,R98}).

CRs of low energies are also expected to be accelerated in OB
associations, through turbulent motions and collective effects of
star winds (e.g. \cite{B92,B92b}). The main acceleration region
for TeV particles would be in the outer boundary of the
supperbubble produced by the core of a given stellar association.
If there is a subgroup of stars located at the acceleration
region, their winds might be illuminated by the locally
accelerated protons, which would have a distribution with a slope
close to the canonical value, $\alpha\sim2$, and produce
detectable $\gamma$-rays \cite{Torres:2003ur}. The  HEGRA
detection in the vicinity of Cygnus OB2, TeV
J2032+4131~\cite{Aharonian:2002ij}, could be, judging from
multiwavelength observations~\cite{Butt:2003xc}, the result of
such a process~\cite{Torres:2003ur}. A nearby EGRET source (3EG
J2033+4118) has also a likely stellar origin
\cite{wc,c,Benaglia:2000uc}.

Truth is, as always, bitter. No astrophysical source of UHECRs,
nor of CRs with energies below the knee, has been ever confirmed.
Out of all SNRs coinciding
with non-variable $\gamma$-ray sources detected by EGRET
\cite{Romero:1999tk,Torres:xa,Torres:zu,Nolan:2003bt}, the
supernova remnant RX J1713.7-3946 is perhaps one the most
convincing cases for a hadronic cosmic-ray accelerator detected so
far in the Galaxy, although yet subject to confirmation
\cite{Butt:2001ff,Enomoto:xk,Reimer:2002ea,Butt:2002im,Uchiyama:2002mt}.
Other excellent candidates include SN1006 (e.g.
\cite{Berezhko:2003ej}) and Cas A (e.g. \cite{Berezhko:2003fb}).
 Kilometer-scale
neutrino telescopes have also been proposed as viable detectors of
hadronic CR sources (e.g.~\cite{Alvarez-Muniz:2002tn,Anchordoqui:2002xu,Romero:2003td,Bednarek:2003cv}),
and will be a welcomed addition to the space- and ground-based
detectors already existing or planned.

\subsection{Plausible sources of UHECRs and the Hillas' plot}

\begin{figure} [t]
\postscript{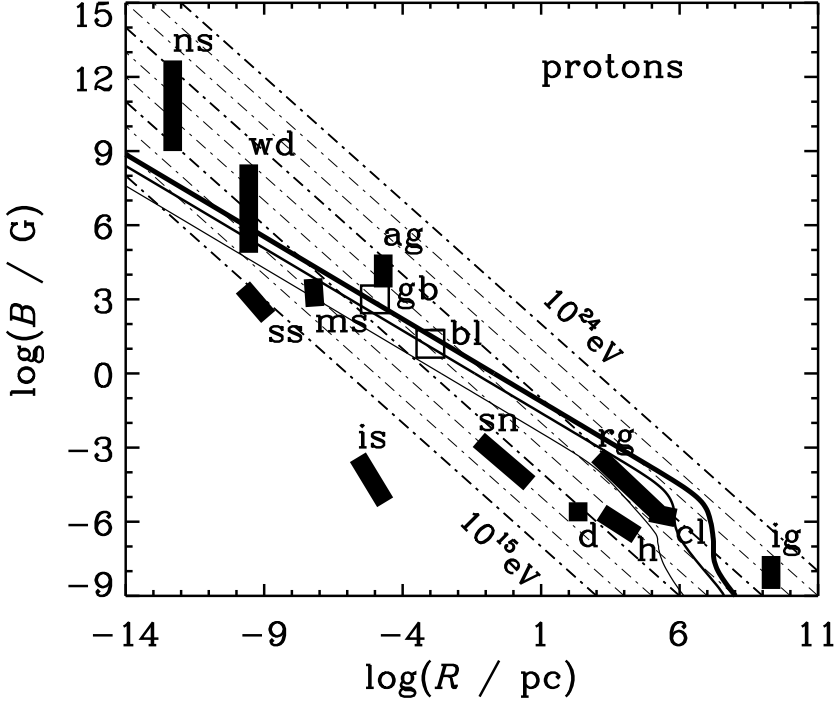}{0.95} \caption{The Hillas
diagram showing (chain curves) magnetic field versus gyroradius for proton
momenta $10^{15}, \, 10^{16}, \, \dots, 10^{24}$~eV$/c$. The solid curves
correspond to different shock-waves velocities: the upper solid curve
indicates
the maximum attainable energy $\beta =1,$ the middle and lower solid curves
indicate plausible less effective acceleration processes. Typical size and
magnetic field of possible acceleration
sites are shown for neutron stars (ns), white dwarfs (wd), sunspots (ss),
magnetic stars (ms), active galactic nuclei (ag), interstellar space (is),
SNRs (sn), radio galaxy lobes (rg), galactic disk (d) and halo (h), clusters
of galaxies (cl) and intergalactic medium (ig). Typical jet-frame parameters
of the synchrotron proton blazar model~\cite{Mucke:ap} and gamma ray burst
model~\cite{Pelletier:2000ch} are indicated by open squares labeled ``bl''
and ``gb'', respectively~\cite{Protheroe:2004rt}.} \label{hillasf} \end{figure}

Following the main ideas behind the concept of Fermi's first order
acceleration, when $r_{\rm L}$ approaches the accelerator
size, it becomes very difficult to  magnetically confine the CR to
the acceleration region, and thus to continue the accelerating
process up to higher energies. If one includes the effect of the
characteristic velocity $\beta c$ of the magnetic scattering
centers\footnote{The size of the accelerating region containing
the magnetic field should be as large as $2r_{_{\rm L}}$. Taking
into account a characteristic velocity $\beta c$ of the scattering
centers this transforms into $2r_{_{\rm L}} /\beta$. }, the above
argument leads to the general condition (sometimes called the
``Hillas criterion'' \cite{Hillas:1984}),
\begin{equation}
E_{\rm max} \sim 2 \beta\, c\, Ze\,B\, r_{_{\rm L}}\,,
\label{hillas}
\end{equation}
for the maximum energy acquired by a particle travelling in a
medium with magnetic field $B$.

In the case of one-shot acceleration scenarios, the maximum
reachable energy turns out to have a quite similar expression to
the shock acceleration case of Eq.~(\ref{hillas}). For instance, a
dimensional analysis suggests that the maximum energy that can be
obtained  from a pulsar is \cite{Hillas:1984}
\begin{equation}
E_{\rm max} = \frac{\omega}{c}\, Ze\, B_s \,R_{s}^2\,,
\label{hillas2}
\end{equation}
where $\omega$ is the pulsar angular velocity, $B_s$ the surface
magnetic field and $R_{s}$ the neutron star radius.
Therefore, if $B_s \sim 10^{12}$~G, $R_{ s} \sim 10$~km, and
$\omega \sim 60 \pi$~s$^{-1}$ (as for the Crab pulsar), a circuit
connected between pole and equator would see an emf $\sim
10^{18}$~V for an aligned or oblique dipole. When realistic models
of acceleration are constructed, however, this ideal dimensional
limit is not fully realized, because the large potential drop
along the magnetic field lines is  significantly short-circuited
by electron and positrons moving in opposite directions along
the field lines~\cite{Venkatesan:1996jw}.

The dimensional arguments of Eqs.~(\ref{hillas}) and
(\ref{hillas2}) are conveniently depicted in the ``Hillas
diagram''~\cite{Hillas:1984} shown in Fig.~\ref{hillasf}. Very few
sites can generate particles with energies $> 10^{20}$~eV: either
this occurs on highly condensed objects with huge $B$ or
enormously extended objects. Some of these potential astrophysical
sources are discussed in what follows. For further
details on the electrodynamical limitations of CR sources see,
e.g.~\cite{Aharonian:2002we,Medvedev:2003sx}.

\subsection{Neutron stars}

\subsubsection{Magnetohydronamic acceleration of iron nuclei in pulsars}
\label{NS}

Following earlier
ideas~\cite{olinto}\footnote{Even earlier ideas relating UHECRs with
neutron stars can be found in Refs. \cite{gunn,bell}, although
these attempts have failed at either reaching the highest
energies, or reproducing the spectrum, or reproducing the apparent
isotropy of the arrival directions of UHECRs.},  Blasi et al.
\cite{BLASI} \footnote{See
\cite{giller,Bednarek:2002nm}, also
\cite{DeGouveiaDalPino:2001ci}, for related research. See
\cite{Litwin:2001ku} for yet another model of UHECR generation in
neutron stars, involving planetoid impacts.} have shown that young
magnetized neutron stars in our own Galaxy may be one such
astrophysical system that satisfies the Hillas' criterion.

Neutron stars --endpoints of stellar evolution-- begin their life
rotating rapidly ($\Omega \sim 3000 {\rm ~rad~s}^{-1}$) and with
large surface magnetic fields ($B_S \gtrsim 10^{13}$ G). The
dipole component decreases as the cube of the distance from the
star's surface $B(r) = B_S (R_S / r)^3$, where the radius of the
star is $R_S \simeq L$. At the light cylinder $R_{lc} = c /\Omega
\sim 10^7 \Omega_{3k}^{-1} $  cm,  where $\Omega_{3k} \equiv
\Omega / 3000 \, {\rm \ rad \ s}^{-1}$, the dipole field cannot be
casually maintained, the field is mostly azimuthal, with field
lines spiraling outwards \cite{Michel}. Inside the light cylinder,
the magnetosphere corotates with the star, and the  density of material 
(mostly iron peak elements formed during the supernova event that
were stripped off the surface due to strong electric fields) has
the Goldreich-Julian value, $n_{GJ}(r) ={B(r) \Omega /( 4 \pi Z e
c} )$, where  $c$ is the speed of light \cite{Goldreich}. The
behavior of the plasma outside the light cylinder is still not yet
fully understood \cite{Gallant,Begelman,Chiueh,Melatos}, although
some analytical and numerical studies show the development of
kinetically dominated relativistic winds (see e.g.,
\cite{Begelman}).

Blasi et al. \cite{BLASI} assumed that the magnetic field in the
wind zone decreases  as $B(r) \lesssim B_{lc} R_{lc} / r$. For
surface fields of  $B_S \equiv 10^{13} \, B_{13}$ G, the field at
the light cylinder is $B_{lc} =10^{10}  \,  B_{13} \Omega_{3k}^3$
G. The maximum energy of particles that can be contained in the
wind near the light cylinder is
\begin{equation} E_{\rm max} = {Z e B_{\rm lc} R_{\rm lc} \over c}\simeq 8 \times
10^{20}  \, Z_{26} B_{13} \Omega_{3k}^2 \, {\rm eV} \ ,
\label{eq:Emax}
\end{equation} where $Z_{26} \equiv Z/26$.
The  typical  energy of the accelerated CRs, $E_{\rm cr}$,
can be estimated by considering the  magnetic energy per ion at
the light cylinder $ E_{\rm cr} \simeq B_{lc}^2/ 8 \pi n_{\rm
GJ}$. At the light cylinder $ n_{\rm GJ} = 1.7 \times 10^{11} \,
{B_{13} \Omega_{3k}^4 / Z}\, {\rm cm}^{-3}  $  which  gives
\begin{equation}
 E_{\rm cr} \simeq  4 \times 10^{20}\,  Z_{26} B_{13}
\Omega_{3k}^2 \, {\rm eV} \  \label{eq:Ecr} .
\end{equation}
In this model, as the  star spins down, the energy of the cosmic
ray particles ejected with the wind decreases.
The total fluence of UHECRs between energy $E$ and $E+dE$  is
\begin{equation} N(E) dE =  \frac{\dot {\cal N}}{\dot \Omega}\,
\frac{d\Omega}{dE}\,  dE \ , \label{eq:spec}
\end{equation}  where the particle  luminosity  is
\begin{equation}
\dot {\cal N} = \xi \, n_{GJ} \, \pi R_{lc}^2 c = 6 \times 10^{34}
\xi {B_{13} \Omega_{3k}^2 \over Z_{26}}   \,  {\rm s}^{-1}
\end{equation} and $\xi<1$  is the efficiency  for accelerating
 particles at the light cylinder. For
a spin down rate dominated by magnetic dipole radiation, given by
$
 I \Omega \dot \Omega =  - {B_S^2  R_S^6 \Omega^4 / 6 c^3}
$ where $I= 10^{45}$ g cm$^2$ is the moment of inertia,  the time
derivative of the spin frequency is $ \dot \Omega =    1.7 \times
10^{-5}  B_{13}^2 \Omega_{3k}^3 \, {\rm s}^{-1} $, and  Eq.
(\ref{eq:Ecr})  gives
\begin{equation}
\frac{dE}{d\Omega}= {1.7 \times 10^{-3} } \ {E \over \Omega_{3k}}
\ .
\end{equation}
Substituting in Eq. (\ref{eq:spec}), the particle spectrum from
each neutron star is
\begin{equation} N(E) = \xi{ \, 5.5 \times 10^{31} \over B_{13}
E_{20}  Z_{26} }  {\rm GeV}^{-1}    \ .
\end{equation}

Taking the confining volume for these particles to be $V_c$ and
the lifetime for confinement to be $t_c$, the UHECR density is
$n(E) = \epsilon N(E) t_c/\tau V_c$, and the flux at the surface
of the Earth is $F(E) = n(E) c/4$. For a characteristic
confinement dimension of $R = 10 ~R_1$  kpc, $V_c = 4 \pi R^3/3$
and  $t_c = QR/c$, where $Q > 1$ is a measure of the how well the
UHECR are trapped. With this in mind, the predicted
UHECR flux at the Earth is \cite{BLASI}\footnote{Note that the predicted spectrum of Eq.
(\ref{eq:spec2}) is very flat, $\gamma =1$,  which would agree
only with the lower end of the plausible range of  $\gamma$
observed at ultra-high energies. Propagation effects can produce
an energy dependence of the  confinement parameter $Q$ and,
correspondingly, a steepening of the spectrum toward the middle of
the observed range $1 \lesssim \gamma \lesssim 2$.  }
\begin{equation}
F(E)  = 10^{-24} {   \, \xi \epsilon Q  \over \tau_2 R_1^2 B_{13}
E_{20}  Z_{26} } \, {\rm GeV}^{-1}  {\rm cm}^{-2} {\rm s}^{-1} \ \
. \label{eq:spec2}
\end{equation}
Here, the fact that neutron stars are produced in our Galaxy at a
rate $1/\tau$, where $\tau \equiv 100 ~\tau_2$ yr, but that not
all them (but rather only a fraction $\epsilon$) have the required
magnetic fields, initial spin rates and magnetic field geometry to
allow efficient conversion of magnetic energy into kinetic energy
of the flow, was taken into account. By comparing with
observations,  the required efficiency factor, $\xi \epsilon$, can
be estimated, and it only needs to be $\xi \epsilon \gtrsim 4
\times 10^{-6} Q^{-1}$.

The condition that a young neutron star could produce the UHECRs\footnote{The gyroradius of these UHECRs  in a Galactic field of $\mu$G
strength  is considerably less than the typical distance to a
young neutron star ($\sim $ 8 kpc). Since the iron arrival
distribution  at $10^{20}$ eV probes similar trajectories to
protons at a few times $10^{18}$ eV, Galactic iron nuclei would
show a nearly isotropic distribution with a slight correlation
with the Galactic center and disk, at higher energies.}
is that $E_{cr}$ exceeds the needed energy when the envelope
becomes transparent (i.e. before the spinning rate of the neutron
star decreases to the level where the star is unable to emit
particles of the necessary energy), $E_{cr}(t_{tr}) > 10^{20}
E_{20}$ eV. This translates into the following  condition
\cite{BLASI}\footnote{A supernova that imparts $E_{SN}= 10^{51}
{\cal E}_{51}$ erg  to the stellar envelope of mass $M_{env}= 10
~M_1 ~{\rm M}_{\odot}$ is considered. Also, that the condition for
iron nuclei to traverse the supernova envelope without significant
losses is that $\Sigma \lesssim 100 \, $ g cm$^{-2}$
\cite{berezinsky-book}.}
\begin{equation}
 \Omega_{i} > {3000\, {\rm s^{-1}} \over  B_{13}^{1/2} \left [ 4
Z_{26} E_{20}^{-1} -  0.13 M_1 B_{13} {\cal E}_{51}^{-1/2}
\right]^{1/2}} \ . \label{tanto}
\end{equation}
Eq.~(\ref{tanto}) translates in turn into upper bounds on the
surface magnetic field strength and the star initial spin period
$P_i = 2\pi/\omega_i$,
\begin{equation}
B_{13} < \frac{31 \,Z_{26}\,E_{51}^{1/2}}{M_1\,E_{20}}\,,
\label{nhg}
\end{equation}
and
\begin{equation}
P_i < 8 \pi B_{13}^{1/2}\,Z_{26}\, E_{20}^{-1}\,.
\label{nhg2}
\end{equation}
For $M_1 = 2$ and $E_{20} = E_{51} = Z_{26} = 1$, Eq~(\ref{nhg})
gives $B_{13} < 15.4$, whereas Eq.~(\ref{nhg2}) leads to $P_i \alt
10$~ms, not very restrictive values for a young neutron star, see
for example the Parkes Multibeam Pulsar Survey.\footnote{The
Parkes multibeam pulsar survey is a large-scale survey of a narrow
strip of the inner Galactic plane ($|b|< 5^\circ$, $260^\circ < l
< 50^\circ$, see \cite{Kramer:2003py} and references therein). It
has much greater sensitivity than any previous survey to young and
distant pulsars along the Galactic plane, and it has resulted in
the detection of many previously unknown young pulsars,
potentially counterparts of unidentified $\gamma$-ray sources,
e.g. \cite{Torres:2001zb,Torres:2002rs}.}

\subsubsection{Magnetars}

Magnetars, neutron stars with surface dipole fields on the order
of $10^{15}$ G \cite{duncan,pac,kou00,kou98,kou99}, were also
proposed as plausible sites for the generation of UHECRs~\cite{Arons}.
Assuming that they occur in all galaxies which
form massive stars (then avoiding the large-distance GZK problem),
and that the UHECR are arriving from outside our own Galaxy,
the luminous infrared galaxies are preferred sites to search for a
magnetar origin of CRs (see \cite{Singh:2003xr} and below the
discussion in Section \ref{LIGS}).

The magnetar model for the acceleration of UHECRs proposed by
Arons \cite{Arons} is a variant of that using neutron stars
outlined in the previous Section. The theory predicts an injection
of charged particles with maximum energy
\begin{equation}
E_{\rm max} = Ze\Phi_{\rm i} = Ze \frac{B_* \Omega_{\rm
i}^2}{R_*^3 c^2} =
        3 \times 10^{22} Z B_{15} \Omega_4^2 \; {\rm eV},
\end{equation}
where $B_*$ is a magnetar's surface magnetic field, $B_{15} =
B_*/(10^{15}$~G), $\Omega_{\rm i} = 10^4 \Omega_4$ s$^{-1}$ is
the initial angular velocity of the neutron star, $R_{*}$ is the
stellar radius, and $c$ is the speed of light. The initial
rotation period is $P_i = 0.64/\Omega_4$ ms (if $ Z = 1-2$, one
requires $P_{\rm i} < 2-3$ ms for the model to be viable).

The ions actually gain their energy in the relativistic wind
electromagnetically expelled from the neutron star at distances
$r$ larger than the radii of the star and its magnetosphere. This
avoids catastrophic radiation losses; the electric potential in
the wind is $rE = rB = \Phi$. As the star spins down, as in the
model by Blasi et al., the
voltage and the maximum particle energies decline. Summing over the
formation and spindown event, one finds a per event injection
spectrum proportional to $f(E) = E^{-1} [1 + (E/E_{\rm g})]^{-1} $
for $E < E_{\rm max} $. Here $E_{\rm g}$ measures the importance
of gravitational wave losses (calculated for a star with static
non-axisymmetric quadrupole asymmetry) in spinning the star down.
When they exert torques larger than the electromagnetic torque,
the star spends less time at the fastest rotation rates (i.e. less
time accelerating the highest energy particles), thus causing a
steepening in the spectral slope at the highest energies.  If the
star has an internal magnetic field even stronger than the already
large surface field, equatorial ellipticities $\epsilon_{\rm e}$
in excess of $10^{-3}$ can exist, in which case $E_{\rm g}$ would
be less than $E_{\rm max}$.  Three cases of energy loss due to
gravitational radiation (GR) were considered in the model: no GR
loss ($\epsilon_{\rm e} =0, E_{\rm g} = \infty$); moderate GR loss
($\epsilon_{\rm e} = 0.01, E_{\rm g} = 3 \times 10^{20}$ eV);
strong GR loss ($\epsilon_{\rm e} = 0.1, E_{\rm g} = 3 \times
10^{18}$ eV). If one assumes that all magnetars have exactly the same
starting voltage ($10^{22.5}$~V), then  the model predicts that the
spectrum $E^3 J(E)$ should rise with $E$ above the energy $E_g =
2.8 \times 10^{20}$~eV, where the GZK loss rate becomes
approximately energy independent (unless the gravitational wave
losses are large)~\cite{Arons}. This prediction will certainly be
testable with the Pierre Auger Observatory.

The total number of particles injected per event is
\begin{equation}
    N_{\rm i} \approx 2 \frac{c^2 R_*^3 I}{ZeB_*} \approx
    \frac{10^{43}}{ZB_{15}},
\label{Ni}
\end{equation}
for a stellar radius of 10 km and a moment of inertia $I =
10^{45}$ cgs. The rate at which galaxies inject UHECR into the
universe in this model then is $\dot{n}_{\rm cr} = \nu_{\rm
mag}^{\rm fast} N_{\rm i} n_{\rm galaxy}$, where $n_{\rm galaxy}
\approx 0.02$ Mpc$^{-3}$ \cite{Blanton}, $N_{\rm i}$ is given by
Eq.~(\ref{Ni}), and $\nu_{\rm mag}^{\rm fast}$ is the birth rate
of rapidly rotating magnetars per galaxy. Multiplication of the
source spectrum $q(E) \propto \dot{n}_{\rm cr} f(E)$ by the energy
dependent GZK loss time yields a spectrum received at the Earth in
reasonable accord with the existing observations of UHECR
if $\nu_{\rm mag}^{\rm fast} \approx 10^{-5}$ yr$^{-1}$~\cite{Arons}.
That fast magnetar birth rate lies
between 1\% and 10\% of the total magnetar birth rate inferred for
our galaxy, and about 0.1\% of the total core collapse supernova
rate in average star forming galaxy,
$\sim 10^{-2}$ yr$^{-1}$~\cite{Cappellaro}.

\subsubsection{UHECRs from a pulsar in Cygnus OB2?}

As discussed in Sec.~\ref{vanilla1}, some evidence may be emerging
for a CR accelerator in the Cygnus spiral arm. The HEGRA
experiment has detected an extended TeV $\gamma$-ray source in the
Cygnus region with no clear counterpart and a spectrum not easily
accommodated with leptonic  radiation~\cite{Aharonian:2002ij}. The
difficulty in accommodating the spectrum by conventional
electromagnetic mechanisms has been exacerbated by the failure of
Chandra and VLA to detect significant levels of X-rays or
radiowaves signaling acceleration of any
electrons~\cite{Butt:2003xc}. Especially intriguing is the
possible association of this source with part of Cygnus OB2 itself
\cite{Torres:2003ur}, a cluster of several thousands young, hot OB
stars with a total mass of $\sim 10^4
\,M_\odot$~\cite{Knodlseder:2000vq}. At a relatively small
distance to Earth, $\approx 1.7$~kpc, this is the largest massive
Galactic association. It has a diameter of $\approx 60$~pc and a
core radius of $\sim 10$~pc. The typical main sequence evolution
lifetime of massive O stars is ${\cal O}$ (Myr) and a few tens Myr
for massive B stars. Since the O-star population should pass
through the Wolf-Rayet phase and explode as supernovae, very fast
pulsars are expected to be born in explosions of these massive
stars at a rate of about one every ten thousand years.

Apart from the mentioned interpretation of a hadronic production
of the TeV radiation within the winds of  outlying OB stars of Cyg
OB2 \cite{Torres:2003ur}, it was recently put forward that the TeV
emission reported by HEGRA  and the CR anisotropy observed at
about $10^{18}$~eV in the direction of the Cygnus region can be
related to a young pulsar and its pulsar wind nebulae (PWN), born
in the Cygnus OB2 association a few ten thousands years
ago~\cite{Bednarek:2003cx}. The TeV $\gamma$-ray emission would
originate in the PWN as a result of interactions of high energy
hadrons and/or leptons, whereas there would be a directional CR
signal due to neutrons that are dissolved from heavy nuclei
accelerated by the pulsar.\footnote{A similar scenario would
explain the anisotropy spot towards the Galactic
Center~\cite{Bednarek:2001ir}.} Within this model, however,
it is hard to explain the absence of counterparts at lower (EGRET)
energies at the location of the TeV source, as well as the absence
of a stronger X-ray source.\footnote{Indeed, location problems may
arise as well: the PWN size (given that the TeV source size is 6
pc if indeed located at 1.7 kpc, and that the source is diffuse)
would make the PWN substantially larger than both Vela's PWN
($\sim $0.1 pc at 250 pc) and Crab's ($\sim $1 pc). Additionally,
if one were to associate the nearby EGRET source 3EG 2033+4118
with the putative pulsar itself, it is unclear whether the PWN
hypothesis for the TeV source would imply that it is only a {\it
one sided PWN} (Y. Butt, private comunication). } However,
disregarding these difficulties, an interesting consequence for CR
physics can be inferred.

 As already discussed
in Sec.~\ref{NS}, heavy nuclei can attain ultra high energies
through magnetic sling shots inside the neutron star wind zone.
The small part of nuclei, which escaped from the pulsar wind
nebula, are likely to be captured by strong magnetic fields of
dense regions of the OB association. Magnetic field strengths in
dense molecular clouds are expected to be $\sim
1$~mG~\cite{Crutcher}. Thus, nuclei with $E/Z \sim 1$~EeV
propagate attaining Bohm diffusion~\cite{Bednarek:2003cx}. The
resulting time delay of several thousand
years~\cite{Bednarek:2001av} produces a steepening of the
characteristic power law injection spectrum, $\propto E^{-2},$ of
the pulsar nebula.\footnote{Note that once diffusion has been
established, additional Rayleigh steps in the Galactic magnetic
field do not change the spectral index ($\sim -3$) significantly.}
In their random traversal of the OB association, the nuclei
undergo photodisintegration on the far infrared thermal photon
population and liberate neutrons.

The thermal photon density at the source reads,
\begin{equation}
n_{\rm IR} = \frac{T^3}{\pi^2} \,\, \int_{x_1}^{x_2} \,
\frac{x^2\, dx}{e^x -1} \, \approx \frac{(4.37\,\,T)^3}{\pi^2}\,  \int_{x_1}^{x_2} \,
\frac{x^2\, dx}{e^x -1}\,\,{\rm cm^{-3}},
\label{hhggh}
\end{equation}
where $T$ is the kinetic temperature of the
molecular cloud, $x \equiv E_\gamma/T,$ and
$E_{\gamma}$ is the target photon energy in degrees K.
The relevant photon energy range is established via standard kinematics
of the photodisintegration
process,
\begin{equation}
E_{\gamma}=\frac{E^*\ m_N}{(E_A/A)}\ = 10\, \frac{E^*_{\rm
MeV}}{E_{N,{\rm PeV}}}\, {\rm K} \ , \label{kinem}
\end{equation}
where $5 <E^*_{\rm MeV} <25$ is the energy region for giant dipole resonance
contribution in the nucleus rest frame, and $E_A/A \equiv
E_{N,{\rm PeV}}$ is the lab frame energy/nucleon. For
$E_{N,{\rm PeV}}\sim 10^3,$ Eq. (\ref{kinem})
leads to  $x_1 = 50/T$ and $x_2 = 250/T,$ with $T$ in degree Kelvin.
Molecular clouds with HII regions have temperatures between 15 and
100~K~\cite{Wilson:1997ud}, thus taking an average photodisintegration cross section of 40~mb,\footnote{The photoabsorption cross  section roughly obeys the Thomas-Reiche-Kuhn
sume rule, i.e., $\Sigma_d = 60 NZ/A$ mb-MeV~\cite{Stecker:1998ib}.} we find
that the nucleus
mean free path lies between 0.80 and 380 pc. This corresponds to a
reaction time between 4 to 1500 yr $ \ll$ time delay, allowing
sufficient neutron production to explain the
anisotropy~\cite{Anchordoqui:2003vc}.

The Galactic anisotropy observed by the various collaborations
spans the energy range 0.8 to 2.0 EeV. The lower cutoff specifies
that only neutrons with EeV energies and above have a boosted
$c\tau_n$ sufficiently large to serve as Galactic messengers.
The decay mean free path of a neutron is
$c\,\Gamma_n\,\overline\tau_n=10\,(E_n/{\rm EeV})$~kpc,
the lifetime being boosted from its rest-frame value
$\overline\tau_n=886$~seconds to its lab value via
$\Gamma_n=E_n/m_n$. Actually, the broad scale anisotropy from the
direction of the GP
reported by Fly's Eye Collaboration~\cite{Bird:1998nu} peaks in the energy bin $0.4 - 1.0$ EeV, but
persists with statistical significance to energies as low as 0.2 EeV.
This implies that if neutrons are the carriers of the anisotropy,
there {\em needs to be} some contribution from at least one source
closer than 3 - 4 kpc. Interestingly, the full Fly's Eye data include a
directional signal from the Cygnus region which was somewhat lost in
unsuccessful attempts~\cite{Cassiday:kw,Teshima:1989fc} to
relate it to $\gamma$-ray emission from Cygnus X-3. The upper cutoff
reflects an important feature
of  photodisintegration at the source: heavy nuclei with energies
in the vicinity of the ankle will fragment to neutrons with
energies about an order of magnitude smaller. To account for the
largest neutron energies, it may be necessary to populate the heavier nucleus
spectrum in the region above the ankle.\footnote{To produce the highest
energy neutrons (with energy $\alt 10^{18.2}$~eV) via
photodisintegration of medium mass (say, $A= 10 - 20$) nuclei, one needs
primary particles with energies $\alt 10^{19.2}$~eV. For a falling spectrum
$\propto E^{-3},$ the
medium mass nucleus population at $10^{19.2}$~eV is roughly 3 orders of
magnitude smaller than the diffuse flux at $10^{18.2}$~eV. This means that
it is about 1.6 orders of magnitude smaller than the reported CR excess
(which is about 4\% of the diffuse flux, see Sec.~\ref{DoAD}). However, since
each nucleus produces roughly $A-Z$  neutrons, the average number of
liberated neutrons is on the same order of magnitude,
$1.6 - \log_{10} (A-Z)$, than the obseved neutron
population at $\alt 10^{18.2}$~eV.} This is not a problem --
one fully expects the emerging harder extragalactic spectrum to
overtake and hide the steeply falling galactic population. It is
not therefore surprising that in order to fit the spectrum in the
anisotropy region and maintain continuity to the ankle region
without introducing a cutoff, the AGASA Collaboration required a
spectrum $\propto E^{-3}$ or steeper~\cite{Hayashida:1998qb}.

For every surviving neutron at $\sim$~EeV,
there are many neutrons at lower energy that decay via
$n\rightarrow p+e^- + \nuebar.$ The proton is bent by the Galactic magnetic
field, the electron quickly loses energy via synchrotron radiation,
and the $\nuebar$ travels
along the initial neutron direction, producing a directed TeV energy
beam~\cite{Anchordoqui:2003vc}. The sensitivity of forthcoming neutrino telescopes to this signal is
discussed in Sec.~\ref{AoI}.

\subsection{Radio Galaxies and Active Galactic Nuclei}

\subsubsection{Definitions}

\begin{figure}[t]
\centering
\includegraphics[width=14cm, height=12cm]{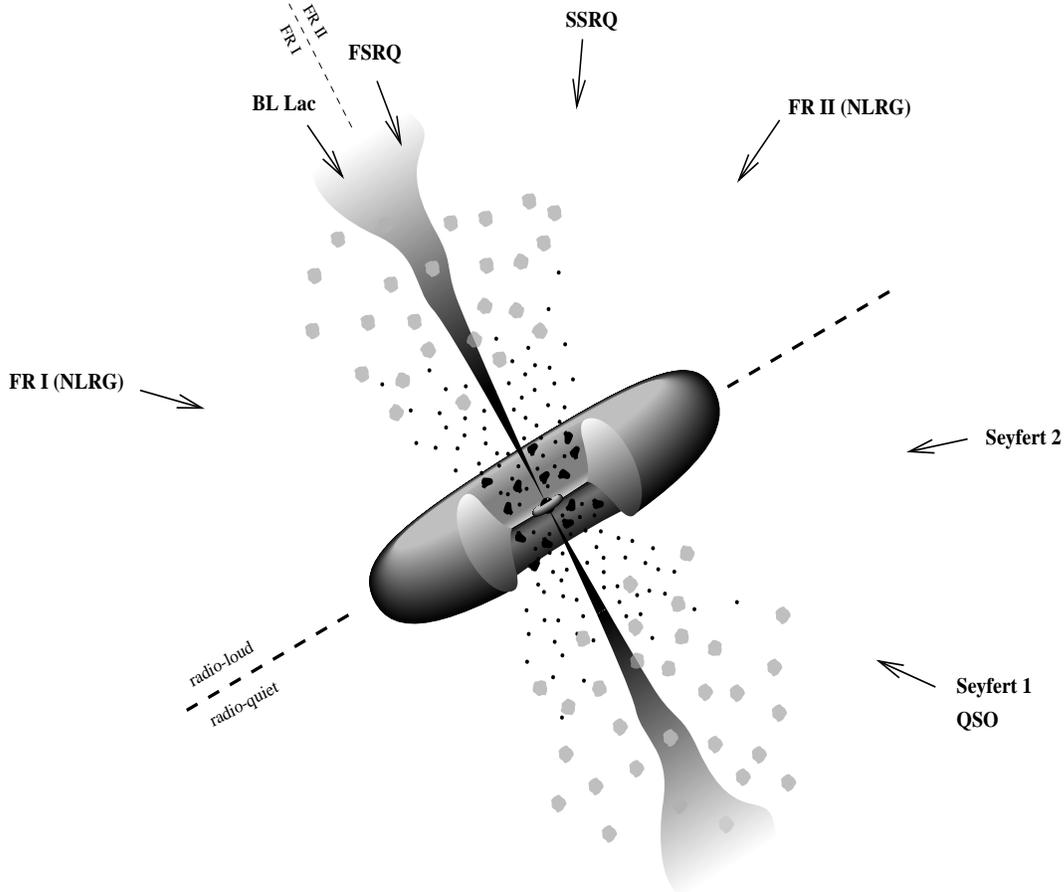}
\caption{The unification model for AGNs. The components of the
figure are discussed in the text. Blazars are those AGNs for which
the jets are close to line of sight. A regular quasar or a Seyfert
1 galaxy is observed if the orientation angle is $\sim 30^{{\rm
o}}$, where the narrow-line and broad-line regions are visible. At
larger angular offsets, the broad-line region will be hidden by
the torus, the corresponding class being Seyfert 2 galaxies.
Perpendicular to the jet axis, the full extent of the jets may be
seen particular at low frequencies, giving rise to a morphology
typical of radio galaxies. The figure is adapted from Refs.
\cite{Urry,Padovani,Collmar,Torres:2003zb}. \label{uni}}
\end{figure}

Blazars are active galactic nuclei (AGNs) with a) strong flat spectrum
radio emission [the
power law index $\alpha > -0.5$, with $S(\nu) \propto \nu^\alpha$]
and/or b) significant optical polarization, and/or c) significant
flux variability in the optical and in other wavelengths. When the
optical variability occurs on short timescales, the objects are
referred to as optically violently variable --OVV-- quasars. The
blazar classification also includes BL Lacertae (BL Lac) objects,
which present a complete or nearly complete lack of emission
lines, and highly polarized quasars (HPQs). It also refers,
sometimes, to flat spectrum radio quasars (FSRQs), although these
are generally more distant, more luminous, and have stronger
emission lines. Within the unification model, the underlying
scenario for all AGNs is intrinsically similar. At the very center
of the galaxy there is a supermassive black hole ($\sim$10$^6$ to
$\sim$10$^{10}\,$M$_{\odot}$) which accretes galactic matter
forming an accretion disk. Broad emission lines are produced in
clouds orbiting above the disc at high velocity, the broad line
region (BLR) and this central region is surrounded by an
extended, dusty, molecular torus. A hot electron corona populates
the inner region, probably generating continuum X-ray emission.
Narrower emission lines are produced in clouds moving much farther
from the central black hole.  Two-sided jets of relativistic
particles emanate perpendicular to the plane of the accretion
disc, the generation of which is still not fully understood.
Unification of different AGN classes is achieved taking into
account the intrinsic anisotropy of the phenomenon, is shown in
Fig.~\ref{uni} (see Refs. \cite{Urry,Padovani,Collmar,Torres:2003zb} for
further and more detailed discussions).

For example, Seyfert galaxies possess a dusty torus of gas at distances
intermediate between the BLR and NLR (narrow line region). An observer whose line of sight to the
black hole intercepts this torus would see a heavily reddened (or completely
extinguished) BLR and central continuum radiation but an
unreddened NLR. This would be identified with a Seyfert 2 galaxy. If the
line-of-sight does not intercept the torus, the central regions of the
nucleus can be observed directly, leading to a Seyfert 1 classification.
Radio loud quasars are then objects in which the line-of-sight
is close to the jet cone of the source. In the cases in which we are not
directly looking into the jet cone --blazars where relativistic effects
produce highly variable and continuum dominated emission-- emission from the
BLR can be observed. Objects with larger inclinations have a less dominant
central continuum flux, resulting in Fanaroff-Riley II (FRII) galaxies. If
the torus surrounding
the black hole obscures the BLR, a narrow line radio galaxy (NLRG) can be
observed. It is not clear how FRI radio galaxies fit into such a scheme.
Clearly, some (as yet unknown) physical
mechanism, probably related to source power, produces different radio
morphologies in FRI and FRII sources.\footnote{The Faranoff-Riley
classification is based on one parameter, $R_{\rm
FR}$, the ratio of the distance between the regions of highest surface
brightness on opposite sides of the central galaxy to the total extent of
the source. Objects with $R_{\rm FR} < 0.5$ are classified as FRI,
whereas those with $R_{\rm FR}>0.5 $ are classified as FRII. It is found
that the brighter sources are all FRII class, although the distinction
between classes is not clear cut
in luminosities (for further details see page 220 of Ref.~\cite{Narlikar}).}
 Some blazars may
be beamed FRI objects, but there is a lack of broad-line FRI radio galaxies
\cite{Urry}. This make the classification within the unified scheme harder to
achieve.


\subsubsection{Radiogalaxies}

FRII galaxies~\cite{Fanaroff} are the largest
known dissipative objects (non-thermal sources) in the Universe.
Localized regions of intense synchrotron emission, known as ``hot
spots'', are observed within their lobes. These regions are
presumably produced when the bulk kinetic energy of the jets
ejected by a central active nucleus (supermassive black hole +
accretion disk) is reconverted into relativistic particles and
turbulent fields at a ``working surface'' in the head of the jets
\cite{Blandford}. Specifically, the speed $v_{\rm h}$ with which
the head of a jet advances into the intergalactic medium of
particle density $n_{\rm e}$ can be obtained by balancing the
momentum flux in the jet against the momentum flux of the
surrounding medium. Measured in the frame comoving with the
advancing head, $v_{\rm h}\approx \;v_{\rm j}\,[ 1 + ( n_{\rm e}
/n_{\rm j})^{1/2}]^{-1}$, where $n_{\rm j}$ and $v_{\rm j}$ are
the particle density and the velocity of the jet flow,
respectively. $v_{\rm j}> v_{\rm h}$ for $n_{\rm e} \geq n_{\rm
j}$, and the jet will decelerate. The result is the formation of a
strong collisionless shock, which is responsible for particle
reacceleration and magnetic field amplification~\cite{Begelman2}.
The acceleration of particles up to ultrarelativistic energies in
the hot spots is the result of repeated scattering back and forth
across the shock front \cite{Biermann:ep}. Dimensional arguments
suggest that the energy density per unit of wave number of MHD
turbulence is of the Kolmogorov type~\cite{Kolmogorov}, and so for
strong shocks the acceleration time for protons is~\cite{Drury}
\begin{equation}
\tau_{\rm acc} \simeq \frac{40}{\pi}\, \frac{1}{c\,\beta_{\rm
jet}^2} \,\frac{1}{u}\,\left(\frac{E}{eB}\right)^{1/3}\, R^{2/3}
\label{acc}
\end{equation}
where $\beta_{\rm jet}$ is the jet velocity in units of $c,$ $u$
is the ratio of turbulent to ambient magnetic energy density in
the region of the shock (of radius $R$), and $B$ is the total
magnetic field strength. The acceleration process will be
efficient as long as the energy losses by synchrotron radiation
and photon--proton interactions do not become dominant. The
subtleties surrounding the conversion of a particle kinetic energy
into radiation provide ample material for
discussion~\cite{Biermann:ep,Mannheim:ac,Mannheim:1998fs,Mannheim:1999hf,Mannheim:jg,Aharonian:2000pv}. The proton blazar model relates $\gamma$-ray emission
to the development of electromagnetic cascades triggered by
secondary photomeson products that cool instantaneously via
synchrotron
radiation~\cite{Biermann:ep,Mannheim:ac,Mannheim:1998fs,Mannheim:1999hf,Mannheim:jg}.
The synchrotron loss time for protons is given by~\cite{Rybicki}
\begin{equation}
\tau_{\rm syn} \sim \frac{6\, \pi\, m_p^3\,c}{\sigma_{\rm
T}\,m_e^2\,\Gamma\,B^2}\,, \label{tausyn}
\end{equation}
where $m_e$, $m_p$, $\sigma_{\rm T}$ and $\Gamma$ are the electron
mass, proton mass, Thomson cross section, and Lorentz factor,
respectively. The characteristic single photon energy in
synchrotron radiation emitted by an electron is
\begin{equation}
E_\gamma = \left(\frac{3}{2}\right)^{1/2} \frac{h\,e\,E^2\,B}{2
\,\pi\, m_e^3 \, c^5} \sim 5.4\times 10^{-2}\, B_{\mu{\rm G}}\,
E_{20}^{2} \, {\rm TeV}\ \ .\label{synch}
\end{equation}
For a proton this number  is $(m_p/m_e)^3 \sim 6 \times 10^9$
times smaller. High energy $\gamma$-ray production through proton
synchrotron radiation requires very large, ${\cal O}(100\ {\rm
G}),$ magnetic fields. Considering an average cross section
$\bar{\sigma}_{\gamma p}$ for the three dominant pion--producing
interactions~\cite{Armstrong:1971ns}, $ \gamma p \rightarrow p
\pi^0\,, \gamma  p \rightarrow n  \pi^+\,, \gamma p\rightarrow p
\pi^+  \pi^- \,, $ the time scale of the energy losses, including
synchrotron and photon interaction losses,
reads~\cite{Biermann:ep}
\begin{equation}
\tau_{\rm loss} \simeq \frac{6\pi\ m_p^4\ c^3}{\sigma_{\rm T}\
m_e^2\ B^2\ (1+Aa)}\ E^{-1}\ = \frac{\tau_{\rm syn}}{1 + Aa} \,,
\end{equation}
where $a$ stands for the ratio of photon to magnetic energy
densities and $A$ gives a measure of the relative strength of
$\gamma p $ interactions versus the synchrotron emission. Note
that the second channel involves the creation of ultrarelativistic
neutrons (but $\Gamma_n \alt  \Gamma_p$) with mean free path in
the observer rest frame given by $\lambda_n = \Gamma_n c \tau_n$,
where $\tau_n \sim 900$~s, is the neutron lifetime. Since
$\lambda_n > \lambda_p$ for $\Gamma_n \alt \Gamma_{p \ {\rm
max}}$, such neutrons can readily escape the system, thereby
modifying the high end of the proton spectrum. Biermann and
Strittmatter~\cite{Biermann:ep} have estimated that $A \approx
200$, almost independently of the source parameters. The most
energetic protons injected in the intergalactic medium will have
an energy that can be obtained by balancing the energy gains and
losses~\cite{Anchordoqui:2001bs}
\begin{equation}
E_{20}=1.4\times 10^5\,\,B_{\mu{\rm G}}^{-5/4}\,\,\beta_{\rm
jet}^{3/2} \,\,u^{3/4}\,\,R_{\rm kpc}^{-1/2}\,\,(1+Aa)^{-3/4}\ ,
\label{ab}
\end{equation}
where $R_{\rm kpc}\equiv R/1\,{\rm kpc}$.

For typical hot-spot conditions ($B \sim 300~\mu$G, $u \sim 0.5$,
and $\beta_{\rm jet} \sim 0.3$) and assuming that the magnetic
field of the hot spot is limited to the observable region, one
obtains $E < 5 \times 10^{20}$~eV for
$a<0.1$~\cite{Rachen:1992pg}.\footnote{The shock structure in hot
spots is likely to be much more extended than the visible region
in the non-thermal radioemission, as suggested by
magnetohydrodynamical modeling~\cite{Rachen:1992pg}.} Particles
can also attain ultrahigh energies ($E \agt 10^{20}$~eV) within
the jets or the AGNs themselves. For instance, the knot A in the
M87 jet, with a length scale $l_{87} \sim 2 \times 10^{20}$ cm,
has a magnetic field strength $B_{87} \sim
300~\mu$G~\cite{Stocke}.  Typical AGN sizes are $l_{\rm AGN} \sim
10^{15}$~cm, and $B_{\rm AGN} \sim 1$~G~\cite{Angel}.
Observational evidence suggests that in the jets $a\ll 1$, whereas
$a \sim 1$ for AGNs~\cite{Biermann:ep}.

\subsubsection{Cen A: The source of most UHECRs observed at Earth?}
\label{scena}

Centaurus A (Cen A) is the nearest active galaxy, $\sim
3.4$~Mpc~\cite{Israel}. It is a complex FRI radio-loud source
identified at optical frequencies with the galaxy NGC 5128.
Different multi-wavelength studies have revealed that it is comprised of
a compact core, a jet also visible at $X$-ray frequencies, a weak
counterjet, two inner lobes, a kpc-scale middle lobe, and two
giant outer lobes. The jet would be responsible for the formation
of the northern inner and middle lobes when interacting with the
interstellar and intergalactic media, respectively. There appears
to be a compact structure in the northern  lobe, at the
extrapolated end of the jet. This structure resembles the hot
spots such as those existing at the extremities of FRII galaxies.
However, at Cen A, it lies at the side of the lobe rather than at
the most distant northern edge, and the brightness contrast (hot
spot to lobe) is not as extreme~\cite{Burns}.

Low resolution polarization measurements in the region of the
suspected hot spot give magnetic fields as high as $25\ \mu{\rm
G}$ \cite{Burns}. However, in certain regions where measurements
at both high and low resolution are available, the $B$-field
amplitude at high resolution can be seen to be twice that at low
resolution. The higher resolution can reveal amplification in the
post-shock region \cite{Landau}, yielding $B$-fields possibly as
high as $50-60\ \mu{\rm G}$~\cite{Romero:ed,Romero:1995tn}. The
radio-visible size of the hot spot can be directly measured from
the large scale map~\cite{Junkes}, giving $R_{\rm HS}\simeq 2$
kpc. The actual size can be larger by a factor $\sim 2$ because of
uncertainties in the angular projection of this region along the
line of sight.\footnote{For example, an explanation of the
apparent absence of a counterjet in Cen A via relativistic beaming
suggests that the angle of the visible jet axis with respect to
the line of sight is at most 36$^{\circ}$~\cite{Burns}, which
could lead to a doubling of the hot spot radius. It should be
remarked that for a distance of 3.4 Mpc, the extent of the entire
source has a reasonable size even with this small angle.} Then, if
the magnetic field of the hot spot is confined to the visible
region, the limiting energy imposed by the Hillas' criterion is
$E_{\rm max} \sim 10^{20.6}$ eV.

Estimates of the radio spectral index of synchrotron emission in
the hot spot and the observed degree of linear polarization in the
same region suggests that the ratio of turbulent to ambient
magnetic energy density in the region of the shock is
$u \sim 0.4$~\cite{Combi}. The jet velocity is model dependent: possible values
range from $\sim 500$~km s$^{-1}$ to $0.99\, c$~\cite{Burns}. For FRI
galaxies, the ratio of photon to magnetic energy densities, $a$,
is expected to be $\ll 1$. Now, by replacing these numbers into
Eq.~(\ref{ab}), one can easily see that Cen A can accelerate particles
to energies~$\agt 10^{20}$~eV, with a maximum attainable energy set
by the Hillas' criterion.\footnote{It was recently 
suggested~\cite{Schopper:2001xt} that the 
25 kpc jet (with $B \approx 10~\mu$G) of Cen A  could be another promising 
region for proton acceleration to ultra high energies.}

Recent observations of the $\gamma$ ray flux for energies $>100$ MeV
by EGRET~\cite{Sreekumar:1999xw} allow an estimate $L_{\gamma}
\sim 10^{41}\ \es$ for the source.\footnote{Note that the received
radiation is negligibly affected by interactions with the various
radiation backgrounds~\cite{Aharonian:2000pv}.} This value of
$L_{\gamma}$ is consistent with an earlier observation of photons
in the TeV-range during a period of elevated activity
\cite{Grindlay}, and is considerably smaller than the estimated
bolometric luminosity $L_{\rm bol}\sim 10^{43}\es$\cite{Israel}.
Data across the entire $\gamma$ ray bandwidth of Cen A is given in
Ref.\cite{Steinle}, reaching energies as high as
150~TeV~\cite{Clay:uy}, though data at this energy await
confirmation. For  values of $B$ in the $\mu$G range, substantial
proton synchrotron cooling is suppressed, allowing the production
of high energy electrons through photomeson processes. The average
energy of synchrotron photons scales as $\overline{E}_\gamma
\simeq 0.29 E_\gamma$ \cite{Ginzburg}. With this in mind, it is
straightforward to see that to account for TeV photons, Cen A
should harbor a population of ultra-relativistic electrons with $E
\sim 6 \times 10^{18}$ eV. We further note that this would require
the presence of protons with energies between one and two orders
of magnitude larger, since the electrons are produced as
secondaries.\footnote{Consecutive factors of $\sim 2$ energy loss
occur in the processes $p\gamma\rightarrow N\pi^0,\
\pi^0\rightarrow \gamma\gamma,\ \gamma\rightarrow e^+e^-.$
Eq.(\ref{synch}) then implies proton energies of $\sim 10^{20}$ eV
for 100 TeV photons.}

There are plausible physical
arguments~\cite{Mannheim:jg,Waxman:1995vg} as well as some
observational reasons \cite{Rawlings} to believe that when proton
acceleration is being limited by energy losses, the CR luminosity
$L_{\rm CR}\approx L_{\gamma}$. Defining $\epsilon$, the
efficiency of UHECR production compared to high
energy $\gamma$ production -- from the above, $\epsilon\simeq 1$
-- and using equal power per decade over the interval $(\emin,\
\emax)$, the source luminosity is found to be~\cite{Farrar:2000nw}
\begin{equation}
\frac{E^2 \,dN^{p+n}_0}{dE\,dt} \, \approx \frac{6.3\,\epsilon\lf
\,10^{52} {\rm eV/s}}{\ln(\emax/\emin)}\,, \label{2}
\end{equation}
where $\lf \equiv$ luminosity of Cen A$/10^{41}\es$ and the
subscript ``0'' refers to quantities at the source.

For fiducial values,
$B=0.5~\mu$G, $\ell = 0.5$~Mpc, the diffusive
distance traveled by CRs with $E=10^{19}$\ eV, is
$c \tau_D=50$~Mpc $\gg d =3.4$~Mpc. Moreover, one can easily check
that for 3.4~Mpc the diffusion
time of any proton with energy above the photopion production threshold is
always less than the GZK-time, and consequently energy losses can be safely
neglected. This implies that
the density of protons at the present time $t$
of energy $E$
at a distance $r$ from Cen A (which is assumed to be continuously emitting at
a constant spectral rate $dN^{p+n}_0/dE\,dt$ from time $t_{\rm on}$ until
the present) can be obtained by solving the Kolmogorov-diffusive-equation,
and is found to be~\cite{Anchordoqui:2001nt}
\begin{equation}
\frac{dn(r,t)}{dE}  =  \frac{dN^{p+n}_0}{dE\,dt}
\frac{1}{[4\pi D(E)]^{3/2}} \int_{t_{\rm on}}^t dt'\,
\frac{e^{-r^2/4D(t-t')}}{(t-t')^{3/2}}
  =  \frac{dN^{p+n}_0}{dE\,dt}
\frac{1}{4\pi D(E)r} \,\,I(x)\,,
\label{22}
\end{equation}
where $D(E)$ is the diffusion coefficient given in Eq.~(\ref{D(E)}),
$x = 4D\ton/r^2 \equiv \ton/\tau_D$, $\ton=t-t_{\rm on},$ and
\begin{equation}
I(x) = \frac{1}{\sqrt{\pi}} \int_{1/x}^\infty
\frac{du}{\sqrt{u}} \,\, e^{-u}\ \ . \label{I}
\end{equation}
For $\ton\rightarrow \infty$, the density approaches its time-independent
equilibrium value $n_{\rm eq}$, while for $\ton= \tau_D$,
$n/n_{\rm eq} = 0.16$.

To estimate the power of Cen A,
one can  evaluate the energy-weighted approximately isotropic
proton flux at $1.5\times 10^{19}$ eV, which lies in the center of
the flat ``low'' energy region of the spectrum,
\begin{equation}
E^3 J_p(E)  =
 \frac{Ec}{(4\pi)^2d\,D(E)} \frac{E^2 \,dN^{p+n}_0}{dE\,dt}\,
I(t/\tau_D)  \approx  7.6 \times 10^{24} \, \epsilon \lf\, I \,\,
{\rm eV}^2 \, {\rm m}^{-2} \, {\rm s}^{-1} \, {\rm sr}^{-1}.
\label{jpt}
\end{equation}
In Eq.~(\ref{jpt}) we have used the fiducial values of $B\
\mbox{and}\ \ell$ as given in the previous paragraph, and set
$\emin = 1 \times 10^{19}$~eV, $\emax = 4 \times 10^{20}$~eV. As
noted by Farrar and Piran~\cite{Farrar:2000nw}, by stretching the
source parameters the ``low'' energy flux from Cen A could be
comparable to that of all other sources in the Universe.
To this end, first fix $\epsilon\, \lf\, I =0.40$, after
comparing Eq.~(\ref{jpt}) to the observed CR-flux by AGASA:
$E^3 J_{\rm obs}(E)  = 10^{24.5}$ eV$^2$ m$^{-2}$ s$^{-1}$
sr$^{-1}$~\cite{Hayashida:2000zr}. Next,
$\epsilon \lf\simeq 1,$ determines $I\simeq 0.40,$ and
consequently the required age of the source $\ton$ to be about
400~Myr, which appears plausible \cite{Begelman2,Rawlings}.
To maintain flux at the ``ankle'' for the same $\ton$, one
requires an approximate doubling of $L_{\rm CR}$ at $5\times
10^{18}$ eV. Because of the larger diffusive time delay at this
energy, this translates into an increased luminosity in the early
phase of Cen A. From Eq.~(\ref{synch}), the associated synchrotron
photons are emitted at energies $< 30$ MeV. The increase in
radiation luminosity in this region is not inconsistent with the
flattening of the spectrum observed at lower
energies\cite{Kinzer,Steinle:98}.

Having identified Cen A to plausibly be a powerful source of UHECRs, we
now explore whether $B$-field deflections provide
correct directional properties, i.e., sufficient isotropy.
This can be found by computing the incoming current flux density
$D\nabla n$ as viewed by an observer on Earth, and one finds for a
continuously-emitting source
a distribution $\sim(1+\alpha \cos\theta)$
about the direction of the source, where $\theta$ is the angle to
the zenith and
\begin{equation}
\alpha = \frac{2D(E)}{cr}\, \frac{I\pri}{I} \,, \hspace{3cm}
I\pri(x) = \frac{1}{\sqrt{\pi}} \int_{1/x}^\infty du\ \sqrt{u}
\,\, e^{-u},
\label{anisotropy}
\ee
with $x=\ton/\tau_D$, and $I$ as defined in
Eq.~(\ref{I})~\cite{Anchordoqui:2001nt}. For our choices of $B$ and
$\ell,$ and $T_{\rm on}=400$~Myr, we
find for $E=10^{19}$~eV $(E=10^{20}$~eV) that $\alpha = 0.04\ (\alpha = 0.07).$
This is in complete agreement with the upper bounds on dipole anisotropies
recently reported by  HiRes Collaboration~\cite{Abbasi:2003tk}. One caveat is
that the large deflection angle of
the highest energy Fly's Eye event with respect to the line of sight to
Cen A must be explained as a
$2\sigma$ fluctuation~\cite{Isola:2001ng}. Additionally, Monte Carlo
simulations~\cite{Isola:2002ei} show the predicted auto-correlation function
is not consistent with the clustering at small scale reported  by
AGASA Collaboration~\cite{Hayashida:2000zr}. Therefore, if the hypothesis of
CR pairing proposed by AGASA Collaboration is confirmed by future data, it
will constitute a serious objection to the model outlined above.
On the other hand, an interesting observational feature for a Cen A origin
of UHECRs is the possible detection of neutrons, which at the highest
energies could survive decay and produce a spike in the direction of the
source~\cite{Anchordoqui:2001nt}. The estimated event rate at PAO is about
2 direct events per year, against negligible background.
Thus, in a few years of running, the hypothesis of Cen A as the source of
most UHECRs observed at Earth can be directly tested.

\subsubsection{M87: The end of all roads?}

M87 is a giant radio galaxy for which there has been a recent
report of a TeV excess at a level of 4$\sigma$
\cite{Aharonian:2003tr}. It is also expected to be a source for
GLAST, having an EGRET upper limit of $2.8\times 10^{-8}$ photons
cm$^{-2}$ s$^{-1}$ above 100 MeV (Reimer, private communication,
see also the limit imposed in Ref. \cite{sre}), and comparable
theoretical flux predictions \cite{DR,Protheroe:2002gv}.

M87 was thought as a high-energy CR emitter since quite long ago
\cite{Ginzburg:63,Wdowczyk:sg}. At a distance of 16.3 Mpc
\cite{Cohen}, it is the dominant radio galaxy in the Virgo cluster
($l = 282^\circ, \, b=74^\circ$)~\cite{Blandford:1999}. The
emission of synchrotron radiation with a steep cutoff at
frequencies about $3 \times 10^{14}$~Hz from its radiojets and hot
spots~\cite{Heinz,Bicknell} implies an initial turbulence
injection scale having the Larmor radius of protons at
$10^{21}$~eV.

The major difficulty with a M87 generation of UHECRs is the
observation of the nearly isotropic distribution of the CR arrival
directions. One can again argue that the orbits are bent. However,
the bending cannot add substantially to the travel time, otherwise
the energy would be GZK-degraded. An interesting explanation to
overcome this difficulty relies on a Galactic wind, akin the solar
wind, that would bend all the orbits of the highest energy CRs
towards M87~\cite{Ahn:1999jd,Biermann:fd}. Indeed, it has long
been expected that such a kind of wind is active in our
Galaxy~\cite{Burke,Johnson,Mathews}. In the analysis
of~\cite{Ahn:1999jd}, it was assumed that the magnetic field in
the Galactic wind has a dominant azimuthal component, with the
same sign everywhere. This is because in a spherical wind the
polar component of the magnetic field becomes negligible rather
quickly, decaying like $1/r^2$, and thus the azimuthal part of the
magnetic field quickly becomes dominant, with $B_\phi \sim \sin
\theta/r$ in polar coordinates~\cite{Parker}. Under these
considerations one is left with two degrees of freedom: the
strength of the azimuthal component at the location of the Sun,
and the distance to which this wind extends. Recent estimates
suggest that the magnetic field strength near the Sun is $\sim
7~\mu$G~\cite{Beck}. The second parameter is more uncertain. Our
Galaxy dominates its near environment well past our neighbor, M31,
the Andromeda galaxy, and might well extend its sphere of
influence to half way to M81. This implies an outer halo wind of
$\sim 1.5$~Mpc. With this in mind, the mean flight time of the
protons in the Galaxy is $\sim 5.05 \times 10^{6}$~yr $\ll
\tau_s$, the time for straight line propagation from
M87 (Medina Tanco, private communication). The directions where the 13
highest energy CR
events point towards when they leave the halo wind of our Galaxy
is consistent with an origin in the Virgo region~\cite{Ahn:1999jd}: {\it (i)} for CR protons, except for the two highest
energy events, all other events can be traced back to within less than about
$20^\circ$ from Virgo; {\it (ii)} if one assumes that the two highest events
are helium nuclei, all 13 events point within
$20^\circ$ of Virgo. Arguably, the super-Galactic plane sheet can focus
UHECRs along the sheet. Hence, the particles would
arrive at the boundary of our Galactic wind with the arrival
directions described by an elongated ellipse along the
super-Galactic plane sheet~\cite{Stanev:2001rr}.
This would allow a bending of 20$^\circ$ to be accomodated.

Additionally, in order to account for most of the CRs observed above the
ankle, the power requirement of Virgo cluster~\cite{Anchordoqui:1999aj}
needs a fine-tuning of the source direction relative to the symmetry axis
of the wind, so as to turn on magnetic lensing effects~\cite{Harari:2000he}.
In such a case, M87
could be as high as $> 10^{2}$ times more powerful than if unlensed at
energies below $E/Z \sim 1.3 \times
10^{20}$~eV. Criticisms of this model~\cite{Billoir:2000wi} have been
addressed in~\cite{Biermann}.


\subsubsection{Other powerful nearby radiogalaxies}

Apart from Cen A (which would provide the most energetic particles
detectable on Earth), the CR-sky above PAO, if populated by
radiogaxies, should be dominated by Pictor A (a strong source with a flat
radio spectrum) which would contribute with the larger CR
flux~\cite{Rachen:1992pg}, and PKS 1333-33~\cite{Anchordoqui:jn}. Other
two southern candidates would be Fornax A ($z=0.057$) and PKS
2152-69 ($z=0.027$), which
could provide contributions to the CR flux above the cutoff. For other
powerful sources and their properties see~\cite{Rachen:1992pg,Rachen:1993gf}.

There are two additional EGRET sources, one of them at
high latitude, for which a possible radio galaxy counterpart has
been suggested. One such source is 3EG J1621+8203 $(l=115.5^{\rm
o}, b=31.8^{\rm o})$~\cite{Mukherjee:2002ef}. 3EG J1621+8203
observations in individual viewing periods yielded near-threshold
detections by EGRET, as for Cen A. However, in the cumulative
exposure, it was clearly detected and the measured flux above 100
MeV was $1.1 \times 10^{ -7}$ photon cm$^{ -2}$ s$^{-1}$. The
photon spectral index for this source is 2.27$\pm$0.53, steeper
than the usual blazar-like spectrum. Mukherjee et al.~\cite{Mukherjee:2002ef}
analyzed
the X-ray and radio field coincident with 3EG J1621+8203 and
concluded that NGC 6251, a bright FRI radio galaxy \cite{Urry} at
a redshift of 0.0234 (implying a distance 91 Mpc for $H_0= 75$ km
s$^{ -1}$ Mpc$^ {-1}$), and the parent galaxy of a radio jet
making an angle of 45$^{{\rm o}}$ with the line of sight
\cite{Sudou:2000wk}, is the most likely counterpart of the EGRET
source. With this identification, the implied $\gamma$-ray
luminosity is also a factor of $10^{-5}$ below that typical of
blazars. Compared with Cen A, the greater distance to NGC 6251
could, perhaps, be compensated by the smaller angle between the
jet and the line of sight.

Combi et al. \cite{Combi:2003hw} have also recently reported the
discovery of a new radio galaxy, J1737$-$15, within the location
error box of the low-latitude $\gamma$-ray source
{3EG~J1735$-$1500}, whose photon index is $\Gamma=3.24\pm0.47$.
The radio galaxy morphology at 1.4~GHz is typical of the
double-sided FRII. The integrated radio flux is $55.6\pm1.5$~mJy
at 1.4~GHz,  the source
is non-thermal 
and it is not detected at 4.8~GHz. Using the relation between
approaching and receding jets: $S_{\rm appr}/ {S_{\rm
rec}}=\left({1+\beta\cos\theta}/{1-\beta\cos\theta}\right)^{2-\alpha},$
as well as the radio fluxes of each jet component, a viewing angle
in the range $79^{{\rm o}}-86^{{\rm o}}$ for a velocity
$\beta=v/c$ between 0.3 and 0.9 and $\alpha=-1$ is derived.
Depending on the jet and ambient medium parameters, most
double-sided radio sources have sizes below $\sim300$~kpc
\cite{Begelman2}. In the case of J1737$-$15, and using standard
Friedmann-Robertson-Walker models, this size translates into a
possible distance smaller than 350~Mpc. If {3EG~J1735$-$1500} is
indeed the result of $\gamma$-ray emission in  {J1737$-$15}, the
intrinsic luminosity at $E>100$~MeV, at the distance quoted,
should then be less than $2\times10^{44}$~erg~s$^{-1}$, also
several orders of magnitude smaller than that of blazars. If both
radiogalaxies are closer than 100~Mpc, they could also be relevant
acceleration sites of the observed UHECRs.



\subsubsection{Correlations of UHECRs with QSOs, BL LACs, and EGRET sources}

Since an alignment beyond random expectations between UHECRs and
QSOs would certainly constitute a great discovery, the possible
correlation between UHECRs and QSOs was subject to a great deal of
scrutiny. In the spring of 1998, Farrar and Biermann pointed out
the existence of a directional correlation between compact
radio-QSOs and UHECRs: all events at the high end of the spectrum
observed by that time, with energy at least 1$\sigma$ above
$10^{19.9}$~eV, were aligned with high redshifted quasars, a
phenomenon with a chance probability of occurrence less than 0.5\%
\cite{Farrar:1998we}. Since then, this correlation has been
analyzed several times. Hoffman stated that one of the 5 events
used in the Farrar and Biermann's study, the highest energy event
observed by the Fly's Eye experiment, should not be
included in the UHECR sample under analysis, because this very
same event was considered to introduce the
hypothesis~\cite{Hoffman:1999ev}.
Without this event, the positive alignment with
random background probability is increased to $ <3\%$, in any case
small enough as to be plausibly significant~\cite{Farrar:1999fw}.
Using an updated event list (twice the size of the previous) from the Haverah
Park~\cite{Ave:2000nd}  and the AGASA~\cite{Hayashida:2000zr} experiments,
Sigl et al.~\cite{Sigl:2000sn} showed that the statistical significance of the
alignment is lowered to 27\%. Other authors,
however, favored the earlier alignment~\cite{Virmani:xk}, but
their correlation signal comes from events with large uncertainty
both in energy and in position: they considered events from the
SUGAR experiment, although it is not clear whether all these events
are above the GZK cutoff. Notwithstanding, after the Haverah
Park energy estimates have been re-assessed~\cite{Ave:2001hq}, the
original correlation has to be dropped altogether: for the cosmic
rays in question, the energy of the 2 events observed by this
array with incident zenith angle $<45^\circ$, that was previously
quoted as $> 10^{19.9}$~eV at 1$\sigma$, is now shifted $\approx
30\%$ downwards, below the energy cut chosen by Farrar and
Biermann. Hence, independently of the statistical test used, when
considering only the highest energy ($> 10^{19.9}$ eV at
1$\sigma$) events the correlation between UHECRs and QSOs is
consistent with a random distribution at the $1\sigma$ level.

Tinyakov and Tkachev~\cite{Tinyakov:2001nr,Tinyakov:2001ir,Tinyakov:2003bi}
reported a correlation between the arrival directions of UHECRs
and BL Lacs. Specifically, the (22) BL Lacs chosen were those
identified as such in the (9th-Edition) Veron-Cetty and Veron
(2000) \cite{Veron} catalogue of Quasars and Active Galactic
Nuclei, with redshift $z> 0.1$ or unknown, magnitude $m < 18$, and
radio flux at 6 GHz $F_6 > 0.17$~Jy. This analysis proposed no
energy buffer against contamination by mis-measured protons piled
up at the GZK energy limit.\footnote{The CR sample of Tinyakov and
Tkachev consists of 26 events measured by the Yakutsk experiment
with energy $> 10^{19.38}$~eV~\cite{Afanasiev}, and 39 events
measured by the AGASA experiment with energy $> 10^{19.68}$~eV
\cite{Hayashida:2000zr}.} The evidence supporting their claim is
based on 6 events reported by the AGASA Collaboration (all with
average energy $< 10^{19.9}$~eV), and 2 events recorded with the
Yakutsk experiment (both with average energy $< 10^{19.6}$~eV),
which were found to be within $2.5^\circ$ of 5 BL Lacs contained
in the restricted sample of 22 sources. The chance probability for
this coincidence set-up was claim to be $2 \times 10^{-5}.$ Here
also the data set used to make the initial assertion is also being
used in the hypothesis testing phase. What is further subject to
critique, is that the imposed cuts on the BL Lac
catalogue were chosen so as to maximize the signal-to-noise ratio,
compensating {\it a posteriori} the different cut adjustments by
inclusion of a penalty factor~\cite{Evans:2002ry}. Without such
arbitrary cuts, the significance of the correlation signal is
reduced at the 1$\sigma$ level. Not to anyone's surprise, even in
acceptance of this approach, the estimated value of the penalty
factor is subject to debate~\cite{Evans:2002ry,Tinyakov:2003bi}.

Recently, in order to test the {\it hypothetical} correlation between UHECRs
and BL Lacs, Torres et al.~\cite{Torres:2003ee} performed a blind
analysis using the Haverah
Park~\cite{Stanev:1995my} and Volcano Ranch~\cite{Linsley:VR} data samples.
Such an analysis shows no positional coincidences between these
two samples up to an angular bin $> 5^\circ,$ an angular scale that
is well beyond the error in arrival determination of these experiments
($\approx 3^\circ$)~\cite{Uchihori:1999gu}. On
the basis of the
strongly correlated sample analyzed by Tinyakov and Tkachev, one
expects the distribution describing the correlation between the
set of BL Lacs and any UHECR data-set with 33 entries to be
Poisson with mean $\approx 4.06.$
This implies a $2\sigma$ deviation effect. Alternatively, the 95\% CL
interval of the distribution which samples the correlation between
the BL Lacs and CRs recorded by Volcano Ranch + Haverah
Park is (0, 3.09)~\cite{Feldman:1997qc}, so that
the probability to measure the expected mean value $\approx 4.06$
is $\ll 5\%$. With this in mind, Torres et
al.~\cite{Torres:2003ee} conclude that
the 8 coincidences found in the Tinyakov and Tkachev's analysis do
not represent a statistically significant effect.

\begin{figure}[t]
\includegraphics[width=6cm,height=8cm]{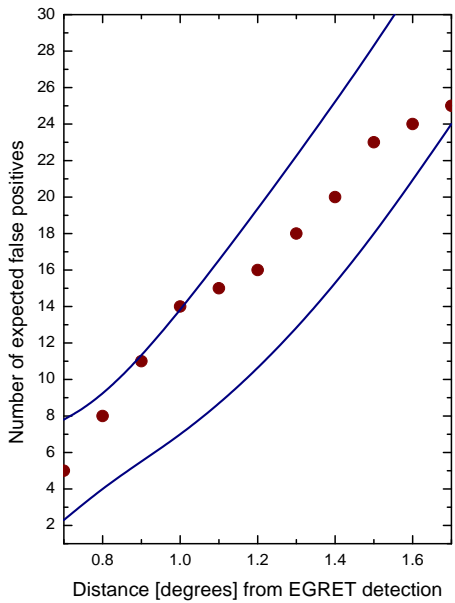} \hspace{1cm}
\caption{The expected distribution of radio-loud quasars (louder
than 0.5 Jy at 5 GHz) to occur by random chance as a function of
the distance from the center of the field for a sample of 114
EGRET detections. Points represent the number of $\gamma$-ray
detections for which the counterparts are beyond the 95\%
confidence contour. The dotted curve are the boundaries of the
68\% confidence band for the hypothesis that the radio sources are
randomly distributed in the EGRET detection fields. Adapted from
Punsly (1997). The number of sources whose possible counterpart
are beyond the 95\% confidence contour is compatible with the
chance expectation.  } \label{sow}
\end{figure}

Additionally, Gorbunov et al.~\cite{Gorbunov:2002hk} claimed that
a set of $\gamma$-ray loud BL Lacs can be selected by intersecting
the EGRET, the UHECR, and the BL Lac catalogs (all conveniently
cut). The only requirement Gorbunov et al. considered for an
object (here, a BL Lac) to be physically associated with an EGRET
source is that the angular distance between the best estimated
position of the pair does not exceed $2\times R_{95}$, where
$R_{95}$ is the 95\% confidence level contour of the EGRET
detection. Torres et al.~\cite{Torres:2003ee} pointed out
that identifying EGRET sources with BL Lacs (or any other
object) just by positional pairing within twice the EGRET error
grossly underestimates the goodness of existing $\gamma$-ray data.
At this stage, it is worth recalling the reader that the typical
$R_{95}$ radius for EGRET sources is 0.5--1$^\circ$. One can then
argue that if the confidence
contours have any
significance at all, a source should appear beyond the 95\%
contour only a few percent of the time. Working with 114 EGRET sources
above $|b|>10^{{\rm o}}$, Punsly~\cite{Punsly} have
estimated the number of random coincidences as a function of the
field radius: $\sim 2$ (10) quasars with more than 1 Jy of 5 GHz
flux are expected to correlate by random chance if the size of the
typical EGRET angular uncertainty is 0.7$^{{\rm o}}$ (1.7$^{{\rm
o}}$), see Fig.~\ref{sow}.

In our opinion, available statistics on the arrival directions of the UHECRs
reveals no significant correlations above random with BL Lacs nor
with any other type of quasars, including EGRET blazars.

\subsection{Remnants of quasars}

\subsubsection{What is a quasar remnant and how would they accelerate particles?}

Interestingly, the absence of powerful radio emitting objects in
the direction of several UHECRs led some colleagues to think that dead, faint
objects,  yet ones sufficiently active as to accelerate particles
up to relativistic energies, are responsible for the UHECRs
observed. Such is the idea behind the concept of quasar remnants
(QRs) as UHECR emitters \cite{Boldt:1999ge,Boldt:2000dx}: a
spinning supermassive black hole, threaded by magnetic fields
generated by currents flowing in a disc or torus, induces an emf
which, if vaccum breakdown is prevented and in the absence of
severe energy losses, accelerate a particle near the full voltage.

Although in the present epoch there is a paucity of luminous
quasars ($L\gtrsim 10^{47}$ ergs s$^{-1}$), those which appear at
large redshifts, the expected local number of dead quasars
associated with the same parent population is expected to be large
\cite{1a,1b,1c}. Supermassive black holes are now to be found in
the relatively dormant nuclei of giant elliptical galaxies,
generally regarded as QRs.\footnote{Indeed, the term ``quasar
remnants" was introduced by Chokshi and Turner \cite{CT} to
describe the present-epoch population of dead quasars harboring
supermassive black hole nuclei.} The compact dynamo model has been
proposed as a natural mechanism for accelerating CRs in
such environments \cite{Boldt:1999ge}. In this model
\cite{BZ1977}, if $B$ is the ordered poloidal field near the hole,
$V\sim aB$, where $a$ is the hole's specific angular momentum
(e.g., $a=M$ for an extreme Kerr hole of mass $M$). In an
appropriate astrophysical scaling \cite{znajek78}
\begin{equation}
V \sim 9\times 10^{20}(a/M)B_4M_9\ {\rm V},
\end{equation}
where $B_4\equiv B/(10^4 {\rm G})$ and $M_9\equiv M/(10^9{\rm
M}_{\odot})$. Assuming that the energy density of the magnetic
field near the event horizon is in equipartition with the rest
mass energy density of accreting matter \cite{krolik99}, it is
possible to introduce the accretion rate in the voltage
expression. In terms of an Advection Dominated Accretion Flow
(ADAF) model (e.g., Ref. \cite{dimateo99}), for example,
\begin{equation}
B_4 \sim 1.4\, {M_9}^{-1}{\dot{ M}}^{1/2},
\end{equation}
where $\dot{ M}$ is the accretion rate $dM/dt$ in ${ \rm M}_{\odot}\
{\rm yr}^{-1}$, and then, the maximum emf ($a/M \sim 1$) is
\begin{equation}
V \sim 1.2\times 10^{21}{\dot{ M}}^{1/2}\ {\rm V}.
\end{equation}
This potential is not, however, the maximum obtainable energy.

The rate of energy loss through curvature radiation by a particle
of energy $E=mc^2\Gamma$ can then be expressed as
\begin{equation}
P=\frac{2}{3}\frac{e^2c\Gamma^4}{\rho^2},
\end{equation}
see \cite{Levinson:ea} for a detailed explanation. Here,
 $\rho$ is
the average curvature radius of an accelerating ion, assumed to be
independent of the ion energy. The energy change per unit length
of an accelerating ion having charge $Z$ and mass $m_i=\mu m_p$ is
given by
\begin{equation}
d\epsilon/ds=eZ\Delta V/h - P/c,
\end{equation}
where $h$ is the gap height. After integration from $s=0$ to $h$,
the maximum acceleration energy is obtained
\begin{equation}
E_{\rm max}=3\times10^{19}\mu Z^{1/4}M_9^{1/2}B_4^{1/4}
(\rho^2h/R_g^3)^{1/4}\ \ \ {\rm eV}.
\end{equation}
Consequently, only a fraction
\begin{equation}
\eta=0.1 \mu M_9^{-1/2}(ZB_4)^{-3/4}(\rho/R_g)^{1/2}(h/R_g)^{-7/4};\ \ \ \ \eta\le1,
\label{eta}
\end{equation}
of the potential energy available will be released as UHECRs; the
rest will be radiated in the form of curvature photons.

For a proton, the suppression ratio is $ E/[e(V)]\approx
[(50M_9)^{-1/2}{B_4}^{-3/4}]r^{1/2}, $ where $r$ denotes the
magnetic field curvature in units of the Schwarzschild radius ($h
\sim R_g$) \cite{Boldt:2000dx}.  For $r\approx 1$ and ${\dot{
M}}\approx (0.1-10)\ {\rm M}_{\odot}\ {\rm yr}^{-1}$, and using
the previous equations, \be E_{\rm max}=(1.0-1.8)\times
10^{20}{M_9}^{1/4} \; {\rm eV}. \ee

Heavier nuclei would reach higher energies, but are subject to
photo-disintegration. For highly energetic protons, energy losses
due to photo-pion production in collisions with ambient photons
also becomes a relatively important effect. A lower limit to the
radiation length ($\Lambda_{\rm min}$) for the proton energy loss
associated with photo-pion production is estimated by considering
the population of target photons within the source region $\left[
R {\rm (source \; radius)} \geq 2GM/c^2 \right]$ at radio
frequencies $\nu \geq 360(\Gamma/10^{11})^{-1}$ GHz
 is given by~\cite{Boldt:2000dx}
\begin{eqnarray}
\Lambda/R & = & c\pi R/(\langle K\sigma_{p\gamma}\rangle Q) \\
          & = & (278/\langle K\sigma_{p\gamma,\mu{\rm b}}\rangle)(R/R_S)M_9(Q/10^{53}\
{\rm s}^{-1})^{-1}, \nonumber
\end{eqnarray}
where $\sigma_{p\gamma,\mu{\rm b}} \equiv
\sigma_{p\gamma}/10^{-30}$~cm$^2$, $R_S$ is the Schwarzschild
radius (twice the gravitational radius $R_g=GM$), $K\equiv \langle
E({\rm loss})\rangle/E({\rm initial})$ is the inelasticity in a
single collision \cite{stecker68} and $Q$ is the core emission
rate (photons s$^{-1}$) for electromagnetic radiation at $\nu>360$
GHz, given by $ Q=h^{-1}\int\nu^{-1}L_{\nu}d\nu $ where $h$ is the
Planck constant and $L_{\nu}=4\pi D^2F_{\nu}$ for a source of
spectral density $F_{\nu}$ at distance $D$. A useful approximation
is that \cite{Boldt:2000dx} $ \langle K \sigma_{p\gamma,\mu{\rm
b}} \rangle\equiv \left[\int(K \sigma_{p\gamma,\mu{\rm b}}
(dQ/d\nu)d\nu\right]/Q<120\ \mu{\rm b}.$ Those quasar remnants
with $\Lambda_{\rm min} > R$ are expected to successfully
accelerate protons up to $\sim E_{\rm max}$. Noteworthy, the
observed flux of CRs would apparently drain only a negligible
amount of energy from the black hole dynamo, since replenishing
the particles ejected at high energies ($> 10^{20}$~eV) would only
require a minimal mass input: a CR luminosity of $10^{42}$~ergs/s
in such particles (if protons) corresponds to a rest mass loss
rate of less than  $10^{-5} {\rm M}_\odot$ in a Hubble time.

\subsubsection{Correlations of UHECRs with QRs}

The first analysis of possible correlations between UHECRs and QRs
was carried out in Ref. \cite{Torres:2002bb}, where, imposing very
restrictive selection criteria --those necessary for obtaining
candidate objects providing the most favorable setting for a black
hole based compact dynamo model of UHECR production-- a group of
galaxies from the Nearby Optical Galaxy (NOG) catalog of Giuricin
et al. \cite{giu} was a priori selected as individual plausible
sources of CRs.\footnote{The latter catalog is a complete
magnitude-limited (corrected blue total magnitude $B\leq14$),
distance-limited (redshift $z \leq 0.02$) sample of several
thousand galaxies of latitude $\left| b \right|
> 20^\circ$.}
It was found that nearby QR candidates present an above-random
positional correlation with the sample of UHECRs. Surprisingly,
this correlation appears on closer angular scales than that
expected when taking into account the deflection caused by
typically assumed intergalactic or Galactic magnetic
fields.\footnote{Deflections due to the magnetic fields would of
course be avoided if the primary were a photon, generated in the
neighborhood of the QR via an accelerated charged particle
interaction.}

As one can see using Eq.~(\ref{t2}), scattering in large scale
magnetic irregularities ${\cal O}$ (nG) are enough to bend the
orbits of super-GZK protons by about $4$ deg in a 50 Mpc
traversal (see, however, Ref.~\cite{Dolag:2003ra}).  The CR 
angular offsets observed for the quasar
remnants in Ref. \cite{Torres:2002bb} are much smaller than
$\theta$ whereas, for a variety of assumed magnetic field
scenarios, $\theta$ is often substantially larger than the
estimated AGASA measurement error of 1.6 degrees. Should the
apparent clustering of correlated pairs be supported by future
data, perhaps a viable scenario under which this could occur is
that in which the intergalactic medium between Earth and the three
apparently `clustered' quasar remnants is sufficiently different
from the intergalactic medium in front of the remaining nine
objects that are much more uniformly distributed on the accessible
sky.  This appears at least plausible judging from the 100 micron
map of the region shown in Fig.~\ref{map}.

\begin{figure}[t]
\postscript{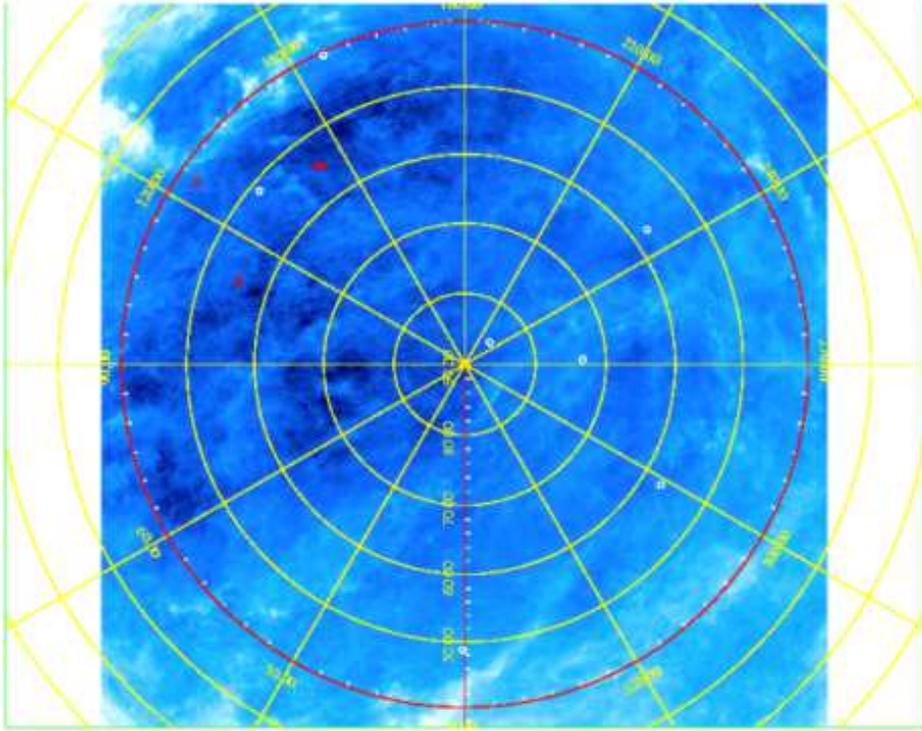}{0.80} \caption{ IRAS 100 micron
view of the North Galactic Pole (white=high flux, dark blue=low
flux), towards the region of several QRs that appear coinciding
with UHECRs. Small circles mark the 12 galaxies (candidate quasar
remnants) in the sample of Ref. \cite{Torres:2002bb}.  Red circles
mark galaxies perhaps associated with UHECRs; white circles mark
galaxies without associated UHECRs. Note that the palusible UHECR
sources tend to be found in directions of lower 100 micron flux,
[UHECR-coinciding source directions present IRAS fluxes 0.614 $\pm$ 0.022
MJy/sr whereas    the directions towards those galaxies non-coinciding with
UHECRs presnet 1.68  $\pm$ 0.63  MJy/sr.]
To the extent that the 100 micron flux traces the Galactic dust
and magnetic field, then CRs are likely to be better
aligned with their sources in directions of low flux. Figure
courtesy of Tim Hamilton.} \label{map}
\end{figure}

For the deflection of an energetic (60 EeV) proton in traversing
34 Mpc (the mean of the QRs distances in Table~3 of~\cite{Torres:2002bb}) 
to be less than a degree, the magnetic field
must fulfill $B < 2\times10^{-10}\;\ell^{-1/2}$~G, which appears
to be not drastically different from the canonical nG
$B$ field and the usually assumed coherence length $\sim 1$~Mpc. Also, in 
some directions, the magnetic field of our own galaxy
could well lead to a deflection of up to several degrees for
primaries with energies below 60 EeV; see, for instance, Table 1
of Ref.~\cite{Stanev:1996qj}. A recent study of this issue was presented
by Alvarez-Mu\~niz et al.~\cite{Alvarez-Muniz:2001vf}. However, a
possible filamentary topology of the Galaxy's magnetic field would
likely allow some directional windows, albeit narrow, where the
deflection of an UHECR could be much less than typical.

In a very recent paper, Isola et al. \cite{Isola:2003jk} have
studied predictions for large and small scale UHECR arrival
direction anisotropies in a scenario where the particles are
injected with a mono-energetic spectrum (all particles coming from
a source are emitted with the maximal energy of acceleration for
that source as derived in that same paper) by a distribution of
QRs. They find that the sample of (37) QRs they considered is
distributed too anisotropically to explain the isotropic ultra
high energy CR flux except in the case where extragalactic
magnetic fields of $\simeq 0.1\mu$G extend over many Mpc. As
statistical quantities for this analysis  spherical
multi-poles and the autocorrelation function were used. For a weak magnetic
field, of order  $1\,$nG, the predictions appear to be
inconsistent with the observed distribution of arrival directions
of UHECRs, because the magnetic field is too weak to isotropize
the distribution coming from a limited number of non-uniformly
distributed sources, as already pointed out in
Ref.~\cite{Isola:2002ei}. Isola et al. also found that the
contribution from the farthest sources is completely negligible
even for this weak magnetic field. None of the objects that were
found close to UHECR positions in the Torres et al. previous set
were included in the Isola et al. sample. Little further progress
regarding possible correlation of sources can be made until a much
more larger set of UHECRs is recorded, something that will have to
wait to the operation of PAO.

\subsubsection{TeV emission from QRs}

A concomitant effect of UHECR emission from QRs is that, as shown
by Levinson~\cite{Levinson:nx}, the dominant fraction of the
rotational energy extracted from the black hole is radiated in the
TeV band. He showed that the spectrum produced by the curvature
radiation of a single ion will peak at an energy
\begin{eqnarray}
E_{\gamma \; {\rm max}} &=&1.5\Gamma^3\hbar c/\rho \\
&=& 1.6\times10^{-7} E_{\rm max}\mu^{-1}
(ZB_4)^{1/2}(h/R_g)^{1/2} \\
&=& 5 M_9^{1/2}(ZB_4)^{3/4}(\rho^2 h^3/R_g^5)^{1/4} {\rm TeV},
\end{eqnarray}
and is a power law $I(E_{\gamma})\propto (E_{\gamma})^{1/3}$
below the cutoff.
 The overall spectrum of curvature photons would depend
on the energy distribution of the accelerating particles, and is
expected to be somewhat softer below the peak.  For $E_{\rm
max}=3\times 10^{20}$ eV and $h\sim R_g$, $E_{\gamma \; {\rm
max}}\simeq50 \mu^{-1}(ZB_4)^{1/2}$ TeV. Then, provided that
vaccum breakdown does not occur, and that TeV photons can escape
the system, QRs should emit $\gamma$-rays.

Indeed, it was recently noted by Neronov et al.
\cite{Neronov:2004ga} that the concomitant TeV radiation would be
at a level sufficiently high as to be (for many combination of the
system parameters) ruled out by bounds imposed by HEGRA/AIROBICC.

Recent numerical simulations additionally suggest that the
accretion process and magnetic field structure in the vicinity of
the horizon can be non-stationary, owing to rapid magnetic field
reconnection \cite{Hawley:2000mg}. This would probably lead to
appreciable complications of the model, as the location of the
gap, the injection of seed particles into the gap, and, perhaps,
the voltage drop across it might change with time. How should this
affect the picture described above is unclear at present.

\subsection{Starbursts}

\subsubsection{What are they?}

Starbursts are galaxies (sometimes, the term also refers only to
particular regions of galaxies) undergoing a large-scale star
formation episode. They feature  strong infrared emission
originating in the high levels of interstellar extinction, strong
HII-region-type emission-line spectrum (due to a large number of O
and B-type stars), and  considerable radio emission produced by
recent SNRs. Typically, starburst regions are located close to the
galactic center, in the central kiloparsec. This region alone can
be orders of magnitude brighter than the center of normal spiral
galaxies. From such an active region, a galactic-scale superwind
is driven by the collective effect of supernovae and particular
massive star winds. The enhanced supernova explosion rate creates
a cavity of hot gas ($\sim10^8$ K) whose cooling time is much
greater than the expansion time scale. Since the wind is
sufficiently powerful, it can blow out the interstellar medium of
the galaxy, preventing it from remaining trapped as a hot bubble.
As the cavity expands, a strong shock front is formed on the
contact surface with the cool interstellar medium. The shock
velocity can reach several thousands of kilometers per second and
ions like iron nuclei can be efficiently accelerated in this
scenario, up to ultrahigh energies, by Fermi's mechanism
\cite{Anchordoqui:1999cu}. If the super-GZK particles are heavy
nuclei from outside our Galaxy, then the nearby ($\sim 3$ Mpc
\cite{heckman}) starburst galaxies M82 ($l=141^\circ, b=41^\circ$)
and NGC 253 ($l=89^\circ, b= -88^\circ$) are prime candidates for
their origin.

\subsubsection{M82 and NGC253}

M82 is probably the best studied starburst galaxy, located at only
3.2 Mpc. The total star formation rate in the central parts is at
least $\sim10$ M$_{\odot}$ yr$^{-1}$ \cite{OM}. The far infrared
luminosity of the inner region within 300 pc of the nucleus is
$\sim 4\times 10^{10}$ L$_{\odot}$ \cite{rieke}. There are $\sim
1\times 10^7$ M$_{\odot}$ of ionized gas and $\sim 2 \times 10^8$
M$_{\odot}$ of neutral gas in the IR source \cite{rieke,SA}. The
total dynamical mass in this region is $\sim (1-2) \times 10^9$
M$_{\odot}$ \cite{SA}. The main observational features of the
starburst can be modelled with a Salpeter IMF extending from 0.1
to 100 M$_{\odot}$. The age of the starburst is estimated in $\sim
(1-3) \times 10^7$ yr \cite{rieke}. Around $\sim 2.5\times 10^8$
M$_{\odot}$ (i.e. $\sim 36$ \% of the dynamical mass) is in the
form of new stars in the burst \cite{SA}. The central region,
then, can be packed with large numbers of early-type stars.

NGC 253 has been extensively studied from radio to $\gamma$-rays
(e.g. \cite{beck1,paglione,ptak}). A TeV detection was reported by
CANGAROO~\cite{Itoh:2003yg}, but has been yet unconfirmed by other
experiments. More than 60 individual compact radio sources have
been detected within the central 200 pc \cite{ulvestad}, most of
which are supernova remnants (SNRs) of only a few hundred years
old. The supernova rate is estimated to be as high as $0.2-0.3$
yr$^{-1}$, comparable to the massive star formation rate, $\sim
0.1$M$_\odot$ yr$^{-1}$ \cite{ulvestad,forbes}. The central region
of this starburst is packed with massive stars. Four young
globular clusters near the center of NGC 253 can account for a
mass well in excess of 1.5$\times 10^6 M_\odot$~\cite{watson,keto}.
Assuming that the star formation rate has been
continuous in the central region for the last 10$^9$ yrs, and a
Salpeter IMF for 0.08-100 $M_\odot$, the bolometric luminosity of
NGC 253 is consistent with 1.5 $\times 10^8 M_\odot$ of young
stars \cite{watson}. Based on this evidence, it appears likely
that there are at least tens of millions of young stars in the
central region of the starburst. These stars can also contribute
to the $\gamma$-ray luminosity at high
energies~\cite{Romero:2003tj,Torres:2003ur}. Physical, morphological, and
kinematic evidence for the existence of a galactic superwind has
been found for NGC 253~\cite{mccarthy}. Shock interactions with
low and high density clouds can produce X-ray continuum and
optical line emission, respectively, both of which have been
directly observed.

A region about 1 kpc of the M82 galactic center appears to be a fossil
starburst, presenting
a main sequence stellar cutoff corresponding to an age of 100-200 Myr
and a current average extinction of 0.6 mag (compare with the extinction
of the central and current starburst region, 2.2 mag) whereas, nearby
globular glusters age estimations are between $2 \times 10^8$ and
$10^9$~yr~\cite{deGrijs:1999hc}.
It appears possible for this galaxy, then, that a starburst
(known as M82 ``B'') of similar
amplitude than the current one was active in the past.

\subsubsection{Two-step acceleration-process in starbursts}

The acceleration of particles in starburst galaxies is thought to
be a two-stage process \cite{Anchordoqui:1999cu}. First, ions are
thought to be diffusively accelerated at single SNRs within the
nuclear region of the galaxy. Energies up to $\sim 10^{14-15}$ eV
can be achieved in this step (see, e.g. \cite{Lagage}). Due to the
nature of the central
region, and the presence of the superwind, the escape of the iron
nuclei from the central region of the galaxy is expected to be
dominated by convection.\footnote{The relative importance of
convection and diffusion in the escape of the CRs from a
region of disk scale height $h$ is given by the dimensionless
parameter, $q={V_0\,h}/{\kappa_0}, $ where $V_0$ is the
convection velocity and $\kappa_0$ is the CR diffusion
coefficient inside the starburst~\cite{new3}. When $q \alt
1$, the CR outflow is difussion dominated, whereas when $q
\agt 1$ it is convection dominated. For the central
region of NGC 253 a convection velocity of the order of the
expanding SNR shells $\sim$ 10000 km s$^{-1}$, a scale
height $h \sim 35$ pc, and a reasonable value for the diffusion
coefficient $\kappa_0 \sim 5 \times 10^{26}$ cm$^2$ s$^{-1}$
\cite{berezinsky-book}, lead to $q \sim 216$. Thus, convection
dominates the escape of the particles. The residence time of the
iron nuclei in the starburst results $t_{\rm RES} \sim h / V_0
\approx 1 \times 10^{11}$ s.} Collective plasma motions of several
thousands of km per second and the coupling of the magnetic field
to the hot plasma forces the CR gas to stream along from
the starburst region. Most of the nuclei then escape through the
disk in opposite directions along the symmetry axis of the system,
being the total path travelled substantially shorter than the mean
free path.

Once the nuclei escape from the central region of the galaxy  they
are injected into the galactic-scale wind and experience further
acceleration at its terminal shock. CR acceleration at superwind
shocks was first proposed in Ref. \cite{bebito} in the context
of our own Galaxy. The scale length of this second shock is of the
order of several tens of kpc (see Ref. \cite{heckman}), so it can
be considered as locally planar for calculations. The shock
velocity $v_{\rm sh}$ can be estimated from the empirically
determined superwind kinetic energy flux $\dot{E}_{\rm sw}$ and
the mass flux $\dot{M}$ generated by the starburst through: $
\dot{E}_{\rm sw}={1}/{2} \dot{M} v_{\rm sh}^2. $ The shock radius
can be approximated by $r\approx v_{\rm sh} \tau$, where $\tau$ is
the starburst age. Since the age is about a few tens of million
years, the maximum energy attainable in this configuration is
constrained by the limited acceleration time arising  from the
finite shock's lifetime. For this second step in the acceleration
process, the photon field energy density drops to values of the
order of the cosmic background radiation (we are now far from the
starburst region), and consequently, iron nuclei are safe
from photodissociation while energy increases to $\sim 10^{20}$
eV.

To estimate the maximum energy that can be reached by the nuclei,
consider the superwind terminal shock propagating in a homogeneous
medium with an average magnetic field $B$. If we work in the frame
where the shock is at rest, the upstream flow velocity will be
${\bf v_1}$ ($|{\bf v_1}|=v_{\rm sh}$) and the downstream
velocity, ${\bf v_2}$. The magnetic field turbulence is assumed to
lead to isotropization and consequent diffusion of energetic
particles which then propagate according to the standard transport
theory \cite{jokipii}. The acceleration time scale is then
\cite{drury}: $ t_{\rm acc}=\frac{4 \kappa}{v_1^2}  \label{t} $
where $\kappa$ is the upstream diffusion coefficient which can be
written in terms of perpendicular and parallel components to the
magnetic field, and the angle $\theta$ between the (upstream)
magnetic field and the direction of the shock propagation: $
\kappa=\kappa_{\parallel} \cos^2\theta + \kappa_{\perp}
\sin^2\theta . $ Since strong turbulence is expected from the
shock we can take the Bohm limit for the upstream diffusion
coefficient parallel to the field, i.e. $
\kappa_{\parallel}=\frac{1}{3}{E}/{ZeB_1} ,$ where $B_1$ is the
strength of the pre-shock magnetic field and $E$ is the energy of
the $Z$-ion. For the $\kappa_{\perp}$ component we shall assume,
following Biermann \cite{birmanncr1}, that the mean free path
perpendicular to the magnetic field is independent of the energy
and has the scale of the thickness of the shocked layer ($r/3$).
Then, $ \kappa_{\perp}={1}/{3} \; r (v_1-v_2) $ or, in the strong
shock limit, $ \kappa_{\perp}={r v_1^2}/{12}.$  The upstream
time scale is $t_{\rm acc}\sim r/(3 v_1)$,
${r}/{3v_1}={4}/{v_1^2}\left({E}/({3ZeB_1}) \cos^2\theta + {r
v_1^2}/{12} \sin^2\theta\right). $ Thus, using $r=v_1\tau$ and
transforming to the observer's frame one obtains
\begin{equation}
E_{\rm max}\approx\frac{1}{4} ZeB v_{\rm sh}^2 \tau
\approx\frac{1}{2} ZeB \frac{\dot{E}_{\rm sw}}{\dot{M}} \tau .
\end{equation}

The predicted kinetic energy and mass fluxes of the starburst of
NGC 253 derived from the measured IR luminosity are
$2\times10^{42}$ erg s$^{-1}$ and 1.2 M$_{\odot}$ yr$^{-1}$,
respectively \cite{heckman}. The starburst age is estimated from
numerical models that use theoretical evolutionary tracks for
individual stars and make sums over the entire stellar population
at each time in order to produce the galaxy luminosity as a
function of time \cite{rieke}. Fitting the observational data
these models provide a range of suitable ages for the starburst
phase that, in the case of NGC 253, goes from $5\times 10^7$ to
$1.6\times 10^8$ yr (also valid for M82) \cite{rieke}. These
models must assume a given initial mass function (IMF), which
usually is taken to be a power-law with a variety of slopes.
Recent studies has shown that the same IMF can account for the
properties of both NGC 253 and M82 \cite{engelbracht}.  Finally,
the radio
and $\gamma$-ray emission from NGC 253 are well matched by models
with $B\sim50\mu$ G \cite{paglione}. With these figures, already
assuming a conservative age $\tau=50$ Myr, one obtains
a maximum energy for iron nuclei of $ E_{\rm max}^{\rm Fe} >
3.4\times 10^{20}\;\;\;\;{\rm eV}.$

\subsubsection{The starburst hypothesis: UHECR-luminosity and correlations}

For an extragalactic, smooth, magnetic field of $\approx 15
-20$~nG, diffusive propagation of nuclei below $10^{20}$~eV
evolves to nearly complete isotropy in the CR arrival
directions~\cite{Anchordoqui:2001ss,Bertone:2002ks}. Thus, we
could use the rates at which starbursts inject mass,
metals and energy into superwinds to get an estimate of the
CR-injection spectra. Generalizing the
procedure discussed in
Sec.~\ref{scena} --using equal power per decade over the
interval $10^{18.5}\,{\rm eV}< E< 10^{20.6}\,{\rm eV}$ -- we obtain
a source CR-luminosity
\begin{equation}
\frac{E^2 \,dN_0}{dE\,dt} \, \approx 3.5 \,\varepsilon \,10^{53}
{\rm eV/s} \label{1crluminosity}
\end{equation}
where  $\varepsilon$ is the
efficiency of ultra high energy CR production by the superwind
kinetic energy flux. With this in mind, the energy-weighted, 
approximately isotropic nucleus flux at  $10^{19}$ eV
is given by~\cite{Anchordoqui:2001ss}
\begin{equation}
E^3 J(E)  =  \frac{Ec}{(4\pi)^2d\,D(E)} \frac{E^2
\,dN_0}{dE\,dt}\,
I_\star
  \approx  2.3 \times 10^{26} \, \epsilon \,
I_\star \, {\rm eV}^2 \, {\rm m}^{-2} \, {\rm s}^{-1} \, {\rm
sr}^{-1}, \label{jp}
\end{equation}
where $I_\star = I_{\rm M82} + I_{\rm NGC\ 253}$. To estimate the
diffusion coefficient we used $B_{\rm nG} = 15$, $\ell_{\rm Mpc} =
0.5$, and an average $Z=20$.
We fix
\begin{equation}
\epsilon\, I_\star =0.013,
\label{mimi}
\end{equation}
after comparing Eq.(\ref{jp}) to the observed CR-flux.
Note that the contribution of $I_{\rm M82}$ and
$I_{\rm NGC\ 253}$ to $I_\star$ critically depends on the age of
the starburst. The relation
``starburst-age/superwind-efficiency'' derived from Eq.~(\ref{mimi}),
leads to $\epsilon \approx 10\%$, if both M82 and NGC 253 were
active for $115$~Myr. The power requirements may be reduced
assuming contributions from M82 ``B''~\cite{Anchordoqui:2001ss}.

Above $> 10^{20.2}$~eV iron nuclei do not propagate diffusively.
Moreover, the CR-energies get attenuated
by photodisintegration  on the CMB and the
intergalactic infrared background photons. However, the energy-weighted
flux beyond the GZK-energy
due to a single M82 flare
\begin{equation}
E^3 J(E)  =
 \frac{E}{(4 \pi d)^2} \frac{E_0^2 \,dN_0}{dE_0\,dt}\,e^{-R\,t/56}
 \approx  2.7 \times 10^{25} E_{20}
\epsilon \,e^{-R\,t/56}\,
{\rm eV}^2 \, {\rm m}^{-2} \, {\rm s}^{-1} \, {\rm sr}^{-1},
\end{equation}
is easily consistent with observation~\cite{Anchordoqui:2001ss}.
Here,
$R$ is the effective nucleon loss rate of the nucleus on the
CBM~\cite{Anchordoqui:1997rn}.

In the non-diffusive regime (i.e., $10^{20.3}~{\rm eV} \alt E \alt
10^{20.5}~{\rm eV}$), the accumulated deflection angle
from the direction of the source in the extragalactic $B$-field is
roughly $10^\circ \alt \theta
\alt 20^\circ$~\cite{Bertone:2002ks}. The nuclei suffer
additional deflection in the Galactic magnetic field. In particular,
if the Galactic field is of the ASS type, the arrival direction of the
4 highest energy CRs can be traced backwards to one of the
starbursts~\cite{Anchordoqui:2002dj}.
Figure~\ref{magnetica} shows the extent to which  the observed
arrival directions of the highest energy CRs deviate from their
incoming directions at the Galactic halo because of bending in the
magnetic field given in Eq.~(\ref{oko}). The incoming CR
trajectories are traced backwards  up to distances of 20 kpc away
from the Galactic center, where the effect of the magnetic field
is negligible. The diamond at the head of each solid line denotes
the observed arrival direction, and the points along these lines
indicate the direction from which different nuclear species (with
increasing mass) entered the Galactic halo. In particular, the tip
of the arrows correspond to incoming directions at the halo for
iron nuclei, whereas the circles correspond to nuclei of neon.
Regions within the dashed lines comprise  directions lying within
$20^\circ$ and $30^\circ$ degrees of  the starbursts. It is  seen
that trajectories for CR nuclei  with  $Z\ge 10$  can be further
traced back to one of the starbursts, within the uncertainty of
the extragalactic deviation.

\begin{figure}
\begin{center}
\includegraphics[width=6cm,height=7cm]{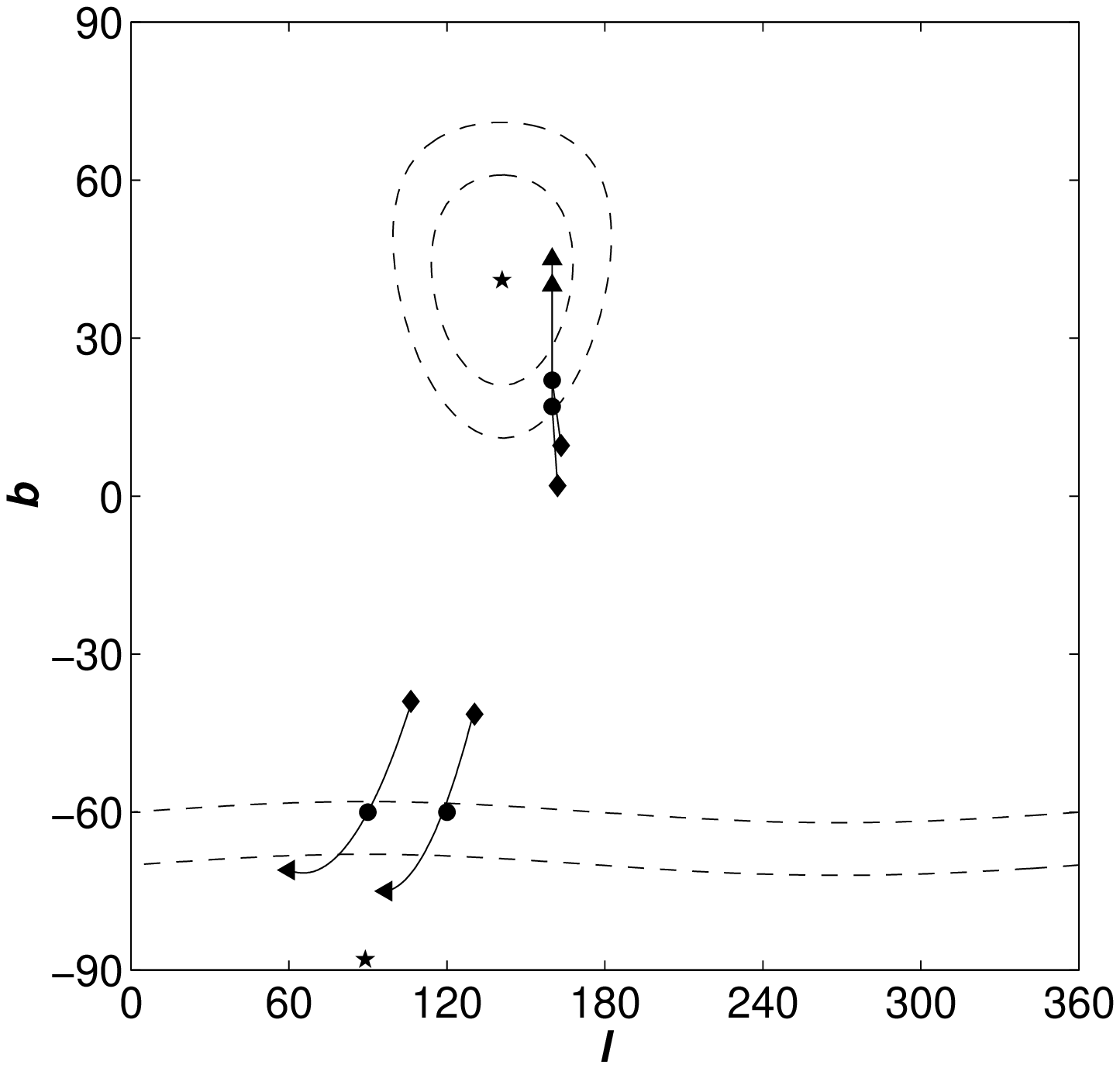}
\includegraphics[height=7cm]{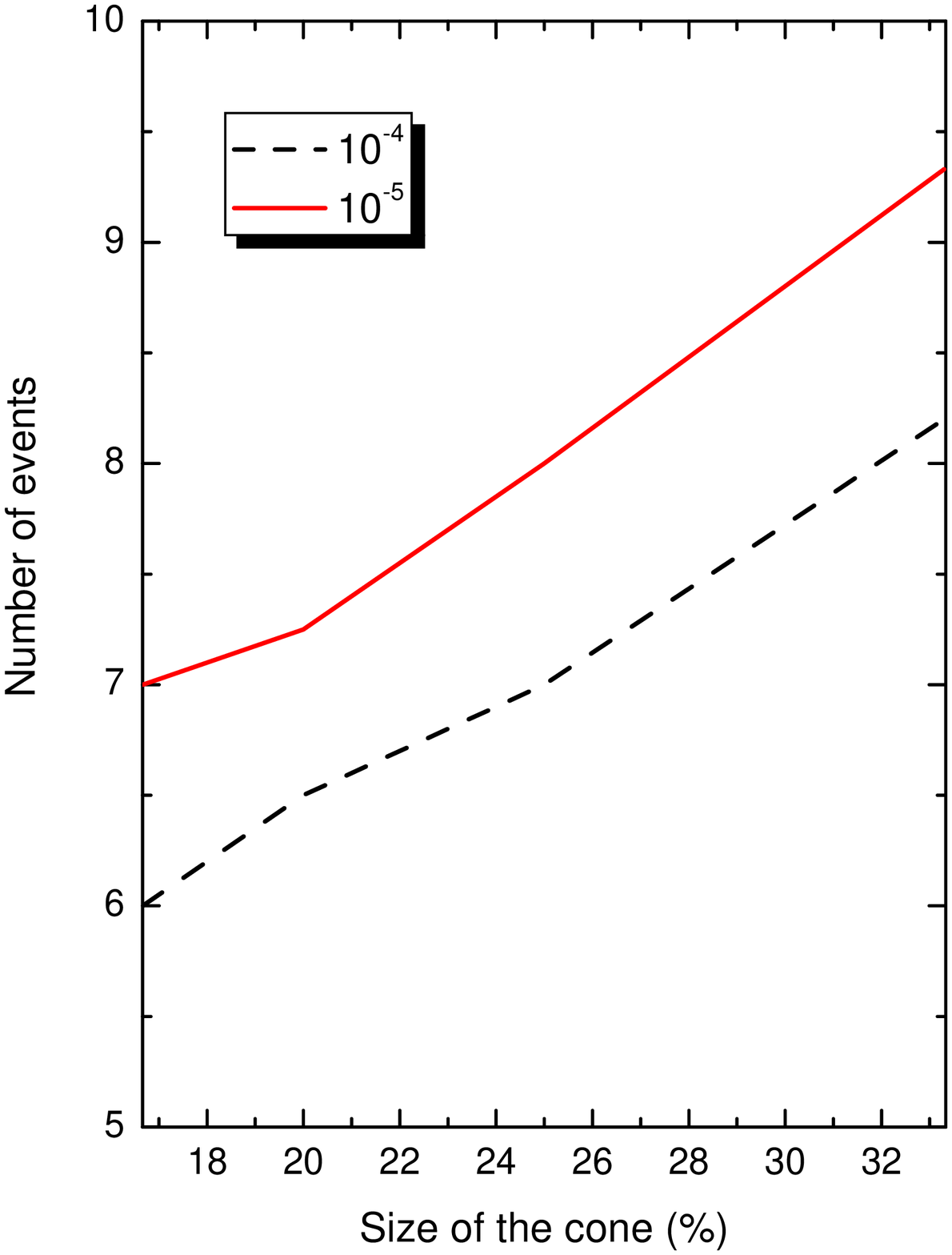}
\caption{Left: Directions in Galactic coordinates of the four
highest energy CRs at the boundary of the Galactic halo.
The diamonds represent the observed incoming directions. The
circles and arrows show the directions of neon and iron nuclei,
respectively, before deflection by the Galactic magnetic field.
The solid line is the locus of  incoming directions at the halo
for other species with intermediate atomic number. The stars
denote the positions of M82 and NGC253. The dashed lines are
projections in the $(l,b)$ coordinates of angular directions
within $20^{\circ}$ and $30^{\circ}$ of the starbursts. Right:
Curves of constant probabilities in the two-dimensional parameter
space defined by the size of the cone and the minimum number of
events originating within the resulting effective solid
angle~\cite{Anchordoqui:2002dj}.} \label{magnetica}
\end{center}
\end{figure}

The effects of the BSS configuration are completely different.
Because of the averaging over the frequent field reversals, the
resulting deviations  of the CR trajectories are markedly smaller,
and in the wrong  direction for correlation of current data with
the starburst sources. We note that the energy-ordered 2D
correlation distribution of the AGASA data is in disagreement with
expectations for positively charged particles and the BSS
configuration~\cite{Alvarez-Muniz:2001vf}.

We now attempt to assess to what extent these correlations are
consistent with chance coincidence.  We
arrive at the effective angular size of the source in a two-step
process. Before correcting for bias due to the coherent structure
of the Galactic magnetic field, the deflections in the
extragalactic and Galactic fields (regular and random components)
may be assumed to add in quadrature, so that the angular sizes of
the two sources are initially taken as cones with opening
half-angles between 40$^{\circ}$ and 60$^{\circ}$, which for the
purpose of our numerical estimate we  approximate to
50$^\circ$.
However, the global structure of the field will introduce a strong
bias in the CR trajectories, substantially diminishing the
effective solid angle. The combined deflections in the $l$ and $b$
coordinates mentioned above concentrate the effective angular size
of the source to a considerably smaller solid
angle. As a conservative estimate, we
retain 25\% of this cone as the effective source size. A clear
prediction of this consideration is then that the incoming flux
shows a strong dipole anisotropy in the harmonic decomposition.

Now, by randomly generating
four CR positions in the portion of the sky accessible to the
existing experiments (declination range $\delta > -10^\circ$), an
expected number of random coincidences can be obtained. The term
``coincidence'' is herein used to label a synthetic CR whose
position in the sky lies within an effective solid angle
$\Omega_{\rm eff}$ of either starburst. $\Omega_{\rm eff}$ is
characterized by a cone with opening half-angle reduced from
$50^{\circ}$ to $24^{\circ}$ to account for the 75\% reduction in
effective source size due to the magnetic biasing discussed above.
Cosmic ray positional errors were considered as circles of
$1.6^\circ$ radius for AGASA. For the other experiments the
asymmetric directional uncertainty was represented by a circle
with radius equal to the average experimental error.
The random prediction for the mean number of coincidences is $0.81\pm
0.01$. The Poisson probability~\footnote{Because of constraints
inherent in partitioning events among clusters, the distributions are
very close to, but not precisely Poisson~\cite{Goldberg:2000zq}.}
for the real result to be no more than the tail of the random
distribution is $1\%$.  Alternatively,
we may analyze this in terms of confidence intervals. For the 4
observed events, with zero background, the Poisson signal mean 99\%
confidence interval is $0.82-12.23$~\cite{Feldman:1997qc}. Thus our
observed mean for random
events, $0.81 \pm 0.01$, falls at the lower edge of this
interval, yielding a 1\% probability for a chance occurrence.
Of course, this is not
compelling enough to definitively rule out chance probability as
generating the correlation of the observed events with the
candidate sources, but it is suggestive
enough to deserve serious attention in analyses of future data.

Assuming an
extrapolation of AGASA flux ($E^3 J_{\rm obs} (E)$) up
to $10^{20.5}$~eV, the event rate at Pampa Amarilla\footnote{The
Southern Site of PAO has been christened Pampa Amarilla. Recall that it
has an aperture $A \approx 7000$~km$^2$ sr for showers with incident
zenith angle less than $60^\circ$.}
is given by
\begin{equation}
\frac{dN}{dt}  =  A\, \int_{E_1}^{E_2}\, E^3 J(E)\, \frac{dE}{E^3}
   \approx  \frac{A}{2} \,\langle E^3\, J(E) \rangle\,
  \left[ \frac{1}{E_1^2} - \frac{1}{E_2^2} \right]
 \approx  5.3 \,\,{\rm yr}^{-1}\,\,,
\end{equation}
where $E_1 = 10^{20.3}$~eV and $E_2 = 10^{20.5}$~eV. Considering
a 5-year sample of 25 events and that for this energy range
the aperture of PAO is mostly receptive to cosmic
rays from NGC253, we allow for different possibilities of the
effective reduction of the cone size because of the Galactic
magnetic field biasing previously discussed. In
Fig.~\ref{magnetica} we plot contours of constant probabilities
($P= 10^{-4},\ 10^{-5}$) in the two-dimensional parameter space of
the size of the cone (as a fraction of the full $50^{\circ}$
circle) and the minimum number of events originating within the
resulting effective solid angle.  The model predicts that after 5
years of operation, all of the highest energy events would be
observed in the aperture described above. Even if 7 or 8 are
observed, this is sufficient to rule out a random fluctuation at
the $10^{-5}$ level. Thus, a clean test of the starburst hypothesis can
be achieved at a very small cost: $< 10^{-5}$ out of a total
$10^{-3}$ PAO probability budget~\cite{Clay:2003pv}.

\subsection{Luminous Infrared Galaxies
\label{LIGS}}

\subsubsection{Definition}

In carrying to the extreme the concept of a starburst, we find the
powerful luminous infrared galaxies (LIGs). At luminosities above
10$^{11}$ L$_\odot$, LIGs ($L_{\rm FIR}
> 10^{11}$ L$_\odot$) are the dominant extragalactic objects in
the local universe ($z<0.3$) with such high luminosities. Some,
having $L_{\rm FIR}
> 10^{12}$ L$_\odot$, are the most luminous local objects (see
\cite{SM} for a review). These galaxies possess very large amounts
of molecular material (e.g., \cite{DS1,DS2,S1,S2,S3}). They
present large CO luminosities, but also a high value for the ratio
$L_{\rm FIR}/L_{\rm CO}$, both being about one order of magnitude
greater than for spirals, implying a higher star formation rate
per solar mass of gas.\footnote{For stars to form, a large mass
fraction at high density, $dM/d(\log n)$, is required. The Milky
Way has an order of magnitude more gas at less than 300 cm$^{-3}$
than what it has at 10$^4$ cm$^{-3}$, contrary to LIGs, which have
nearly equal mass per decade of density between 10$^2$ and 10$^4$
cm$^{-3}$ (see, e.g., Ref. \cite{S2,S3})} LIGs are generally
regarded as recent galaxy mergers in which much of the gas of the
colliding objects has fallen into a common center (typically less
than 1 kpc in extent), triggering a huge starburst phenomenon
\cite{SM}. There is evidence for the existence of even more
extreme enviroments within LIGs (see, e.g., \cite{DS1}): These,
larger than giant molecular clouds but with densities found only
in small cloud cores, appear to be the most outstanding star
formation places in the universe. They are well traced by HCN
emission, i.e., they produce a substantial fraction of the whole
HCN emission observed for the whole galaxy \cite{DS1}, see also
\cite{G1,G2}. The CR enhancement factor in these small but massive
regions can well exceed the average value for the galaxy. In Arp
220, for instance, two such regions were discovered to contain
about $ 2 \times 10^9$ M$_\odot$ \cite{DS1}. If the CR enhancement
in these regions is significantly larger than the starburst
average, these extreme environments could be the main origin for
any $\gamma$-ray emission observed from this galaxy~\cite{Torres:2004wf}.

\subsubsection{Propagation and further studies}

Using the PSCz catalogue \cite{sa00}, Smialkowski et al.
\cite{Smialkowski:ty} constructed all-sky maps of UHE proton
intensities plausibly originating in LIGs, taking into account
effects of particle propagation through the extragalactic medium
and the possible influence of the regular galactic magnetic field.
The PSCz catalogue consists of almost 15000 IR galaxies with known
redshifts, covering 84\% of the sky. Finding correlations with
such an overdistributed sample might be thought as a risky
business. There are, however, some phenomenological reasons
---apart of LIGs being super-starbursts--- to search for a
possible UHECR origin in LIGs.

Arp 299, one of the the brightest infrared source within 70 Mpc
and a system of colliding galaxies showing intense starburst,
appeared earlier as VV 118 in the list of candidates for the AGASA
triplet presented by in Ref.
\cite{Takeda:1998ps,Hayashida:2000zr,Takeda:2002at}. Indeed,  for
Arp 299, even when a hidden AGN was observed, it can not account
for the whole FIR luminosity \cite{cecca} and a strong starburst
activity is required to explain it. It might be considered as a
prime candidate for the origin of the triplet
\cite{Smialkowski:ty}, if such is not a statistical fluctuation.

Equatorial maps of the expected intensity of the UHE protons
originating in LIGs for the energy region 40-80 EeV with sky
coverage and declination dependent exposure for the AGASA
experiment were presented in Ref.~\cite{Smialkowski:ty}.
Expected proton intensities for 40-80 EeV appear to show
good correlation with the distribution of the experimental events
from AGASA, especially, from the region of the sky near Arp 299
and the AGASA triplet of events (RA$\approx170^{\circ}$,
$\delta\approx60^{\circ}$). No correlation of the data events
above 80 EeV with the expected CR intensities was reported.

The latter result was confirmed in Ref. \cite{Singh:2003xr}, where
application of the Kolmogorov-Smirnov (KS) to the 11 highest AGASA
events above $10^{20}$ eV yields a KS probability of $< 0.5$\%,
rejecting the possible association at $>99.5$\% significance
level. This was used to argue that
 the existing CR events above
$10^{20}$ eV do not owe their origin to long burst GRBs, rapidly
rotating magnetars, or any other events associated with core
collapse supernovae (although it would yet be too early to
completely rule out such a possibility, based only in a 2$\sigma$
deviation).\footnote{Following \cite{Singh:2003xr}, the core
collapse supernova rate per galaxy, summed over all galaxy types,
is $\dot{S}_{\rm sn} \approx 0.011$ SN per galaxy-year
\cite{Cappellaro}, which yields a volume averaged rate of
supernovae of $\dot{S}_{\rm sn} n_{\rm g} \approx 2.2 \times
10^{-4} $ SN/Mpc$^3$-year. The supernova rate for galaxies with
far infrared emission is $\dot{S}_{\rm sn}^{\rm fir} = 2.5 \times
10^{-4} L_{10}$ SN/(FIR galaxy)-year \cite{Mannucci}. Integrating
this supernova rate over the FIR galaxy luminosity function yields
$\dot{S}_{\rm sn}^{\rm fir} n_{\rm firg} \approx 0.7 \times
10^{-4} K_{\rm obsc} (L_{10}^{\rm max}/300)$ SN/Mpc$^3$-year.
$K_{\rm obsc}$, the correction for supernovae missed in the
existing optical and near infrared supernova detection surveys,
might be as large as 10, and probably is at least as large as 3
\cite{Mannucci}. Thus, the luminous infrared galaxies contribute
at least 28\% ($K_{\rm obsc} = 1$) of the total supernova rate,
with a total space density only 25\% that of all normal galaxies.
Furthermore, since the brightest infrared galaxies ($L_{\rm FIR} >
10^{12} \; L_\odot$) dominate the contribution from all FIR
galaxies, and these are quite rare, with space density $\sim 4
\times 10^{-8}$ Mpc$^{-3}$ \cite{Soifer}, correlations of UHECR
arrival directions with the sky positions of the luminous IRAS
galaxies offers a promising opportunity to test the hypothesis
that UHECR acceleration has something to do with core collapse
supernovae, as is implied by the GRB shock and magnetar unipolar
inductor models for the acceleration sites. }

A more sensitive approach to study a possible LIG origin of
UHECRs, perhaps, is not to look for possible correlations  between
the whole PSCz calatog and UHECRs, but rather to select {\it a priori}
which LIGs are most likely to be detectable by their possible
UHECR emission. There are several LIGs for which reasonable values
of CR enhancements, comparable to, or lower than, the ratio
between their SFR and the Milky Way's, can provide a $\gamma$-ray
flux above GLAST sensitivity, and if the CR spectrum is
sufficiently hard, also above the sensitivities of the new
\v{C}erenkov telescopes. These LIGs are then most likely to appear
as new point-like $\gamma$-ray sources. Even when it is natural to
expect that a LIG will emit $\gamma$-rays, only the more gaseous,
nearby, and CR-enhanced galaxies are the ones which could be
detected as $\gamma$-ray point sources~\cite{Torres:2004wf}.

Out of the HCN  just presented in Refs. \cite{G1,G2}\footnote{This
survey is a systematic observation (essentially, all galaxies with
strong CO and IR emission were chosen for HCN survey observations)
of 53 IR-bright galaxies, including 20 LIGs with $L_{\rm
FIR}>10^{11}$L$_\odot$, 7 with $L_{\rm FIR}>10^{12}$L$_\odot$, and
more than a dozen of the nearest normal spiral galaxies. It also
includes a literature compilation of data for another 9 IR-bright
objects. This is the largest and most sensitive HCN survey (and
thus of dense interstellar mass) of galaxies to date.} and the
larger Pico dos Dias Survey (PDS, \cite{Coziol})\footnote{The PDS
survey consists of relatively nearby and luminous starbursts
galaxies selected in the FIR. PDS galaxies have a lower mean IR
luminosity $\log({\rm L}_{\rm IR}/{\rm L}_\odot) = 10.3 \pm 0.5$,
redshifts smaller than 0.1, and form a complete sample limited in
flux in the FIR at $2\times10^{-10}$ erg cm$^{-2}$ s$^{-1}$.}, the
most likely $\gamma$-ray sources are listed in Table
\ref{lig-table}, together with EGRET fluxes upper limits, obtained
in~\cite{Torres:2004wf}. In our opinion,  this group, as well as  the
HCN and the PDS  should be separately taken into account when
searching for possible UHECR correlations with new sets of data.
\footnote{A stacking procedure with EGRET data is to be reported
by Cillis et al. \cite{Cillis}.}

\begin{table*}[t]
\caption{Powerful local LIGs (all with interferometric
measurements) likely to be detected by GLAST, with EGRET upper
limits. The luminosity distance in an standard universe ($
D_L={c}/{{H_0{q_0}^2}} \bigl[
1-q_0+q_0z+(q_0-1)(2q_0z+1)^{1/2}\bigr], $  $H_0$ ($\sim$75 km
s$^{-1}$ Mpc$^{-1}$) is the Hubble parameter, $q_0$ ($\sim$0.5) is
the deceleration parameter, and $z$ is the redshift), the
central-sphere radius from which the line emission was detected,
the FIR luminosity and gas mass are also given. The minimum
average value of CR enhancement $k$ for which the $\gamma$-ray
flux above 100 MeV is above 2.4 $\times 10^{-9}$ photons cm$^{-2}$
s$^{-1}$, i.e. GLAST sensitivity is given. The smaller the value
of $\langle k\rangle$, the higher the possibility for these
Galaxies to appear as $\gamma$-ray sources.}
\begin{center}
\begin{tabular}{lcccccc} \hline
Name                 & $D_L$        & $R$      & {$\log (L_{\rm FIR}/$L$_{\odot})$ } & {$\log M$(H$_{\rm 2}/$M$_{\odot}$) } & {$\langle  k\rangle$} & $F_{>100\, {\rm MeV} }^{\rm EGRET}$ \\
                     & [Mpc]        & [pc] &                                     &                                      &                       & [$10^{-8}$ photons cm$^{-2}$ s$^{-1}$]\\
\hline
NGC3079              & 15           & 0.26     & 10.52     & 9.56         &  62 & $<$4.4   \\ %
NGC1068              & 15           & 0.10     & 10.74     & 9.46         &  78 & $<$3.6    \\
NGC2146              & 20           & 0.33     & 10.78     & 9.43         &  149& $<$9.7    \\
NGC4038/9            & 22           & 0.49     & 10.65     & 9.07         &  412& $<$3.7     \\
NGC520               & 29           & 0.38     & 10.58     & 9.47         &  285& $<$4.6   \\ %
IC694                & 41           & 0.30     & 11.41     & 9.59         &  432& $<$2.2    \\
Zw049.057            & 52           & 0.40     & 10.95     & 9.67         &  578& $<$6.9    \\
NGC1614              & 64           & 0.60     & 11.25     & 9.78         &  680& $<$5.0     \\
NGC7469              & 65           & 0.82     & 11.26     & 9.89         &  544& $<$3.2  \\ %
NGC828               & 72           & 0.92     & 11.03     & 10.09        &  421& $<$6.1   \\
Arp220               & 72           &  0.16    & 11.91     & 10.43        &  193& $<$6.1   \\
VV114                & 80           & 0.93     & 11.35     & 10.03        &  597& $<$3.9    \\
Arp193               & 94           &  0.25    & 11.34     & 10.22        &  532& $<$5.2  \\ %
NGC6240              & 98           & 1.65     & 11.52     & 10.03        &  896& $<$6.4   \\
Mrk273               & 152          &  0.13    & 11.85     & 10.33        & 1081& $<$2.3   \\
IRAS 17208$-$0014    & 173          &  0.75    & 12.13     & 10.67        & 640 & $<$7.5    \\
VIIZw31              & 217          &  1.27    & 11.66     & 10.70        & 940 & $<$3.2 \\
\hline \hline
\end{tabular}
\label{lig-table}
\end{center}
\end{table*}


\subsection{Gamma-ray bursts}

\subsubsection{Basic phenomenology}

Gamma ray bursts (GRBs) are flashes of high energy radiation that
can be brighter, during their brief existence, than any other
gamma ray source in the sky. The bursts present an amazing variety
of temporal profiles, spectra, and timescales that have puzzled
astrophysicists for almost three decades \cite{Fishman95}. In
recent years, our observational insight of this phenomenon has
been dramatically improved by the huge amount of data collected by
the Burst and Transient Source Experiment (BATSE): several
thousand GRB observations were obtained. New breakthrough results
are the expected outcome of HETE-2 and Swift.

The temporal distribution of the bursts is one of the most
striking signatures of the GRB phenomenon. There are at least four
classes of distributions, from single-peaked bursts, including the
fast rise and exponential decaying (FREDs) and their inverse
(anti-FREDs), to chaotic structures~(e.g.
\cite{Link:1996ss,Romero:1999mg}).  Burst timescales go through
the 30~ms scale to hundreds of seconds.

The GRB photon spectrum is well fitted in the BATSE detectors
range, 20~keV to 2~MeV \cite{Fishman95}, by a combination of two
power-laws,
$dn_\gamma/d\epsilon_\gamma\propto\epsilon_\gamma^{(\alpha-1)}$
($\alpha$ is the flux density spectral index, $F_\nu \propto
\nu^{+\alpha}$) with different values of $\alpha$ at low and high
energy \cite{Band}. Here, $dn_\gamma/d\epsilon_\gamma$ is the
number of photons per unit photon energy. The break energy (where
$\alpha$ changes) in the observer frame is typically
$\epsilon_{\gamma b}\sim1~{\rm MeV}$, with $\alpha \simeq 0$ at
energies below the break and $\alpha \simeq -1$ above the break.
In several cases, the spectrum has been observed to extend to
energies $>100$~MeV \cite{Fishman95,high}.

The angular distribution of these bursts is isotropic, and the
paucity of comparatively faint bursts implies that we are seeing
to near the edge of the source population~\cite{Meegan:xg}. Both
effects, isotropy and non-homogeneity in the distribution,
strongly suggest a cosmological origin of the phenomenon,
confirmed by the detection of afterglows, delayed low energy
emission of GRBs that allowed the measurement of the distance to
the burst via a redshift determination of several GRB
host-galaxies~(e.g. \cite{Metzger:1997wp,Kulkarni}).

\subsubsection{The fireball model}

The most popular interpretation of the GRB-phenomenology is that
the observable effects are due to the dissipation of the kinetic
energy of a relativistic expanding plasma wind, a ``fireball'',
whose primal cause is not yet
known~\cite{Cavallo:78,Paczynski:1986px,Goodman,Meszaros:1,Meszaros:2,Piran:kx,Piran:1999bk}
(see \cite{Waxman:2003vh} for a detailed review).  The rapid rise
time and short duration, $\sim1$~ms of the burst imply that the
sources are compact, with a linear scale comparable to a light-ms,
$r_0\sim10^7$~cm. If the sources are so distant, the energy
necessary to produce the observed events by an intrinsic mechanism
is astonishing: about $10^{52}$ erg of $\gamma$ rays must be released
in less than 1 second. Compactness and high $\gamma$-ray
luminosity implied by cosmological distances result in a very high
optical depth to pair creation, since the energy of observed
$\gamma$-ray photons is above the threshold for pair production.
The number density of photons at the source $n_\gamma$ is
\begin{equation}
L_\gamma = 4\pi r_0^2 cn_\gamma\epsilon,
\end{equation}
where $\epsilon\simeq1$MeV is the characteristic photon energy.
Using $r_0\sim10^7$cm, the optical depth for pair production at
the source is
\begin{equation}
\tau_{\gamma\gamma}\sim r_0 n_\gamma\sigma_T\sim{\sigma_TL_\gamma
\over4\pi r_0 c\epsilon}\sim10^{15}\, . \label{eq:tau-pair}
\end{equation}
The high optical depth creates the fireball: a thermal plasma of photons,
electrons, and positrons. The radiation pressure on the optically
thick source drives relativistic expansion, converting internal
energy into the kinetic energy of the inflating shell
\cite{Paczynski:1986px,Goodman}. As the source expands, the
optical depth is reduced. If the source expands with a Lorentz
factor $\Gamma$, the energy of photons in the source frame is
smaller by a factor $\Gamma$ compared to that in the observer
frame, and most photons may therefore be below the pair production
threshold.

Baryonic pollution in this expanding flow can trap the radiation
until most of the initial energy has gone into bulk motion with
Lorentz factors of $\Gamma \ge 10^2 -
10^3$~\cite{Waxman:2001tk,Waxman:2003vh}. The kinetic energy,
however, can be partially converted into heat when the shell
collides with the interstellar medium or when shocks within the
expanding source collide with one another. The randomized energy
can then be radiated by synchrotron radiation and inverse Compton
scattering yielding non-thermal bursts with timescales of seconds,
at large radius $r=r_d>10^{12}{\rm cm}$, beyond the Thompson
sphere. Relativistic random motions are likely to give rise to a
turbulent build up of magnetic fields, and therefore to Fermi
acceleration of charged particles.

Coburn and Boggs \cite{NATURE} recently reported the detection of
polarization,  a particular orientation of the electric-field
vector, in the $\gamma$-rays observed from a burst.
 The radiation released through synchrotron
emission is highly polarized, unlike in other previously suggested
mechanisms
 such as thermal emission or energy loss by
relativistic electrons in intense radiation fields. Thus,
polarization in the $\gamma$-rays from a burst provides direct
evidence in support of synchrotron emission as the mechanism of
$\gamma$-ray production (see also \cite{Nakar:2003qc}).

\subsubsection{Fermi acceleration in dissipative wind models of
GRBs}

Following Hillas' criterion,  the Larmor radius $r_{_{\rm L}}$
should be smaller than the largest scale $l_{\rm GRB}$ over which
the magnetic field fluctuates, since otherwise Fermi acceleration
will not be efficient. One may estimate $l_{\rm GRB}$ as follows.
The comoving time, i.e., the time measured in the fireball rest
frame, is $t = r/\Gamma c$. Hence, the plasma wind properties
fluctuate over comoving scale length up to $l_{\rm GRB} \sim
r/\Gamma$, because regions separated by a comoving distance larger
than $r/\Gamma$ are causally disconnected. Moreover, the internal
energy is decreasing because of the expansion and thus it is
available for proton acceleration (as well as for $\gamma$-ray
production) only over a comoving time $t$. The typical
acceleration time scale is then~\cite{Waxman:1995vg}
\begin{equation}
\tau_{\rm acc}^{\rm GRB} \sim \frac{r_{_{\rm L}}}{c \beta^2}\,,
\label{accgrb}
\end{equation}
where $\beta c$ is the Alfv\'en velocity. In the GRB scenario
$\beta \sim 1$, so Eq.~(\ref{accgrb}) sets a lower limit on the
required comoving magnetic field strength, and the Larmor radius
$r_{_{\rm L}} = E'/eB = E/\Gamma eB$, where $E' = E / \Gamma$ is
the proton energy measured in the fireball frame.

This condition sets a lower limit for the required comoving
magnetic field strength \cite{Waxman:1995vg},
\begin{equation}
\left({B\over B_{e.p.}}\right)^2>0.15\Gamma_{300}^2
E_{20}^2L_{51}^{-1},\label{larmor}
\end{equation}
where $E=10^{20}E_{20}{\rm eV}$, $\Gamma=300\Gamma_{300}$,
$L=10^{51}L_{51}{\rm erg}{\rm\ s}^{-1}$ is the wind luminosity,
and $B_{e.p.}$ is the equipartition field, i.e. a field with
comoving energy density similar to that associated with the random
energy of the baryons.

The dominant energy loss process in this case is synchrotron
cooling. Therefore, the condition that the synchrotron loss time
of Eq.~(\ref{tausyn}) be smaller than the acceleration time sets
the upper limit on the magnetic field strength
\begin{equation}
B<3\times10^5\Gamma_{300}^{2}E_{20}^{-2}{\rm G}.\label{sync}
\end{equation}

Since the equipartition field is inversely proportional to the
radius $r$, this condition may be satisfied simultaneously with
(\ref{larmor}) provided that the dissipation radius is large
enough, i.e.
\begin{equation}
r_d>10^{12}\Gamma_{300}^{-2}E_{20}^3{\rm cm}.\label{dis}
\end{equation}
The high energy protons lose energy also in interaction with the
wind photons (mainly through pion production). It can be shown,
however, that this energy loss is less important than the
synchrotron energy loss \cite{Waxman:1995vg}.

A dissipative ultra-relativistic wind, with luminosity and
variability time implied by GRB observations, satisfies the
constraints necessary to accelerate protons to energy $>
10^{20}$~eV, provided that $\Gamma > 100$, and the magnetic field
is close to equipartition with electrons. We stress that the
latter must be satisfied to account for both $\gamma$-ray emission
and afterglow observations~\cite{Waxman:2001tk}. At this stage, it
is worthwhile to point out that for the acceleration process at
shocks with large $\Gamma$ the particle distributions are
extremely anisotropic in shock, with the particle angular
distribution opening angles $\sim \Gamma^{-1}$ in the upstream
plasma rest frame. Therefore, when transmitted downstream the
shock particles have a limitted chance to be scattered efficiently
to re-eneter the shock~\cite{Bednarz:1998pi}. However, in this
particular case, the energy gain by any ``successful'' CR can be
comparable to its original energy, i.e., $\langle \Delta E \rangle
/ E \sim 1$.

\subsubsection{UHECRs and GRBs: connections}

In the GRB model for UHECR production described above\footnote{
An alternative model for the GRB phenomenon has been recently put 
forward~\cite{Preparata:1998rz}. In such a model, the GRB explosion occurs when
a massive star collapses into an electromagnetic black hole which readily
discharges due to the annihilation of pairs produced by vacuum 
polarization~\cite{Damour:1974qv}. The model explains the time variability, 
the spectra, and the GRB 
afterglows to a very high level of 
accuracy~\cite{Ruffini:2002vr,Bianco:2001fw,Ruffini:1999ta}. Interestingly,
within this set up one is able to accelerate the ionized hydrogen atoms 
sorrounding the death star to ultra high energies~\cite{Mattei} 
(see also~\cite{Anchordoqui:2004xj}).}, the high
energy CRs are protons accelerated by Fermi's mechanism in
sources that are distributed throughout the
universe~\cite{Waxman:1995vg,Vietri95}. It is
therefore possible to compare the UHECR spectrum with the
prediction from a homogeneous cosmological distribution of
sources, each generating a power law differential spectrum 
$\propto E^{-2.2}$ of high energy protons as typically expected from Fermi 
acceleration. Under the assumption that the GRB rate evolution is similar 
to the
star-formation rate evolution, the local GRB rate is $\sim0.5~{\rm
Gpc}^{-3}~{\rm yr}^{-1}$ \cite{Schmidt01}, implying a local
$\gamma$-ray energy generation rate of $\approx10^{44}~{\rm
erg}~{\rm Mpc}^{-3}~{\rm yr}^{-1}$.\footnote{The local ($z = 0$)
energy production rate in $\gamma$-rays by GRBs is roughly given
by the product of the characteristic GRB $\gamma$-ray energy,
$E\approx2\times10^{53}$ erg, and the local GRB rate.}

The energy observed in $\gamma$-rays reflects the fireball energy
in accelerated electrons. If accelerated electrons and protons (as
indicated by afterglow observations~\cite{Freedman}) carry similar
energy,
then the GRB
production rate of high energy protons is
\begin{equation}
\epsilon_p^2 (d\dot n_p/d\epsilon_p)_{z = 0}\approx 10^{44}~{\rm
erg}~{\rm Mpc}^{-3}~{\rm yr}^{-1}. \label{eq:cr_rate}
\end{equation}
%
The generation rate
(Eq.~\ref{eq:cr_rate}) of high energy protons is remarkably
similar to that required to account for the flux of $>10^{19}$~eV
CRs, whereas in this model, the suppression of model flux above
$10^{19.7}$~eV is due to
the GZK cutoff. Stecker and Scully~\cite{Stecker:1999wh,Scully:2000dr} have raised doubts on the possibility of this generating
a very strong cutoff at the highest CR energies, since if the GRB
redshift distribution follows that of the
star formation rate in the universe, a rate which is higher at larger
redshift, most of the GRBs would be just too far and CR with energies
above $3\times 10^{19}$ eV would be strongly attenuated by the CMB.
For a HiRes-shape spectrum, a common origin between GRBs and
ultrahigh energy CRs~\cite{Loeb:2002ee} is favored.\footnote{In addition,
dispersion of magnetic fields in the
intergalactic medium can make the number of UHECR-contributing
GRBs to grow above the burst rate within the GZK sphere. The
latter, within 100~Mpc from Earth, is in the range of $10^{-2}$ to
$10^{-3}$ yr$^{-1}$. Assuming a dispersion tiemscale, $\Delta
t\sim 10^7$ yr, the number of sources contributing to the flux at
any given time may be as large as $\sim 10^4$~\cite{Waxman:1995vg}.}
For appraisals of this and other general
criticisms made to the GRB-UHECR connection see
\cite{Vietri:2003te,Waxman:2002vf}.

Two of the highest energy CRs come from directions that are within
the error boxes of two remarkable GRBs detected by BATSE with a
delay of ${\cal O} (10)$ months after the
bursts~\cite{Milgrom:1995um}. However, a rigorous analysis shows
no correlation between the arrival direction of UHECRs and GRBs 
from the third BATSE catalog~\cite{Stanev:1996qc}.
No correlations were found either between a pre-CGRO burst
catalog and the Haverah Park shower set that covered approximately
the same period of time. These analysis, however, could have been
distorted by the angular resolution ($\Delta \theta \sim 3^\circ$)
of the GRB measurements. A sensitive anisotropy analysis between
UHECRs and GRBs will be possible in the near future,
using PAO, HETE-2 and Swift. Preliminary results (if one
assumes that GRBs are most likely to happen in infrared luminous
galaxies) do not seem to indicate any strong correlation (see
above Section \ref{LIGS}).

\subsubsection{A GRB origin for CRs below the ankle?}

Wick, Dermer and Atoyan \cite{Wick:2003ex} have recently proposed
a model for the origin of all CRs above $\sim 10^{14}$ eV/nucleon.
In this model, GRBs are assumed to inject CR protons and ions into
the interstellar medium of star-forming galaxies --including
ours-- with a power-law spectrum extending to a maximum energy
$\sim 10^{20}$ eV. In addition to the more energetic,
extragalactic spectrum of CR, the CR spectrum near the knee was
also shown to be plausibly fitted with CRs trapped in the Galactic
halo that were accelerated and injected by an earlier Galactic
GRB.\footnote{Arguably, from the evidence for beaming and the association of
GRBs with star-forming galaxies like the Milky Way, GRB events are
estimated to occur once every $10^4 - 10^{6}$ yrs in the inner
Galaxy~\cite{Piran:2001ga,Frail:2001qp}.} For power-law CR proton injection 
spectra $\propto E^{-2.2}$ 
and low and high-energy cutoffs,
normalization to the local time- and space-averaged GRB luminosity
density implies that if this model is correct, the nonthermal
content in GRB blast waves is hadronically dominated by a factor
$\approx 60$-200, limited in its upper value by energetic and
spectral considerations. Neutrinos to be detected in
kilometer-scale neutrino detectors such as IceCube (See Sec.~\ref{MR})
provide a clean signal of this model~\cite{Dermer:2003zv}.

The last GRB in the Galaxy has been also proposed as the possible progenitor 
of the CR
anisotropy observed in the direction of the GC~\cite{Biermann:2004hi}.
In order to estimate the remaining traces of any CR activity produced 
by the last GRB in the galaxy, one has to take into account several 
considerations: 

\noindent{\it (i)} The UHECRs escaping the GRB fireball
\begin{equation}
N_0(E>10^{18}~{\rm eV}) \sim 10^{-2}\,\, N_0 
(E> E_{\rm min}) 
\end{equation}
are mostly neutrons, because protons are captive in
the magnetic field and suffer extensive adiabatic losses on the
way out~\cite{Rachen:1998fd}.\footnote{We remind the reader that the differential injection 
spectrum of GRBs $\propto E^{-2.2}.$}  Some of these neutrons will decay into 
protons within the GC thin disk-like ($r \sim 3$~kpc) region of high 
interestellar medium density and high star formation rate. The population of 
secondary protons would then be captured by the strong $B$-field near the GC, 
attaining diffusion with a residence time scale of about $T \sim 10^{5}$~yr. 
At the end of this time, about 1/300 protons are able to avoid  leakeage.
The trapped protons, 
\begin{equation}
N_T (E> 10^{18}~{\rm eV}) = N_0 (E>10^{18}~{\rm eV}) \,\, 
\left(1-e^{-\frac{r\,m_n}{E\,\overline\tau}}\right) / 300\,\,,
\end{equation} 
can then  
be turned back into neutrons by interaction with nuclei in the interstellar 
medium with probability of $5 \times 10^{-2}.$ Here, $m_n$ and $\overline\tau$ are 
the neutron mass and lifetime, respectively.

\noindent{\it (ii)} The formation of the 
$n \rightarrow p$ reservoir depends on the GRB rate in 
the inner Galaxy (taken here as $10^{6}$ yr$^{-1}$) times the 
probability that a GRB 
jet points more o less along the direction of the GP. The latter is estimated to be about 50\%.

\noindent{\it(iii)} The total CR production by a single GRB 
is $\approx 10^{51}$~erg~\cite{Pugliese:1999df}.
 
Now, recalling that the CR excess observed by AGASA 
represents a 
luminosity of particles beyond $10^{18}$ eV of about 
$4 \times 10^{30}$~erg/s~\cite{Hayashida:1998qb}, straightforward calculation 
shows that the observed anisotropy
in the direction of the GC can be easily fitted by the neutrons produced 
in the GRB reservoir, that ultimately travel unscathed to Earth~\cite{Biermann:2004hi}.

\section{Full Throttle: UHECR Generation beyond the Standard Lore}

The astrophysical models discussed present difficulties in
providing a completely satisfactory explanation of the super-GZK
events, if there are any. This may simply reflect our present lack
of statistics, our present ignorance of the true conditions of
processes in some highly energetic regions of the universe, or,
perhaps, may imply that exotic mechanisms are at play. Physics
from the most favored theories beyond the standard model (SM) like
string/M theory, supersymmetry (SUSY), grand unified theories
(GUTs), and TeV-scale gravity have been invoked to explain the
possible flux above the GZK energy limit. This review is not
mainly concerned with beyond--SM scenarios (for more comprehensive
surveys see
e.g.~\cite{Anchordoqui:2002hs,Bhattacharjee:1998qc,Kuzmin:1999zk,Bhattacharjee:2000vh,Sarkar:2003sp}),
but for the sake of completeness, we provide here a brief account
of some of the most relevant exotic explanations.

The most economical among hybrid proposals involves a familiar extension of
the SM, namely, neutrino masses. It was noted many years ago that $\nu$'s
arriving at Earth from cosmologically distant sources have an annihilation
probability on the relic neutrino background of roughly
$3\, h_{65}^{-1}\%$~\cite{Weiler:1982qy}. Inspired on this analysis,
Weiler~\cite{Weiler:1997sh} and Fargion et al.~\cite{Fargion:1997ft} noted
that neutrinos within a few $Z$ widths of the right energy,
\begin{equation}
E^{R_Z}_\nu = \frac{M_Z^2}{2 m_{\nu_i}} = 4 \,\, \left(\frac{{\rm eV}}{m_{\nu_i}}
\right) \times 10^{21}\,\,{\rm eV}\,,
\end{equation}
to annihilate with the relic neutrinos at the $Z$-pole
with large cross section,
\begin{equation}
\langle \sigma_{\rm ann} \rangle^Z \equiv \int \frac{ds}{M_Z^2}\,
\sigma_{\rm ann} (s) = 2 \,\pi\,\sqrt{2}\,G_F  \sim 40.4\, {\rm nb}\,\,,
\end{equation}
may produce a ``local'' flux of nucleons and
photons.\footnote{$G_F = 1.16639(1) \times 10^{-5}$~GeV$^{-2}$ is
the Fermi coupling constant.} Remarkably, the energy of the
neutrino annihilating at the peak of the $Z$-pole has to be well
above the GZK limit. The mean energies of the $\sim $ 2 nucleons
and $\sim$ 20 $\gamma$-rays in each process can be estimated by
distributing the resonant energy among the mean multiplicity of 30
secondaries. The proton energy is given by
\begin{equation}
\langle E_p \rangle\, \sim \frac{M_Z^2}{60\, m_{\nu_j}} \sim 1.3 \,\,\left(\frac{{\rm eV}}{m_{\nu_j}}
\right) \times 10^{20} \,\,{\rm eV},
\end{equation}
whereas the $\gamma$-ray energy is given by
\begin{equation}
\langle E_\gamma\rangle\, \sim  \frac{M_Z^2}{120\,m_{\nu_j}} \sim 0.7 \,\,\left(\frac{{\rm eV}}{m_{\nu_j}} \right) \times 10^{20} \,\,{\rm eV}.
\end{equation}
The latter is a factor of 2 smaller to account for the photon origin in two
body $\pi^0$ decay. This implies that the highly
boosted decay products of the $Z$ could be observed as super-GZK
primaries.\footnote{Similarly,
gravi-burst fragmentation jets can contribute to the super-GZK
spectrum~\protect\cite{Davoudiasl:2000hv}.} However, to reproduce
the observed spectrum, the $Z$-burst mechanism requires very luminous
sources of extremely high energy neutrinos throughout the
universe~\cite{Kalashev:2001sh,Fodor:2002prl,Fodor:2002hy}. The present
limits on these sources are near the threshold of sensitivity to the
required flux~\cite{Gorham:2003da}.

In 1931, Georges Lema\^{\i}tre~\cite{Lemaitre} -- a forerunner of
the Big Bang hypothesis -- introduced the idea that the entire
material filling the universe, as well as the universe's
expansion, originated in the super-radioactive disintegration of a
``Primeval Atom'', which progressively decayed into atoms of
smaller and smaller atomic weight. The CRs were introduced in this
picture as the energetic particles emitted in intermediate stages
of the decay-chain. Echoing Lema\^{\i}tre, in the so-called
``top-down models'', charged and neutral primaries arise in the
quantum mechanical decay of supermassive elementary $X$
particles~\cite{Hill:1982iq,Hill:1986mn,Bhattacharjee:1991zm,Berezinsky:1997kd,Masperi:1997qf,Masperi:2000sp,Berezinsky:1997td,Berezinsky:1997hy,Kuzmin:1997cm,Jaikumar:2003rf,Birkel:1998nx,Fodor:2000za,Coriano:2001mg,Sarkar:2001se,Barbot:2002ep}.

To maintain an appreciable decay rate today, it is necessary to
tune the $X$ lifetime to be longer (but not too much longer) than
the age of the universe, or else ``store'' short-lived $X$
particles in topological vestiges of early universe phase
transitions (such as magnetic monopoles, superconducting cosmic
strings, vortons, cosmic necklaces, etc.). Discrete gauged
symmetries~\cite{Hamaguchi:1998wm,Hamaguchi:1998nj,Hamaguchi:1999cv}
or hidden  sectors~\cite{Ellis:1990iu,Benakli:1998ut} are
generally introduced to stabilize the $X$ particles. Higher
dimensional operators, wormholes, and instantons are then invoked to
break the new symmetry super-softly to maintain the long
lifetime~\cite{Berezinsky:1997hy,Kuzmin:1997cm} (collissional
annihilation has been considered too~\cite{Blasi:2001hr}).
Arguably, these metastable super-heavy relics (MSRs) may
constitute (a fraction of) the dark matter in galactic haloes.

Of course, the precise decay modes of the $X$'s and the detailed
dynamics of the first generation of secondaries depend on the
exact nature of the $X$ particles under consideration. However, in
minimal extensions of the SM, where there are no new mass scales
between $M_{\rm SUSY} \sim 1$~TeV and $m_X,$ the squark and
sleptons would behave like their corresponding supersymmetric
partners, enabling one to infer from the ``known'' evolution of
quarks and leptons the gross features of the $X$ particle decay: 
the strongly interacting quarks
would fragment into jets of hadrons containing mainly pions
together with a 3\% admixture of
nucleons~\cite{Coriano:2001rt,Barbot:2002gt,Barbot:2003cj}.\footnote{In
light of this, the sensitivity of future CR experiments to test
the SUSY parameter space has been
estimated~\cite{Barbot:2002et,Anchordoqui:2004qh}.} This implies
that the injection spectrum is a rather hard fragmentation-type
shape (with an upper limit usually fixed by the GUT scale) and
dominated by $\gamma$-rays and neutrinos produced via pion decay.
Therefore, the photon/proton ratio can be used as a diagnostic
tool in determining the CR origin~\cite{Aharonian:1992qf}. In
light of the mounting evidence that UHECRs are not $\gamma$-rays,
one may try to force a proton dominance at ultrahigh energies by
postulating efficient absorption of the dominant ultrahigh energy
photon flux on the universal and/or galactic radio
background.\footnote{Of course, this is not the case in
traditional scenarios (for an exception
see~\cite{Chisholm:2003bu}) of decaying massive dark matter in the
Galactic halo, which due to the lack of absorption, predict
compositions directly given by the fragmentation function, i.e.,
domination by $\gamma$-rays.}  However, the neutrino flux
accompanying a normalized proton flux is inevitably increased to a
level where it should be within reach of operating
experiments~\cite{Barbot:2002kh}.

It is clear that because of the  wide variety of top down models
the ratio of the volume density of the $X$-particle to its decay
time is model dependent. However, if a top down scenario is to
explain the origin of UHECRs, the injection spectrum should be
normalized to account for the super-GZK events without violating
any observational flux measurements or limits at higher or lower
energies~\cite{Sigl:1998vz}. In particular, neutrino and
$\gamma$-ray fluxes depend on the energy released integrated over
redshift, and thus on the specific top down model. Note that the
electromagnetic energy injected into the Universe above the pair
production threshold on the CMB is recycled into a generic cascade
spectrum below this threshold on a  short time scale compared with
the Hubble time. Therefore, it can have several potential
observable effects, such as modified light element abundances due
to $^4$He photodisintegration, or induce spectral distortions of
universal $\gamma$-ray and neutrino
backgrounds~\cite{Sigl:1995kk,Sigl:1996gm}. Additionally,
measurements of the diffuse GeV $\gamma$-ray
flux~\cite{Sreekumar:1997un}, to which the generic cascade
spectrum would contribute directly, limit significantly the
parameter space in which $X$'s can generate the flux of the
UHECRs~\cite{Sigl:1996im,Protheroe:1996pd,Protheroe:1996zg},
especially if there is already a significant contribution to this
background from conventional sources such as unresolved
$\gamma$-ray blazars. Recently, a possible lower extragalactic
contribution to the diffuse $\gamma$-ray background measured by
EGRET has been pointed out~\cite{Strong:2003ex,Keshet:2003xc}. The
$\sim 50\%$ smaller EGRET flux practically rules out extragalactic
top down and $Z$-burst scenarios~\cite{Semikoz:2003wv}.

If MSRs are the progenitors of the observable UHECRs,  then the
flux will be dominated by the decay or annihilation products of
$X$'s in the Galactic halo, i.e., by sources at distances smaller
than all relevant interaction lengths. A clean signal of this
scenario is the predicted anisotropy due to the non-central
position of the Sun in our
galaxy~\cite{Dubovsky:1998pu,Berezinsky:1998qv}.\footnote{Additionally, 
the arrival directions of UHECRs can show significant deviations 
from a random distribution due to an anisotropic spectrum of relics in 
the early 
universe~\cite{Mavromatos:2003ph}.} Although it was
noted that the predicted anisotropy is consistent with AGASA and
Haverah Park data~\cite{Evans:2001rv}, recent
analyses~\cite{Kachelriess:2003rv,Kim:2003th} of SUGAR data
excludes the MSR hypothesis at the $5\sigma$ level if all events
above $10^{19.6}$~eV are due to metastable $X$ clustered in the
halo. (For the extreme case where the population of MSRs is
responsible only for CRs with energies $\agt 10^{19.8}$~eV, the
annihilation scenario is disfavored at least at the 99\% CL,
whereas decaying MSRs still have a probability $\sim 10\%$ to
reproduce SUGAR data.)

A more exotic explanation postulates that the $X$'s themselves
constitute the primaries: magnetic monopoles easily pick up energy
from the magnetic fields permeating the universe and can traverse
unscathed through the primeval radiation, providing an interesting
candidate to generate extensive air showers~\cite{Kephart:1995bi}.
In particular, a baryonic monopole encountering the atmosphere
will diffuse like a proton, producing a composite
heavy-particle-like cascade after the first
interaction~\cite{Wick:2000yc} with a great number of muons among
all the charged particles~\cite{Anchordoqui:2000mk}. Although this
feature was observed in a poorly understood super-GZK
event~\cite{Efimov,Antonov:kn}, it seems unlikely that a complete
explanation for the UHECR data sample would be in terms of
magnetic monopoles alone. Moreover, any confirmed directional
pairing of events would appear difficult to achieve with the
monopole hypothesis.

A novel beyond--SM--model proposal to break the GZK barrier is to
assume that UHECRs are not known particles but rather a new
species, generally referred to as the uhecron,
$U$~\cite{Farrar:1996rg,Chung:1997rz,Albuquerque:1998va}. The
meager information we have about super-GZK particles allows a
na\"{\i}ve description of the properties of the $U$. The muonic
content in the atmospheric cascades suggests $U$'s should interact
strongly. At the same time, if $U$'s are produced at cosmological
distances, they must be stable, or at least remarkably long lived,
with mean-lifetime $\tau \gtrsim 10^6 \, (m_U/3~{\rm GeV})\, (d/
{\rm Gpc})\,{\rm s},$ where $d$ is the distance to the source and
$m_U$, the uhecron's mass. Additionally, since the threshold
energy increases linearly with $m_U$, to avoid photopion
production on the CMB $m_U \gtrsim 1.5$~GeV. In recent years,
direct searches of supersymmetric
hadrons~\cite{Adams:1997ht,Fanti:1998kx,Alavi-Harati:1999gp,Albuquerque:1994xi}
have severely eroded the attractiveness of the $U$ scenario.
However, adequate fine-tunings leave a small window still
open~\cite{Berezinsky:2001fy,Kachelriess:2003yy}.

On a similar track, it was recently put
forward~\cite{Madsen:2002iw} that strangelets (stable lumps of
quark matter with roughly equal numbers of up, down, and strange
quarks) can circumvent the acceleration problem in a natural way
(due to a high mass and charged, but low charge-to-mass ratio) and
move the expected cutoff to much higher energies.

Another possibility in which super-GZK CRs can reach us from very
distant sources may arise out of photons that mix with light
axions in extragalactic magnetic fields~\cite{Csaki:2003ef}. These
axions would be sufficiently weakly coupled to travel large
distances unhindered through space, and so they can convert back
into high energy photons close to the Earth.\footnote{See 
also~\cite{Gorbunov:2001gc} for a SUSY inspired related scenario.}  
An even more radical
proposal postulates a tiny violation of local Lorentz invariance,
such that some processes become kinematically
forbidden~\cite{Coleman:1998en}. In particular, photon-photon pair
production and photopion production may be affected by Lorentz
invariance violation.\footnote{Existing limits on the violation of Lorentz invariance come
from multi--TeV $\gamma$-ray observations of Mrk 501~\cite{Stecker:2001vb}.} Hence, 
the absence of the GZK-cutoff would
result from the fact that the threshold for photopion production
``disappears'' and the process becomes kinematically not allowed.
This implies that future PAO observations of faraway sources could
provide constraints on, or even a measurement of, the violation of
Lorentz symmetry, yielding essential insights into the nature of
gravity-induced wave dispersion in the
vacuum~\cite{Amelino-Camelia:1996pj,Amelino-Camelia:1997gz,Bertolami:1999da,Aloisio:2000cm,Bertolami:2000qa,Aloisio:2002ed,Alberghi:2003br,Gamboa:2003zk,Dowker:2003hb,Das:2004vc}.

In summary, future UHECR data may not only provide clues to the
particle sources, but could enhance our understanding of
fundamental particle
physics. We are entering  this new High Energy Physics era with 
the Pierre Auger Observatory~\cite{Anchordoqui:2003zk,Cafarella:2003cx,Cafarella:2003nv}.

\section{Minority Report: Neutrino Showers} \label{MR}

Extraterrestrial neutrinos provide a unique window to probe the
deepest reaches of stars, quasars, and exotic structures in the
cosmos. In contrast to all other SM particles, they can escape
from dense astrophysical environments and propagate to the Earth
unscathed, arriving aligned with their source. Even at the highest
energies the $\nu\bar{\nu}$ annihilation mean free path on the
cosmic neutrino background,
\begin{equation}
\lambda_\nu
= (n_\nu \, \sigma_{\nu \bar{\nu}})^{-1} \approx  4
\times 10^{28}\,\,{\rm cm}\,\,,
\end{equation}
is somewhat above the present size of the horizon (recall that
$H_0^{-1} \sim 10^{28}$~cm)~\cite{Roulet:1992pz}.

In addition, the fluxes of ultra-high energy cosmic neutrinos offer clues to the properties of
neutrinos themselves and provide an important probe of new ideas in particle physics. Namely,
their known interactions are so weak that new physics may easily alter neutrino properties. This
is especially relevant for extraterrestrial neutrinos with energies $\agt 10^6$~GeV that interact
with nucleons in the Earth atmosphere with center-of-mass energies above
the electroweak scale, where the SM is expected to be modified by new physics.

In summary, ultra high energy extraterrestrial neutrinos are
inextricably linked with the physics processes discussed
throughout this review. This chapter will therefore focus on the
main features of high energy neutrino astronomy.  For
comprehensive reviews on cosmic neutrinos the reader is referred
to~\cite{Gaisser:1994yf,Halzen:km,Halzen:2002pg,Dolgov:2002wy}.

\subsection{Bounds on the energy spectrum}

The basic operational system of neutrino telescopes is an array of
strings with photo-multiplier tubes (PMTs) distributed throughout
a natural \v{C}erenkov medium such as ice or water. The largest
pilot experiments ($\sim 0.1$~km in size) are: the now defunct
DUMAND (Deep Underwater Muon and Neutrino Detector)
experiment~\cite{Roberts:re} in the deep sea near Hawaii, the
underwater experiment in Lake Baikal~\cite{Belolaptikov:1997ry},
and AMANDA (Antarctic Muon And Neutrino Detector Array
)~\cite{Andres:1999hm} in the South Pole ice. Next generation
neutrino telescopes aim towards an active volume in the range of
1~km$^{3}$ of water. Projects under construction or in the
proposal stage are: two deep sea experiments in the Mediterranean,
the French ANTARES (Astronomy with a Neutrino telescope Abyss
environment RESearch)~\cite{Aslanides:1999vq} and NESTOR (Neutrino
Experiment SouthwesT Of GReece~\cite{Grieder:ha}), and
IceCube~\cite{Alvarez-Muniz:2001gb}, a scaled up version of the
AMANDA detector.

The traditional technique to observe cosmic neutrinos is to look
for muons (along with a visible hadronic shower if the $\nu$ is of
sufficient energy) generated via charged-current interactions,
$(\nu_\mu, \bar\nu_\mu) N \rightarrow (\mu^-, \mu^+) +$ anything,
in the rock below the detector. The \v{C}erenkov light emitted by
these muons is picked up by the PMTs, and is used for track
reconstruction. The muon energy threshold is typically in the
range of $10 - 100$~GeV. However, besides the desired
extraterrestrial neutrinos, below $\sim 10^3$~GeV there is a
significant background of leptons (produced by CRs interacting in
the atmosphere)~\cite{Lipari:hd,Volkova:sw} and so the task of
developing diagnostics for neutrino sources becomes complicated.
With rising energy ($\agt 10^6$~GeV) the three neutrino flavors
can be identified~\cite{Halzen:2001ty}.

The spectacular neutrino fireworks  (in the 10 MeV range) from
supernova 1987A constitute the only
extragalactic source so far observed~\cite{Hirata:1987hu,Bionta:1987qt}.
The Fr\'ejus~\cite{Rhode:es},
Baikal~\cite{Balkanov:2001br}, MACRO~\cite{Ambrosio:2002ma}, and
AMANDA~\cite{Ahrens:2003ee} collaborations reported no excess of neutrinos
above the expected atmospheric
background, enabling significant limts to
be set on the diffuse neutrino flux.\footnote{The Frejus experiment, located
in an underground laboratory, measured the energy of muons produced
by neutrino interactions in the rock above the detector. The detector itself comprised a calorimeter made
from a vertical sandwich of over 900 iron slabs interspersed with flash chambers.  Geiger tubes
embedded in the structure provided the trigger~\cite{Berger:1987ke}. The extensive air shower array on top of the Gran Sasso Laboratory
(EAS-TOP)~\cite{Aglietta:ny} and MACRO~\cite{Ambrosio:mb} used similar
techniques.} These limits are shown in Fig.~\ref{nufluxes}.

\begin{figure}
\postscript{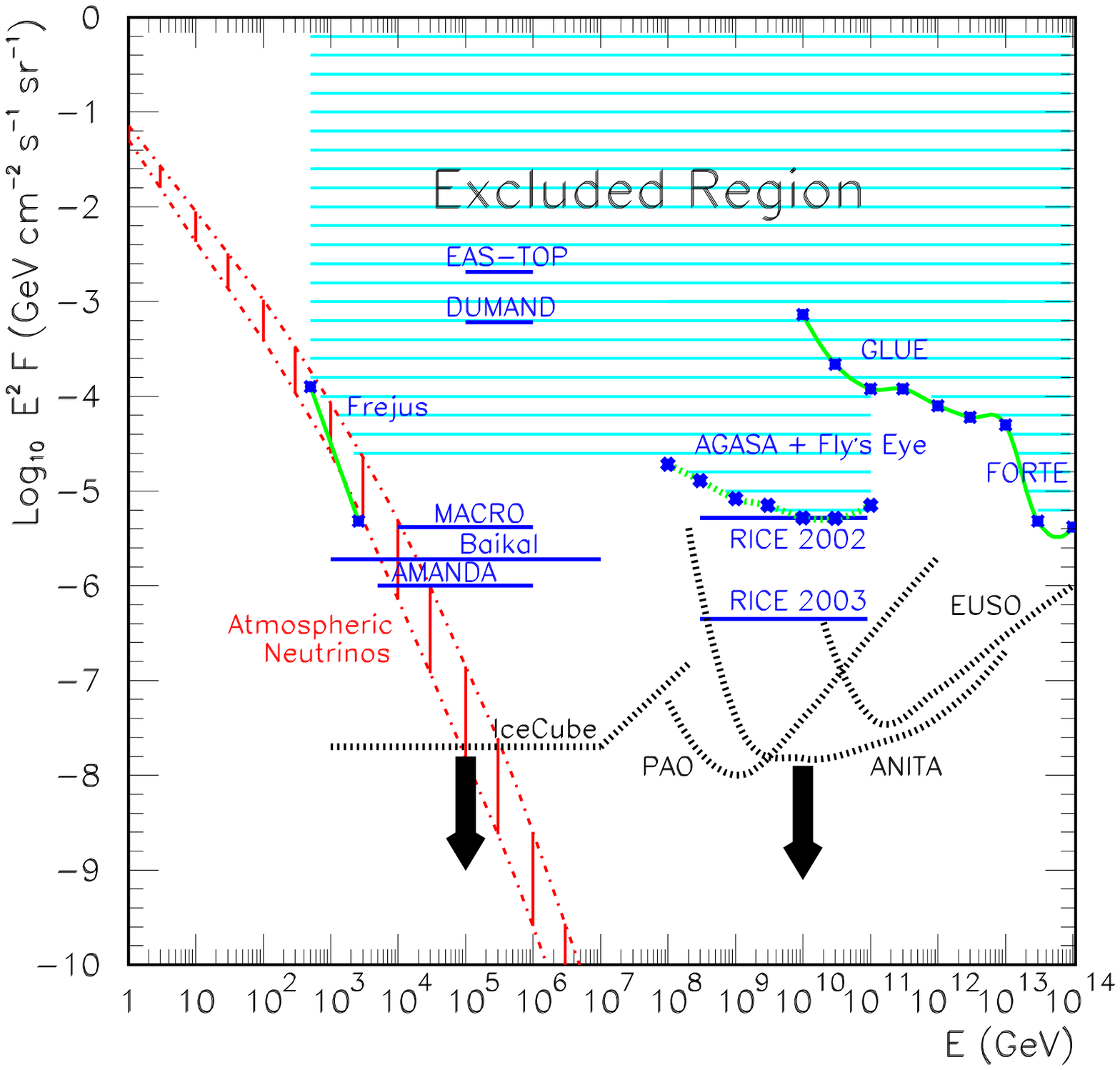}{0.80}
\caption{The horizontal solid lines indicate current 90\% CL upper limits
on the neutrino fluxes $\propto E^{-2}$ as reported by
the EAS-TOP~\cite{Aglietta:1994ge}, DUMAND~\cite{Bolesta:1997va},
Baikal~\cite{Balkanov:2001br}, MACRO~\cite{Ambrosio:2002ma},
AMANDA~\cite{Ahrens:2003ee}, and
RICE~\cite{Kravchenko:2002mm,Kravchenko:2003tc} collaborations. The $*-*-*$ lines indicate
model independent bounds on the neutrino flux 
($d\Phi_{_{\langle \nu_i + \bar{\nu}_i \rangle}}/dE,\, {\rm with} \ i =e, \mu, \tau$)  from AGASA + Fly's Eye data~\cite{Anchordoqui:2002vb} (95\% CL), Frejus~\cite{Rhode:es},
GLUE~\cite{Gorham:2003da,Gorham:2001aj}, and
FORTE searches in the Greenland ice sheet~\cite{Lehtinen:2003xv}. The
horizontal and diagonal thick dotted lines indicate the expected
90\% CL sensitivity of IceCube in 1 yr of operation~\cite{Ahrens:2003ix}. Also
shown (non-horizontal thick dotted lines) are the projected sensitivities of
PAO (1~yr running)~\cite{Bertou:2001vm}, EUSO
(1~yr running)~\cite{Eberle:2004ua} and ANITA (45~days running)~\cite{Eberle:2004ua} corresponding to 1 event per decade. The region between the falling
dashed-dotted lines indicates the flux of atmospheric neutrinos.}
\label{nufluxes}
\end{figure}

Upward going neutrinos with energies $\agt 10^8$~GeV are typically
blocked by the Earth. This shadowing severely restricts the high
energy event rates in underground detectors. However, neutrinos
may also induce extensive air showers, so current and future air
shower experiments might also function as neutrino detectors. The
neutrino interaction length is  far larger than the Earth's
atmospheric depth, which is only $0.36$~km water equivalent (kmwe)
even when traversed horizontally. Neutrinos therefore shower
uniformly at all atmospheric depths.  As a result, the most
promising signal of neutrino-induced cascades are quasi-horizontal
showers initiated  deep in the atmosphere.  For showers with large
enough zenith angles, the likelihood of interaction is maximized
and the background from hadronic cosmic rays is eliminated, since
the latter shower high in the atmosphere. These $\nu$-showers will
appear as hadronic vertical showers, with large electromagnetic
components, curved fronts (a  radius of curvature of a few km),
and signals well spread over time (on the order of microseconds).

The event rate for quasi-horizontal deep showers from ultra-high
energy neutrinos is~\cite{Anchordoqui:2002vb}
\begin{equation}
N = \sum_{i,X} \int dE_i\, N_A \, \frac{d\Phi_i}{dE_i} \, \sigma_{i
N \to X} (E_i) \, {\cal E} (E_i)\ , \label{numevents}
\end{equation}
where the sum is over all neutrino species $i = \nu_e,
\bar{\nu}_e, \nu_{\mu}, \bar{\nu}_{\mu}, \nu_{\tau},
\bar{\nu}_{\tau}$, and all final states $X$. $N_A = 6.022 \times
10^{23}$ is Avogadro's number, and $d\Phi_i/dE_i$ is the source
flux of neutrino species $i$, $\sigma$ as usual denotes the cross
section, and ${\cal E}$ is the exposure measured in cm$^3$ w.e. sr
time. The Fly's Eye and the AGASA Collaborations have searched for
quasi-horizontal showers that are deeply-penetrating, with depth
at shower maximum $X_{\rm max} > 2500$~g/cm$^2$ (see
e.g.~\cite{Baltrusaitis:mt,agasa}). There is only 1 event that
unambiguously passes this cut with 1.72 events expected from
hadronic background, implying an upper bound of 3.5 events at
95\%CL from neutrino fluxes. Note that if the number of events
integrated over energy is bounded by 3.5, then the same limit is
certainly applicable bin by bin in energy. Thus, using
Eq.~(\ref{numevents}) one obtains
\begin{equation}
\sum_{i,X} \int_{\Delta} dE_i\, N_A \, \frac{d\Phi_i}{dE_i} \,
\sigma_{i N \to X} (E_i) \, {\cal E} (E_i)\  < 3.5\ ,
\label{bound}
\end{equation}
at 95\% CL for some energy interval $\Delta$. Here, the sum over
$X$ takes into account charge and neutral current processes.  In a
logarithmic interval $\Delta$ where a single power law
approximation
\begin{equation}
\frac{d\Phi_i}{dE_i} \, \sigma_{i N \to X} (E_i) \, {\cal E} (E_i)
\sim E_i^{\alpha}
\end{equation}
is valid, a straightforward calculation shows that
\begin{equation}
\int_{\langle E\rangle e^{-\Delta/2}}^{\langle E\rangle e^{\Delta/2}}
\frac{dE_i}{E_i} \,
E_i\, \frac{d\Phi_i}{dE_i} \, \sigma_{i N \to X}  \, {\cal
E}  =   \langle \sigma_{i N\rightarrow X}\,
{\cal E} \, E_i\, d\Phi_i/dE_i \rangle \, \frac{\sinh \delta}{\delta}\, \Delta \,,
\label{sinsh}
\end{equation}
where $\delta=(\alpha+1)\Delta/2$ and $\langle A \rangle$ denotes the
quantity $A$ evaluated at the center of the logarithmic interval.  The
parameter $\alpha = 0.363 + \beta - \gamma$, where 0.363 is the
power law index of the SM neutrino cross
sections~\cite{Gandhi:1998ri} and $\beta$ and $-\gamma$ are the power
law indices (in the interval $\Delta$) of the exposure and flux
$d\Phi_i/dE_i$, respectively.  Since $\sinh \delta/\delta >1$, a
conservative bound may be obtained from Eqs.~(\ref{bound}) and
(\ref{sinsh}):
\begin{equation}
N_A\, \sum_{i,X} \langle \sigma_{i N\rightarrow X} (E_i)
\rangle \, \langle {\cal E} (E_i)\rangle\, \langle E_i
d\Phi_i/dE_i \rangle < 3.5/\Delta\ . \label{avg}
\end{equation}
By taking $\Delta =1$ as a likely interval in which the single power law
behavior is
valid (this corresponds  to one $e$-folding of energy), and setting
$\langle E_i d\Phi_i/dE_i \rangle = \frac{1}{6}
\langle E_\nu d\Phi_{\nu}/dE_\nu\rangle$ ($\Phi_{\nu} \equiv$ total neutrino
flux)
from Eq.~(\ref{avg}) it is straightforward to obtain 95\%CL upper limits on the neutrino flux. The limits are shown in Fig.~\ref{nufluxes}.

These bounds will be improved in the near future. In particular, each
site of PAO will observe $\sim 15$~km$^3$we sr of target mass around
$10^{19}$~eV~\cite{Capelle:1998zz}. Moreover, the study of radio pulses
from electromagnetic
showers created by neutrino interactions in ice would provide an increase in
the effective area up to $10^{4}$~km$^2$. A prototype of this technique is
the Radio Ice \v{C}erenkov Experiment (RICE)~\cite{Kravchenko:2001id}.
Similar concepts are  used by  the Goldstone Lunar ultrahigh energy
neutrino Experiment (GLUE),\footnote{This experiment observes
microwave \v{C}erenkov pulses
from electromagnetic showers induced by neutrinos
interacting in the Moon's rim~\cite{Gorham:2001aj}} the ANtartic Impulse
Transient Array (ANITA),\footnote{A balloon-borne payload that will circle the continent
of Antartica at ~35,000 meters, scanning the vast expanses of ice for
telltale pulses of radio emission generated by the neutrino
collisions [{\tt http://www.ps.uci.edu/~anita/}].} and the Fast On-orbit Recording of Transient Events (FORTE).\footnote{The FORTE satellite records bursts of electromagnetic waves arising from near the Earth's surface in the radio frequency range 30 to 300 MHz with a dual polarization antena~\cite{Lehtinen:2003xv}.} Existing limits on $\nu$-fluxes from RICE, GLUE and FORTE, as well as the
projected sensitivities of forthcoming experiments are collected in
Fig.~\ref{nufluxes}.

\subsection{Astronomy on Ice}
\label{AoI}

IceCube is, perhaps, the most promising route for neutrino
detection~\cite{Ahrens:2003ix}. This telescope will consist of 80
kilometer-length strings, each instrumented with 60 10-inch
photomultipliers spaced by 1.7 m. The deepest module is 2.4 km
below the ice surface. The strings are arranged at the apexes of
equilateral triangles 125 m on a side. The instrumented detector
volume is a cubic kilometer.\footnote{Extension of these aperture
is in the proposal stage~\cite{Halzen:2003fi}.} A surface air
shower detector, IceTop, consisting of 160 Auger-style
\v{C}erenkov detectors deployed over 1 km$^2$ above IceCube,
augments the deep-ice component by providing a tool for
calibration, background rejection and air-shower physics. Muons
can be observed from $10^{11}$~eV to $10^{18}$~eV. Cascades,
generated by $\nu_e,$ $\overline\nu_e$, $\nu_\tau,$ and $\overline
\nu_\tau$ can be observed above $10^{11}$~eV and reconstructed at
energies somewhat above $10^{13}$~eV. The angular resolution is
$\approx 0.7^\circ$ at TeV energies.

A variety of neutrino-emitting-sources have been proposed, such 
as supernovae~\cite{Waxman:2001kt} (see also~\cite{Tomas:2003xn}  
for a detailed study on directionality),
GRB fireballs~\cite{Dermer:2003zv,Razzaque:2002kb,Waxman:1997ti,Alvarez-Muniz:2003xi}, microquasars~\cite{Levinson:2001as}, $X$-ray 
binaries \cite{Anchordoqui:2002xu}, AGN~\cite{Neronov:2002xv}, FRI radio galaxies~\cite{Anchordoqui:2004eu}, 
etc.
As an example of the IceCube potential, in what follows we briefly
discuss its sensitivity to probe the neutron hypothesis of UHECRs
via observation of the antineutrino beam $n\rightarrow p+e^- +
\nuebar,$ expected from the Cygnus
direction~\cite{Anchordoqui:2003vc}. To this end, we first
estimate the background signal. As discussed in
Sec.~\ref{vanilla1}, the TeV $\gamma$-ray flux,
\begin{equation}
\frac{d\Fpho}{d\Epho} = 4.7 (\pm 2.1_{\rm stat} \pm 1.3_{\rm sys})
\times 10^{-13}
\left(\frac{E}{1~{\rm TeV}}\right)^{-1.9 (\pm 0.3_{\rm stat} \pm 0.3_{\rm sys})} \,\,{\rm ph\, cm}^{-2} {\rm s}^{-1} {\rm TeV}^{-1} \,\,,
\end{equation}
reported by HEGRA Collaboration~\cite{Aharonian:2002ij} in the
vicinity of Cygnus OB2 is likely due to hadronic processes. Since
$\pi^0$'s, $\pi^+$'s, and $\pi^-$'s are produced in equal numbers,
we expect two photons, two $\nue$'s, and four $\numu$'s per
$\pi^0$. On average, the photons carry one-half of the energy of
the pion, and the neutrinos carry one-quarter.  For $d\Fpi/d\Epi \propto 
\Epi^{-2},$ the energy-bins
$dE$ scale with these fractions, and we arrive
at
\begin{eqnarray}
\frac{d\Fpho}{d\Epho} (\Epho=\Epi/2) & = &
    4\,\frac{dF_{\pi}}{d\Epi}(\Epi) \, \nonumber \\
\frac{d\Fnue}{d\Enue} (\Enue= \Epi/4 ) & = &
    4\,\frac{dF_{\pi}}{d\Epi}(\Epi)\,,  \\
\frac{d\Fnumu}{d\Enumu} (\Enumu= \Epi/4 ) & = &
    8\,\frac{dF_{\pi}}{d\Epi}(\Epi)\,, \nonumber
\end{eqnarray}
for the fluxes at the source, where $\pi$ denotes any one of the
three pion charge-states. Terrestrial experiments (see
e.g.~\cite{Fukuda:1998mi}) have shown that $\numu$ and $\nutau$
are maximally mixed with a mass-squared difference $\sim
10^{-3}{\rm eV}^2$. This together with the known smallness of $|\langle\nu_e|\nu_3\rangle|^2,$
 implies that the $\numu$'s will partition
themselves equally between $\numu$'s and $\nutau$'s on lengths
large compared to the oscillation length $\lambda_{\rm osc}\sim
1.5 \times 10^{-3}\,({E_\nu}/{\rm PeV})$~pc. Here $\nu_3 \simeq (\nu_\mu + \nu_\tau)/\sqrt{2}$ 
is the third neutrino eigenstate. From these remarks,
one finds a nearly identical flux for each of the three neutrino
flavors ($j=e,\mu,\tau$), which is equal to~\cite{Alvarez-Muniz:2002tn}
\begin{equation}
\frac{d\Fnuj}{d\Enuj} (\Enuj= \Epho/2)
= 2\,\frac{d\Fpho}{d\Epho}(\Epho)\,.
\end{equation}

Although TeV neutrinos are copiously produced, because they are
weakly interacting the detection probability  on Earth is tiny,
about $10^{-6}$~\cite{Gaisser:1994yf}.  In particular, the
expected $\nu$ event rate at IceCube associated with the
unidentified HEGRA source is $< 1$~yr$^{-1}$ (D. Hooper, private
communication). Such an event rate is even smaller than the
atmospheric neutrino background.\footnote{For a year of running at
IceCube the expected background from atmospheric neutrinos (with
energy $\geq 1$~TeV) within $1^\circ$ circle centered in the
Cygnus direction (about $40^\circ$ below the horizon) is $< 1.5$
events.} Moreover, existing limits on TeV $\gamma$-ray fluxes in
this region of the sky are near the HEGRA
sensitivity.\footnote{Specifically,
the CASA-MIA experiment observed the Cygnus region over a 5 year period (1990-1995) 
and the collected data place a bound 
$F_\gamma(E_\gamma>115~{\rm TeV}) < 6.3 \times 10^{-15} \,{\rm ph}\, {\rm cm}^{-2}\, {\rm s}^{-1},$ 
at the 90\%CL~\cite{Borione:1996jw}. Extrapolation down to lower energies, assuming a spectrum 
$\propto E_\gamma^{-2},$ 
leads to $F_\gamma (E_\gamma > 1 {\rm TeV}) < 7.2 \times 10^{-13}\, {\rm ph}\, {\rm cm}^{-2} \,{\rm s}^{-1}.$
This is within a factor of 2 of the HEGRA measurements~\cite{Aharonian:2002ij,Rowell}, 
$F_\gamma(E_\gamma> 1 {\rm TeV}) \approx 4.5 \times 10^{-13} \,{\rm ph}\, {\rm cm}^{-2}\, {\rm s}^{-1}.$}
In light of this, we take as
background the atmospheric neutrino event rate, and so Poisson
statistics implies that a signal $\geq 3.7$ events is significant
at the 95\% CL.

Antineutrinos take only a very small part of the energy of the
parent neutron, typically $\sim 10^{-3}$. Hence, to estimate the
event rate of TeV antineutrinos at IceCube, the relevant nucleus
population at the source has an energy per nucleon $E_{N,{\rm
PeV}} \sim 1$~PeV. Nuclei with Lorentz factor $\sim 10^6$ are
synthesized in all supernovae. Hadronic interactions with the HII
population (density $< 30$~cm$^{-3}$~\cite{Butt:2003xc}) and
photodisintegration processes provide the flux of PeV neutrons. In
this energy regime, the target photons at photodisintegration
threshold energies are in the ultraviolet, $\sim 5$ eV. This
includes the entire emission spectrum of the O stars and about
60\% of photons from B stars (with average temperature 28,000~K).
From the photon emission rate $F_{\rm UV}$ the number density
$n_{\rm UV}$ at the surface of a sphere of radius R from the core
center is given by
\begin{equation}
\frac{1}{4} n_{\rm UV}\ c =\frac{F_{\rm UV}}{4\pi R^2}\ . \label{n}
\end{equation}
For the O-star population, the photon emission rate in the Lyman
region is found to be $F_{\rm L} \approx 10^{51}$ photons
s$^{-1}$~\cite{Knodlseder:2000vq}. The Lyman emission corresponds
to 60\% of the entire O star spectrum. Furthermore, as mentioned
above, 60\% of the B star spectrum is also active for
photodisintegration in this energy region, and the B star
population is about 20 times greater than that of the O
stars~\cite{Knodlseder:2000vq}. Now, from the H-R
diagram~\cite{Hanson:2003mf} one can infer that  the energy
luminosity of a B-star is about 0.1 that of an O star.
Additionally, the B star temperature is about 0.5 the O star
temperature, giving a number luminosity ratio of about 0.2. All in
all, for photodisintegration resulting in PeV nucleons, the
relevant photon density in the core of the Cygnus OB2 association
is $n_{\rm UV} \sim 230$ cm$^{-3}.$  The nucleus mean free path is
$\approx 35$~kpc, corresponding to a collision time $\tau=10^5$
yr. Thus, the collision rate for photodisintegration in the core
region is comparable to the hadronic interaction
rate.\footnote{This estimate takes into account a hadronic cross section,
$\sigma_{{\rm Fe}p} \sim A^{0.75} \, \sigma_{pp} \approx 6\times
10^{-25}$ cm$^2$, and the generous upper
limit~\cite{Torres:2003ur} of the nucleon density $\sim
30$~cm$^{-3}$~\cite{Butt:2003xc}.}

Since one is interested in neutrinos, it is still
necessary to compare the production rate for charged pions in the
hadronic case to the overall rate for generating neutrons. To
assess this ratio, we made use of available high energy event
simulations showing spectator nucleon and pion spectra
for Fe-N/$p$-N collisions at $10^{15}$ and $10^{16}$ eV~\cite{Knapp:1996fv}
(summarized in~\cite{Knapp:tv}). The pion rapidity spectra in the central
plateau are roughly energy independent, except for the widening
of the plateau with energy \mbox{--see~\cite{Ranft:2000mz}--,} whereas
there is a slow increase of spectator neutrons as one reaches the
region of interest ($E_N\approx 1$ PeV). Allowing for sizeable differences
in hadronic interaction models, the secondary populations are roughly
35\% $\pi^{\pm}$, 45\% $\gamma$, 10\% nucleons, and 10\% $K$~\cite{Knapp:tv}.
In the energy range yielding PeV neutrons, only about 30\% of the rapidity
plateau contributes charged pions above 2 TeV. Since only half the
nucleons are neutrons, we arrive at a ratio
\begin{equation}
\frac{\pi^{\pm}(> 2\ {\rm TeV})}{n (\sim\ {\rm PeV})}\approx 3
\label{ratio}
\end{equation}
in the hadronic interactions.

However, photodisintegration also takes place in the outer regions of the
OB association  as long as: {\it (i)}  the density of the optical photons
propagating out from the core allows a reaction time
which is smaller than the age of the cluster
$\sim 2.5$~Myr~\cite{Knodlseder:2001eb}
and {\it (ii)}
the diffusion front of the nuclei has passed the region in
question.  From Eq.~(\ref{n}) we estimate
an average photon density $n_{\rm UV} \agt 25$ cm$^{-3}$ out to 30~pc,
which gives a reaction time of $\approx 10^6$ yr. The diffusion time
($\sim 1.2$~Myr) is a bit smaller than the age of the cluster,
and somewhat higher than the reaction time,  allowing about 90\% of the nuclei to interact
during the lifetime of the source. Thus, the production rate of neutrons
via photodisintegration is amplified by a volume factor of 27 over the rate
in the 10 pc core. The net result of all this consideration is that the
PeV neutron population is about an order of magnitude greater than that of the
TeV charged pions~\cite{Anchordoqui:2003vc}.

With this in mind, we now discuss the prospects for a new multi-particle 
astronomy: neutrons as directional pointers + antineutrinos as inheritors 
of directionality. The basic formula that relates the neutron flux at the 
source ($d\Fn/d\En$) to the antineutrino flux observed at 
Earth ($d\Fnu/d\Enu$) is~\cite{Anchordoqui:2003vc}:
\begin{eqnarray}
\frac{d\Fnu}{d\Enu}(\Enu) & = &
\int d\En\,\frac{d\Fn}{d\En}(\En)
\left(1-e^{-\frac{D\,m_n}{\En\,\tbar}}\right)\,
\int_0^Q d\Enubar\,\frac{dP}{d\Enubar}(\Enubar) \nonumber \\
 & \times & \int_{-1}^1 \frac{d\cth}{2}
\;\delta\left[\Enu-\En\,\Enubar\,(1+\cth)/m_n\right]
\,.
\label{nuflux}
\end{eqnarray}
The variables appearing in Eq.~(\ref{nuflux}) are the antineutrino and
neutron energies in the lab ($\Enu$ and $\En$),
the antineutrino angle with respect to the direction of the
neutron  momentum, in the neutron rest-frame ($\thnu$),
and the antineutrino energy in the neutron rest-frame
($\Enubar$).  The last three variables are not observed
by a laboratory neutrino-detector, and so are integrated over.
The observable $\Enu$ is held fixed.
The delta-function relates the neutrino energy in the lab to the
three integration variables.\footnote{Note that $E_{\bar \nu} = \Gamma_n 
(\epsilon_{\bar \nu} + \beta \epsilon_{\bar \nu} \cos \overline 
\theta_{\bar \nu}) = 
E_n \epsilon_{\bar \nu} (1 + \cos \overline \theta_{\bar \nu})/ m_n,$ where $\Gamma_n = E_n/m_n$ is the Lorentz factor, and (as usual) $\beta \approx 1$ 
is the particle's velocity in units of $c$.} 
The parameters appearing in Eq.~(\ref{nuflux}) are the
neutron mass and rest-frame lifetime ($m_n$ and $\tbar$),
and the distance to the neutron source ($D$).
$dF_n/dE_n$ is the neutron flux at the source,
or equivalently, the neutron flux that would be observed
from the Cygnus region in the absence of neutron decay.
Finally, $dP/d\Enubar$ is the
normalized probability that the
decaying neutron in its rest-frame produces a $\nuebar$ with
energy $\Enubar$. Setting the beta-decay neutrino energy $\Enubar$ 
equal to its mean value
$\equiv \eps_0 \sim (m_n-m_p) [1 -m_e^2/(m_n-m_p)^2]/2 \approx 0.55$~MeV, we have
$dP/d\Enubar =\delta(\Enubar-\eps_0).$\footnote{The 
delta-function in the neutron frame gives rise to a flat
spectrum for the neutrino energy in the lab for fixed neutron
lab-energy $\En=\Gamma_n\,m_n$:
\[
\frac{dP}{d\Enu}=\int^1_{-1}\frac{d\cth}{2}\,
\left(\frac{d\Enubar}{d\Enu}\right)\,
\left(\frac{dP}{d\Enubar}\right) = \frac{1}{2\,\Gamma_n\,\eps_0 }
\,,
\]
with $0\le \Enu \le 2\,\Gamma_n\,\eps_0$.} Here, the maximum neutrino 
energy in the
neutron rest frame is $Q\equiv m_n -m_p-m_e= 0.71$~MeV,
and the minimum neutrino energy is zero in the massless limit.\footnote{The 
massless-neutrino approximation seems justifiable here:
even an eV-mass neutrino produced at rest in the neutron
rest-frame would have a lab energy of $m_\nu\,\Gamma_n\alt$~GeV,
below threshold for neutrino telescopes.}
The expression in parentheses in Eq.~(\ref{nuflux}) is the decay
probability for a neutron
with lab energy $\En$, traveling a distance $D$.
In principle, one should consider a source distribution,
and integrate over the volume $\int\,d^3 D$.
Instead, we will take $D$ to be the 1.7 kpc distance
from Earth to Cygnus OB2;
for the purpose of generating the associated neutrino flux,
this cannot be in error by too much.

Putting all this together, normalization to the observed ``neutron'' 
excess at $\sim
10^{18}$~eV leads to about 20 antineutrino events at IceCube per
year~\cite{Anchordoqui:2003vc}. Note that anti-neutrino flux at 1~TeV 
may also originate in the decay of 1~PeV neutrons from
sources whose spectrum cuts off at that energy, and hence are not subject 
to normalization by the anisotropy. Thus, this estimate may be regarded as 
very conservative. A direct TeV $\nuebar$ event in
IceCube will make a showering event, which, even if seen, provides
little angular resolution. In the energy region below 1~PeV,
IceCube will resolve directionality only for $\nu_\mu$ and  $\bar
\nu_{\mu}.$ Fortunately, neutrino oscillations rescue the signal.
Since the distance to the Cygnus region greatly exceeds the
$\nuebar$ oscillation length $\lambda_{\rm osc} \sim
10^{-2}({E_{\bar \nu}}/{\rm PeV})$~pc (taking the solar
oscillation scale $\delta m^2 \sim 10^{-5}{\rm eV}^2$), the
antineutrinos decohere in transit. The arriving antineutrinos are
distributed over flavors, with the muon antineutrino flux $F_{\bar
\nu_\mu}$ given by the factor ${1 \over 4}\, \sin^2
(2\,\theta_\odot) \simeq 0.20$ times the original $F_{\bar \nu_e}$
flux. The $\bar \nu_\tau$ flux is the same, and the $\bar \nu_e$
flux is 0.6 times the original flux. Here we have utilized for the
solar mixing angle the most recent SNO result $\theta_\odot \simeq
32.5^\circ$~\cite{Ahmed:2003kj}, along with maximal mixing for
atmospheric $\nu_\mu$-$\nu_\tau$ neutrinos and a negligible
$\nu_e$ component in the third neutrino eigenstate. All in all, 
for a year of running at IceCube, one conservatively expects 4
$\overline \numu$ showers with energies $\agt 1$~TeV to cluster
within $1^\circ$ of the source direction, comfortably above the
stated CL~\cite{Anchordoqui:2003vc}.

IceCube is not sensitive to TeV neutrinos from the Galactic
Center, as these are above the IceCube horizon, where atmospheric
muons will dominate over any signal. However, other
kilometer-scale neutrino detectors, such as those planned for the
Mediterranean Sea, may see the Galactic Center flux. 

In summary,
in a few years of observation, IceCube will attain  $5\sigma$
sensitivity for discovery of the Fe$ \rightarrow n \rightarrow
\nuebar\rightarrow \overline \nu_\mu$ cosmic beam, providing the
``smoking ice'' for the Galactic Plane neutron hypothesis.

\subsection{Probes of new physics beyond the electroweak scale}

It is intriguing --  and at the same time suggestive -- that the
observed flux of CRs beyond the GZK-energy is well matched by the
flux predicted for cosmogenic
neutrinos~\cite{Berezinsky:gzknu,Engel:2001hd,Kalashev:2002kx,Fodor:2003ph}. 
Of course, this is not
a simple coincidence: any proton flux beyond $E_{p\gamma_{\rm
CMB}}^{\rm th}$ is degraded in energy by photoproducing $\pi^0$
and $\pi^\pm$, with the latter in turn decaying to produce
cosmogenic neutrinos. The number of neutrinos produced in the GZK
chain reaction compensates for their lesser energy, with the
result that the cosmogenic flux matches well the observed CR flux
beyond $10^{20}$~eV. Recently, the prospect of an enhanced
neutrino cross section has been explored in the context of
theories with large compact dimensions.\footnote{A point worth
noting at this juncture: the neutrino-nucleon cross section can
also be significantly enhanced at center-of-mass energies $\agt
100$~TeV (within the SM) via electroweak instanton-induced
processes~\cite{Fodor:2003bn}.} In these theories, the extra
spatial dimensions are responsible for the extraordinary weakness
of the gravitational force, or, in other words, the extreme size
of the Planck mass~\cite{Arkani-Hamed:1998rs,Antoniadis:1998ig}.
For example, if spacetime is taken as a direct product of a
non-compact 4-dimensional manifold and a flat spatial $n$-torus
$T^n$ (of common linear size $2\pi r_c$), one obtains a definite
representation of this picture in which the effective
4-dimensional Planck scale, $M_{\rm Pl} \sim 10^{19}$~GeV, is
related to the fundamental scale of gravity, $M_D$, according to
$M^2_{\rm Pl} = 8 \pi \, M_D^{2 +n} \, r_c^n$. Within this
framework,  virtual graviton exchange would disturb high energy
neutrino interactions, and in principle, could increase the
neutrino interaction cross section in the atmosphere by orders of
magnitude beyond the SM value; namely $\sigma_{\nu N} \sim
[E_\nu/(10^{10})~{\rm
GeV}]$~mb~\cite{Nussinov:1998jt,Domokos:1998ry,Jain:2000pu}.
However, it is important to stress that a cross section of $\sim
100$~mb would be necessary to obtain consistency with observed
showers which start within the first 50 g/cm$^2$ of the
atmosphere. This is because Kaluza--Klein modes couple to neutral
currents and the scattered neutrino carries away 90\% of the
incident energy per interaction~\cite{Kachelriess:2000cb}.
Moreover, models which postulate strong neutrino interactions at
super-GZK energies also predict that moderately penetrating
showers should be produced at lower energies, where the
neutrino-nucleon cross section reaches a sub-hadronic size. Within
TeV scale gravity $\sigma_{\nu N}$ is likely to be sub-hadronic
near the energy at which the cosmogenic neutrino flux peaks, and
so moderately penetrating showers should be copiously
produced~\cite{Anchordoqui:2000uh}. Certainly, the absence of
moderately penetrating showers in the CR data sample should be
understood as a serious objection to the hypothesis of neutrino
progenitors of the super-GZK events.

Large extra dimensions still may lead to significant increases in
the neutrino cross section. If this scenario is true, we might
hope to observe black hole (BH) production (somewhat more masive
than $M_D$) in elementary particle collisions with center-of-mass
energies $\agt {\rm
TeV}$~\cite{Banks:1999gd,Giddings:2001bu,Dimopoulos:2001hw}. In
particular, BHs occurring very deep in the atmosphere (revealed as
intermediate states of ultrahigh energy neutrino interactions)
could trigger quasi-horizontal showers and be detected by cosmic
ray
observatories~\cite{Feng:2001ib,Anchordoqui:2001ei,Emparan:2001kf,Ringwald:2001vk,Dutta:2002ca}.
Additionally, neutrinos that traverse the atmosphere unscathed may
produce BHs through interactions in the ice or water and be
detected by neutrino
telescopes~\cite{Kowalski:2002gb,Alvarez-Muniz:2002ga}.
Interestingly, $\sigma_{\nu N \rightarrow {\rm BH}} \propto
M_D^{(-4+2n)/(1+n)}$. Therefore, the non-observation of the almost
guaranteed flux of cosmogenic neutrinos can be translated into
bounds on the fundamental Planck scale. For $n\ge 5$ extra spatial
dimensions compactified on $T^n,$ recent null results from CR
detectors lead to $M_D > 1.0 -
1.4$~TeV~\cite{Anchordoqui:2003jr,Anchordoqui:2003ur}. These
bounds are among the most stringent and conservative to date. In
the near future, PAO will provide more sensitive probes of
TeV-scale gravity and extra dimensions~\cite{Anchordoqui:2003ug}.
Certainly, the lack of observed deeply-penetrating showers can be
used to place more general, model-independent, bounds on
$\sigma_{\nu
N}$~\cite{Anchordoqui:2002vb,Berezinsky:kz,Morris:1993wg,Tyler:2000gt}.

Up to now we have only discussed how to set bounds on physics
beyond the SM. An actual discovery of new physics in cosmic rays
is a tall order because of large uncertainties associated with the
depth of the first interaction in the atmosphere, and the
experimental challenges of reconstructing cosmic air showers from
partial information. However, a similar technique to that employed
in discriminating between photon and hadron showers can be applied
to search for signatures of extra-dimensions. Specifically, if an
anomalously large quasi-horizontal deep shower rate is found, it
may be ascribed to either an enhancement of the incoming neutrino
flux, or an enhancement in the neutrino-nucleon cross section.
However, these two possibilities may be distinguished by
separately binning events which arrive at very small angles to the
horizontal, the so-called ``Earth-skimming''
events~\cite{Feng:2001ue,Fargion:2000iz}.  An enhanced flux will
increase both quasi-horizontal and Earth-skimming event rates,
whereas a large BH cross section suppresses the latter, because
the hadronic decay products of BH evaporation do not escape the
Earth's crust~\cite{Anchordoqui:2001cg}. For a more detailed
discussion of neutrino interactions in the Earth atmosphere within
TeV scale gravity scenarios see e.g.~\cite{Anchordoqui:2002hs}.

\section{Any Given Sunday: Countdown to Discovery}

With data now at hand, not only there are several interesting,
plausible theoretical models within the standard astrophysical
agenda to explain all the CRs detected so far,  but there could
indeed be too many. Perhaps yet unexpected degeneracy problems
will appear, even with the forthcoming data of the Pierre Auger
Observatory, a topic which till now has not been a subject of
debate. In our view, Occam's razor imposes that {\it all} standard
astrophysical models be eliminated before embarking in the
consequences of explanations involving physics beyond the standard
scenarios. However, should this be the case, the prospect for
encountering a profound scientific revolution are endless. The
puzzle of UHECRs may have something to say about issues as
fundamental as local Lorentz invariance: On the one hand, the
absence of photo-pion production above the GZK-limit would imply
no cosmogenic neutrino flux and possibly undeflected pointing of
the primary back to its source. On the other hand, a significant
correlation of TeV-antineutrinos with directional signals at EeV
energies will validate Special Relativity to an unprecedented
boost factor of $\Gamma \approx 10^{9},$ several orders of
magnitude beyond current limits. Additionally, contrasting the
observed quasi-horizontal  neutrino flux with the expected
neutrino flux can help constrain TeV-scale gravity interactions
and improve current bounds on the fundamental Planck scale. An
optimist might even imagine the discovery of microscopic BHs, the
telltale signature of the Universe's unseen dimensions. At the
time of writing, new data is being collected at Pampa Amarilla.
Whatever happens, in T--2 years we shall witness the lift-off of a new era of
cosmic ray physics.

\section*{Acknowledgments}

During the last few years, we have benefitted from discussions
with several colleagues and collaborators, among them, Felix
Aharonian, John Bahcall, Paula Benaglia, Peter Biermann, Elihu
Boldt, Murat Boratav, Yousaf Butt, Anal\'{\i}a Cillis, Jorge Combi, Jim Cronin, Tom
Dame, Chuck Dermer, Seth Digel, Eva Domingo, Susan and Gabor Domokos, Mar\'{\i}a Teresa
Dova, Luis Epele, Jonathan Feng, Francesc Ferrer, Haim Goldberg,
Francis Halzen, Tim Hamilton, Carlos Hojvat, Dan Hooper, Michael
Kachelriess, Mike Loewenstein, Tom Mc Cauley, Gustavo Medina
Tanco, Pran Nath, Carlos Nu\~nez, Tom Paul, Santiago Perez
Bergliaffa, Brian Punsly, Olaf Reimer, Steve Reucroft, Andreas
Ringwald, Gustavo Romero, Esteban Roulet, Subir Sarkar, Sergio
Sciutto, Dmitri Semikoz, Al Shapere, Paul Sommers, Guenter Sigl, Todor Stanev,
John Swain, Tomasz Taylor, Dave Thompson, Peter Tinyakov, Alan Watson, 
Tom Weiler, and Allan Widom. 
We specially thank Frank Scherb for valuable information on the early 
history of cosmic ray detectors. We would like to thank 
several of the 
collegues mentioned above as well as Ray
Protheroe  for allowing us to use some figures from their papers
in this review. We thank Julianna Gianni for inspiration. The work
of DFT was performed under the auspices of the U.S. Department of
Energy (NNSA) by University of California's LLNL under contract
No. W-7405-Eng-48. The work of LAA has been partially supported by
the US National Science Foundation (NSF) under grant No.
PHY-0140407.

\end{document}